\documentclass[]{aastex701}

\usepackage{bm}
\usepackage{amsmath}
\usepackage{comment}

\newcommand\ltsima{$\; \buildrel <\over\sim \;$}
\newcommand\simlt{\lower.5ex\hbox{\ltsima}}
\newcommand\gtsima{$\; \buildrel >\over\sim \;$}
\newcommand\simgt{\lower.5ex\hbox{\gtsima}}

\shorttitle{KMT-2024-BLG-0211 / KMT-2024-BLG-1522}
\shortauthors{Nunota et al.}

\graphicspath{{./}{figures/}}
\begin{document}

\title{Testing KMTNet--PRIME Optical--Near-Infrared Source-Color Constraints in KMT-2024-BLG-0211 and KMT-2024-BLG-1522}

% =============================================================================
% Lead Authors
% =============================================================================

\author[0009-0005-3414-455X]{Kansuke Nunota}
\affiliation{
Department of Earth and Space Science,
Graduate School of Science,
The University of Osaka,
Toyonaka, Osaka 560-0043, Japan
}
\email{nunota@iral.ess.sci.osaka-u.ac.jp}

\author[0000-0001-9481-7123]{Jennifer C. Yee}
\affiliation{
Center for Astrophysics $|$ Harvard \& Smithsonian,
60 Garden St., Cambridge, MA 02138, USA
}
\email{jyee.astro@gmail.com}

\author[0000-0002-4035-5012]{Takahiro Sumi}
\affiliation{
Department of Earth and Space Science,
Graduate School of Science,
The University of Osaka,
Toyonaka, Osaka 560-0043, Japan
}
\email{sumi@ess.sci.osaka-u.ac.jp}

\author{Ryusei Hamada}
\affiliation{
Department of Earth and Space Science,
Graduate School of Science,
The University of Osaka,
Toyonaka, Osaka 560-0043, Japan
}
\email{hryusei@iral.ess.sci.osaka-u.ac.jp}

\author{Ian A. Bond}
\affiliation{
School of Mathematical and Computational Sciences,
Massey University,
Auckland 0745, New Zealand
}
\email{i.a.bond@massey.ac.nz}

\author{Andrew Gould}
\affiliation{
Department of Astronomy,
Ohio State University,
140 W. 18th Ave.,
Columbus, OH 43210, USA
}
\email{gould.34@osu.edu}

\author[0000-0001-6000-3463]{Weicheng Zang}
\affiliation{Department of Astronomy, Westlake University, Hangzhou 310030, Zhejiang Province, China}
\affiliation{
Center for Astrophysics $|$ Harvard \& Smithsonian,
60 Garden St., Cambridge, MA 02138, USA
}
\email{zangweicheng@westlake.edu.cn}

\nocollaboration{all}

% =============================================================================
% PRIME Collaboration
% =============================================================================

% NOTE:
% Kansuke Nunota, Takahiro Sumi, Ryusei Hamada, and Ian A. Bond
% are already listed above as lead authors and are not duplicated here.

\author{David P.~Bennett}
\affiliation{
Code 667,
NASA Goddard Space Flight Center,
Greenbelt, MD 20771, USA
}
\affiliation{
Department of Astronomy,
University of Maryland,
College Park, MD 20742, USA
}
\email{bennettd@umd.edu}

\author{Aparna Bhattacharya}
\affiliation{
Code 667,
NASA Goddard Space Flight Center,
Greenbelt, MD 20771, USA
}
\affiliation{
Department of Astronomy,
University of Maryland,
College Park, MD 20742, USA
}
\email{abhatta5@umd.edu}

\author{Kotaro Daimon}
\affiliation{
Department of Earth and Space Science,
Graduate School of Science,
The University of Osaka,
Toyonaka, Osaka 560-0043, Japan
}
\email{u407614b@ecs.osaka-u.ac.jp}

\author{Yuki Hirao}
\affiliation{
Institute of Astronomy,
Graduate School of Science,
The University of Tokyo,
2-21-1 Osawa,
Mitaka, Tokyo 181-0015, Japan
}
\email{hirao@iral.ess.sci.osaka-u.ac.jp}

\author{Stela Ishitani Silva}
\affiliation{
Department of Physics,
The Catholic University of America,
Washington, DC 20064, USA
}
\affiliation{
Code 667,
NASA Goddard Space Flight Center,
Greenbelt, MD 20771, USA
}
\email{stela.ishitani@gmail.com}

\author{Shuma Makida}
\affiliation{
Department of Earth and Space Science,
Graduate School of Science,
The University of Osaka,
Toyonaka, Osaka 560-0043, Japan
}
\email{makida@iral.ess.sci.osaka-u.ac.jp}

\author{Shota Miyazaki}
\affiliation{
Institute of Space and Astronautical Science,
Japan Aerospace Exploration Agency,
3-1-1 Yoshinodai,
Chuo, Sagamihara,
Kanagawa 252-5210, Japan
}
\email{miyazaki@ir.isas.jaxa.jp}

\author{Tutumi Nagai}
\affiliation{
Department of Earth and Space Science,
Graduate School of Science,
The University of Osaka,
Toyonaka, Osaka 560-0043, Japan
}
\email{nagai@iral.ess.sci.osaka-u.ac.jp}

\author{Seiya Nakayama}
\affiliation{
Department of Earth and Space Science,
Graduate School of Science,
The University of Osaka,
Toyonaka, Osaka 560-0043, Japan
}
\email{nakayama@iral.ess.sci.osaka-u.ac.jp}

\author{Ryo Ogawa}
\affiliation{
Department of Earth and Space Science,
Graduate School of Science,
The University of Osaka,
Toyonaka, Osaka 560-0043, Japan
}
\email{rogawa@iral.ess.sci.osaka-u.ac.jp}

\author{Ryunosuke Oishi}
\affiliation{
Department of Earth and Space Science,
Graduate School of Science,
The University of Osaka,
Toyonaka, Osaka 560-0043, Japan
}
\email{oishi@iral.ess.sci.osaka-u.ac.jp}

\author{Hideaki Ose}
\affiliation{
Department of Earth and Space Science,
Graduate School of Science,
The University of Osaka,
Toyonaka, Osaka 560-0043, Japan
}
\email{ose@iral.ess.sci.osaka-u.ac.jp}

\author[0000-0001-5069-319X]{Nicholas J. Rattenbury}
\affiliation{
Department of Physics,
University of Auckland,
Private Bag 92019,
Auckland, New Zealand
}
\email{n.rattenbury@auckland.ac.nz}

\author[0000-0002-1228-4122]{Yuki K. Satoh}
\affiliation{
Department of Electrical and Information Engineering,
Nippon Institute of Technology,
Shiraoka, Saitama, Japan
}
\email{sato.yuki@nit.ac.jp}

\author{Daisuke Suzuki}
\affiliation{
Department of Earth and Space Science,
Graduate School of Science,
The University of Osaka,
Toyonaka, Osaka 560-0043, Japan
}
\email{dsuzuki@ess.sci.osaka-u.ac.jp}

\author{Takuto Tamaoki}
\affiliation{
Department of Earth and Space Science,
Graduate School of Science,
The University of Osaka,
Toyonaka, Osaka 560-0043, Japan
}
\email{tamaoki@iral.ess.sci.osaka-u.ac.jp}

\author[0000-0002-6510-0681]{Motohide Tamura}
\affiliation{
Astrobiology Center,
2-21-1 Osawa,
Mitaka-shi, Tokyo 181-8588, Japan
}
\affiliation{
Department of Astronomy,
University of Tokyo,
7-3-1 Hongo,
Bunkyo-ku, Tokyo 113-0033, Japan
}
\email{motohide.tamura@nao.ac.jp}

\author{Sean K. Terry}
\affiliation{
Code 667,
NASA Goddard Space Flight Center,
Greenbelt, MD 20771, USA
}
\affiliation{
Department of Astronomy,
University of Maryland,
College Park, MD 20742, USA
}
\email{skterry@umd.edu}

\author{Chihiro Ueda}
\affiliation{
Department of Earth and Space Science,
Graduate School of Science,
The University of Osaka,
Toyonaka, Osaka 560-0043, Japan
}
\email{ueda@iral.ess.sci.osaka-u.ac.jp}

\author{Aikaterini Vandorou}
\affiliation{
Code 667,
NASA Goddard Space Flight Center,
Greenbelt, MD 20771, USA
}
\affiliation{
Department of Astronomy,
University of Maryland,
College Park, MD 20742, USA
}
\email{katievan@umd.edu}

\author[0000-0001-7692-0581]{Hibiki Yama}
\affiliation{
Department of Astronomy,
Graduate School of Science,
Kyoto University,
Kitashirakawa Oiwake-cho,
Sakyo-ku, Kyoto 606-8502, Japan
}
\email{yama@kusastro.kyoto-u.ac.jp}

\collaboration{all}{(PRIME Collaboration)}

% =============================================================================
% KMTNet Science Team
% =============================================================================

\author[0000-0003-3316-4012]{Michael D. Albrow}
\affiliation{
University of Canterbury,
School of Physical and Chemical Sciences,
Private Bag 4800,
Christchurch 8020, New Zealand
}
\email{michael.albrow@canterbury.ac.nz}

\author[0000-0001-6285-4528]{Sun-Ju Chung}
\affiliation{
Korea Astronomy and Space Science Institute,
Daejeon 34055, Republic of Korea
}
\email{sjchung@kasi.re.kr}

\author[0000-0002-2641-9964]{Cheongho Han}
\affiliation{
Department of Physics,
Chungbuk National University,
Cheongju 28644, Republic of Korea
}
\email{cheongho@astroph.chungbuk.ac.kr}

\author[0000-0002-9241-4117]{Kyu-Ha Hwang}
\affiliation{
Korea Astronomy and Space Science Institute,
Daejeon 34055, Republic of Korea
}
\email{kyuha@kasi.re.kr}

\author[0000-0002-0314-6000]{Youn Kil Jung}
\affiliation{
Korea Astronomy and Space Science Institute,
Daejeon 34055, Republic of Korea
}
\affiliation{
National University of Science and Technology (UST),
Daejeon 34113, Republic of Korea
}
\email{ykjung21@kasi.re.kr}

\author[0000-0001-9823-2907]{Yoon-Hyun Ryu}
\affiliation{
Korea Astronomy and Space Science Institute,
Daejeon 34055, Republic of Korea
}
\email{yhryu@kasi.re.kr}

\author[0000-0002-4355-9838]{In-Gu Shin}
\affiliation{
School of Science,
Westlake University,
Hangzhou, Zhejiang 310030, China
}
\email{ingushin@gmail.com}

\author[0000-0003-1525-5041]{Yossi Shvartzvald}
\affiliation{
Department of Particle Physics and Astrophysics,
Weizmann Institute of Science,
Rehovot 7610001, Israel
}
\email{yossishv@gmail.com}

\author[0000-0003-0626-8465]{Hongjing Yang}
\affiliation{
School of Science,
Westlake University,
Hangzhou, Zhejiang 310030, China
}
\affiliation{
Department of Astronomy,
Tsinghua University,
Beijing 100084, China
}
\email{hongjing.yang@qq.com}

% =============================================================================
% KMTNet Operations Team
% =============================================================================

\author{Dong-Jin Kim}
\affiliation{
Korea Astronomy and Space Science Institute,
Daejeon 34055, Republic of Korea
}
\email{keaton03@kasi.re.kr}

\author[0000-0003-0043-3925]{Chung-Uk Lee}
\affiliation{
Korea Astronomy and Space Science Institute,
Daejeon 34055, Republic of Korea
}
\email{leecu@kasi.re.kr}

\author[0000-0002-6982-7722]{Byeong-Gon Park}
\affiliation{
Korea Astronomy and Space Science Institute,
Daejeon 34055, Republic of Korea
}
\email{bgpark@kasi.re.kr}

\collaboration{all}{(The KMTNet Collaboration)}

\begin{abstract}
We present analyses of two 2024 microlensing events, KMT-2024-BLG-0211 and KMT-2024-BLG-1522, jointly observed by KMTNet in the optical and PRIME in the near-infrared. For KMT-2024-BLG-0211, the KMTNet--PRIME $I-H$ color provides a useful constraint on the angular source radius for a short-timescale finite-source event with a giant source. The event is consistent with a low-mass stellar lens, although the weak parallax constraint leaves the lens mass and distance uncertain.
For KMT-2024-BLG-1522, we compare the results obtained from  $V-I$, $I-H$, $V-H$, and $J-H$ color constraints. 
The microlensing parameters are nearly identical among the four analyses, yielding a robust binary-lens solution with nearly equal masses.
However, the inferred source properties and lens physical parameters depend on the adopted source color because the different color estimates imply different angular source radii. 
Comparing the $(V-I)_{\rm KMT}$ versus $(I-H)_{\rm KMT,PRIME}$ plane shows that the inferred source colors lie off empirical color--color relations but within the scatter of observed stars. The prevalence of such deviations should be tested with a larger sample of KMTNet--PRIME events and this specific case can be further tested with future adaptive-optics follow-up, which can directly measure the lens--source relative proper motion and lens flux.
\end{abstract}


\section{Introduction} \label{sec:intro}

Gravitational microlensing is a unique observational method that can detect a wide variety of dark and faint compact objects, including exoplanets \citep{mao91, bon04}, free-floating planets \citep{mro17,gou22,sum23,kos23}, and even black holes \citep{sah22, lam22}. 
Optical microlensing surveys such as the Microlensing Observations in Astrophysics survey (MOA; \citealt{bon01,sum03}), the Optical Gravitational Lensing Experiment (OGLE; \citealt{uda15}), and the Korea Microlensing Telescope Network (KMTNet; \citealt{kim16}) have been highly successful, discovering more than 2,000 microlensing events per year.

Toward the Galactic center, interstellar extinction becomes increasingly severe at low Galactic latitudes, so previous optical microlensing observations have generally been limited to fields at $|b| \gtrsim 1^\circ$ \citep{mro19,nun25gal}.
The Prime Focus Infrared Microlensing Experiment \citep[PRIME;][]{sum25} was initiated to overcome this limitation by conducting a near-infrared microlensing survey, opening access to inner bulge regions where one can investigate the environmental dependence of planet occurrence and improve our understanding of Galactic structure through measurements of the microlensing event rate.

The scientific potential of PRIME extends beyond simply opening up low-latitude bulge fields that are inaccessible to optical surveys.
Its synergy with optical surveys can also play an important role in the physical characterization of individual microlensing events.
Finite-source effects \citep{wit94} are often detected in microlensing light curves, allowing a measurement of the normalized source size, $\rho \equiv \theta_*/\theta_{\rm E}$, where $\theta_*$ is the angular radius of the source star and $\theta_{\rm E}$ is the angular Einstein radius.
Since $\theta_{\rm E}$ is a fundamental quantity that reflects both the lens mass and the lens--source geometry, a reliable estimate of $\theta_*$ is essential for converting light-curve parameters into physical parameters of the lens system.

In general, $\theta_*$ is estimated from the source color and magnitude measured from the light curve, together with empirical color--surface brightness relations \citep{yoo04, ker04}.
In conventional optical microlensing observations, the source color has often been measured using a primary $I$-band dataset supplemented by a smaller number of $V$-band observations.
However, toward the Galactic bulge, the $V$ band is much more strongly affected by reddening and extinction, and in some cases the source is difficult to detect at all in $V$ \citep[e.g.,][]{ben08}.
In addition, the observing cadence in $V$ is usually much lower than in $I$, so the magnified phase of the event is often not sufficiently sampled.
As a result, especially in highly extincted regions, the source flux in $V$ can be poorly constrained, leaving substantial uncertainty in the estimate of $\theta_*$.

Near-infrared photometry from PRIME is much less affected by extinction and can be combined with optical survey data to construct optical--NIR colors such as $(I-H)$.
Such colors may provide useful constraints on the source properties, particularly for highly extincted events.
At the same time, the use of multiple color combinations raises a practical question: whether different source-color estimates lead to mutually consistent values of $\theta_*$ and hence consistent lens physical parameters.

This question has not yet been thoroughly tested with actual microlensing events observed jointly by optical and near-infrared surveys.
It is therefore important not only to examine whether optical--NIR colors improve the determination of $\theta_*$, but also to test whether source properties inferred from different color indices are mutually consistent.

In this work, we analyze two 2024 microlensing events jointly observed by KMTNet and PRIME: KMT-2024-BLG-0211 and KMT-2024-BLG-1522 (hereafter KB240211 and KB241522, respectively).
KB240211 is a highly extincted event, for which we test whether the combination of KMTNet $I$ and PRIME $H$, i.e., the $(I-H)$ color, can provide a useful constraint on the source properties when optical data alone are insufficient.
KB241522 occurred in a region of more moderate extinction and allows us to compare source and lens properties inferred from several color constraints, including KMTNet $(V-I)$, KMTNet--PRIME $(I-H)$, $(V-H)$, and PRIME $(J-H)$.

Furthermore, in order to evaluate how the choice of color index affects the inferred lens physical parameters, it is important to propagate consistently both the uncertainty associated with determining the source flux from the light curve and the uncertainty involved in deriving $\theta_*$ from the source color and magnitude.
In this study, we therefore use a Bayesian framework that incorporates the uncertainties associated with both the source flux and $\theta_*$, enabling a self-consistent assessment of how different source-color constraints affect the inferred physical properties of the lens system.

This paper is organized as follows.
Section~\ref{sec:obs} first describes the observations of the two events and the corresponding data reduction procedures.
Section~\ref{sec:cmd} then presents the construction of the color--magnitude diagrams, which are used to estimate the reddening and extinction toward the source fields.
In Section~\ref{sec:lc}, we introduce the light-curve modeling and Bayesian framework adopted to infer the microlensing and physical parameters.
The resulting constraints for KB240211 and KB241522 are presented in Section~\ref{sec:result}.
We discuss the implications of these results in Section~\ref{sec:disc} and summarize our conclusions in Section~\ref{sec:concl}.

\begin{figure}[t]
    \centering
    % \fbox{\rule{0pt}{2.2in} \rule{0.9\textwidth}{0pt}}
    \includegraphics[width=0.7\textwidth]{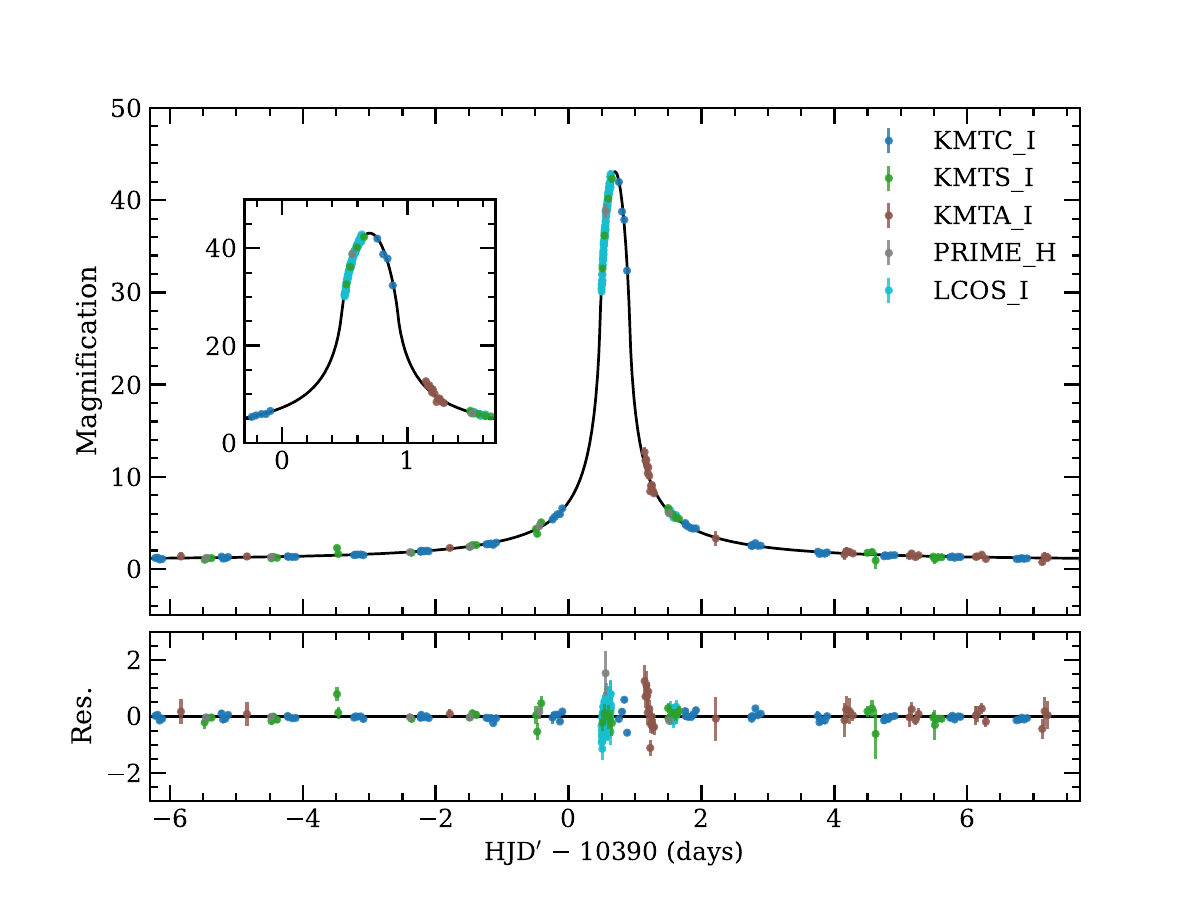}
    \caption{
    Light curve of KB240211.
    The upper panel shows the KMTNet $I$-band photometry from KMTA, KMTC, and KMTS, together with the LCOS $I$-band and PRIME $H$-band data.
    The black curve represents the best-fit finite-source point-lens model, and the lower panel shows the residuals from this model.
    }
    \label{fig:lc_kb240211}
\end{figure}
\begin{figure}[t]
    \centering
    % \fbox{\rule{0pt}{2.2in} \rule{0.9\textwidth}{0pt}}
    \includegraphics[width=0.9\textwidth]{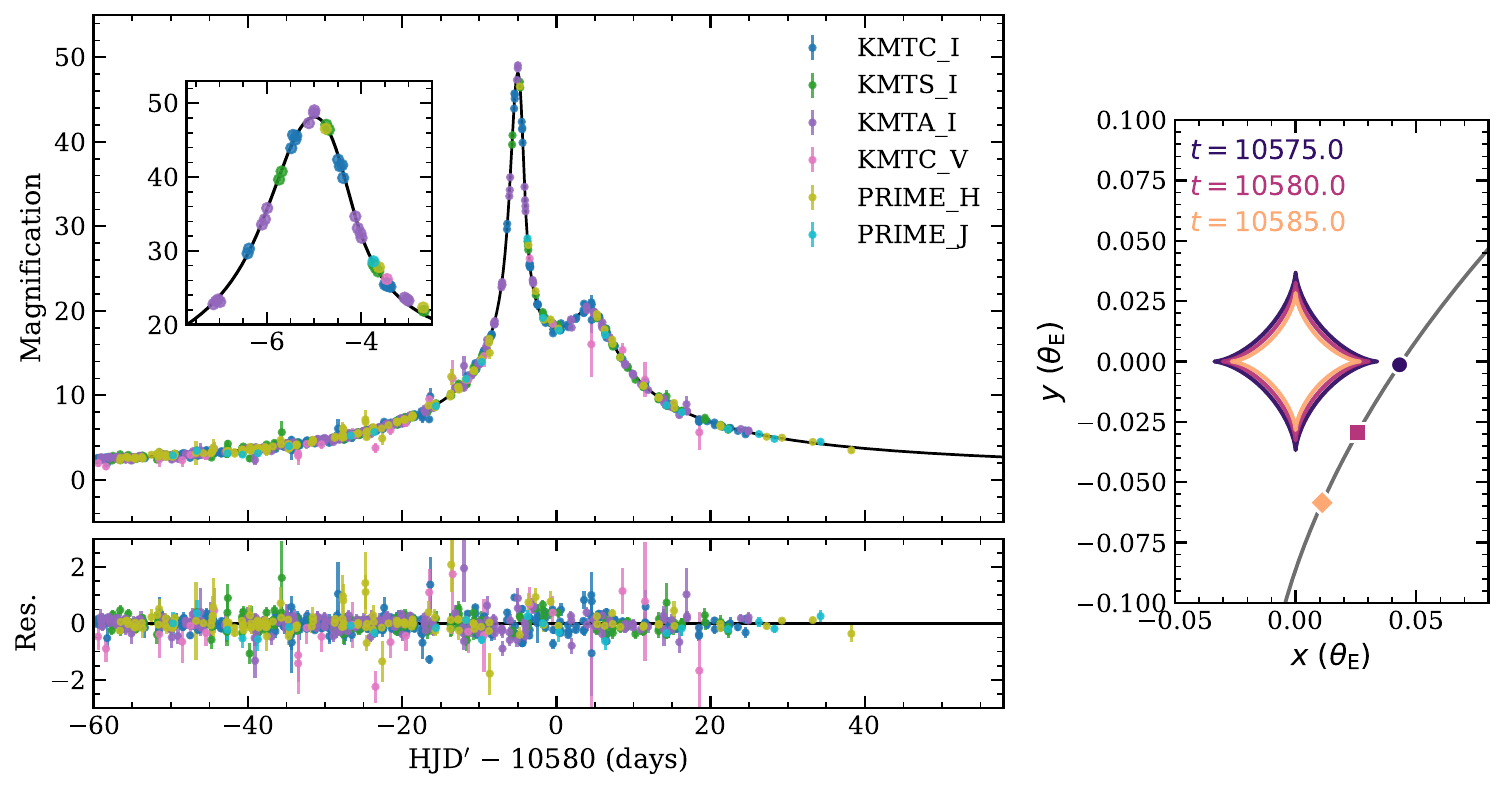}
    \caption{
        Light curve of KB241522.
        The upper panel shows the KMTNet $I$- and $V$-band photometry together with the PRIME $H$- and $J$-band data.
        The black curve represents the best-fit binary-lens model.
        The inset highlights the sharp peak region; the KMTA points near the peak lie slightly above the best-fit model but are consistent with the scatter of the surrounding KMTA data.
    }
    \label{fig:lc_KB241522}
\end{figure}
\section{Observations and Data Reduction} \label{sec:obs}
The microlensing events KB240211 and KB241522 were observed by two major microlensing survey projects: the Korea Microlensing Telescope Network \citep[KMTNet;][]{kim16} and the Prime Focus Infrared Microlensing Experiment \citep[PRIME;][]{sum25}.

The KMTNet conducts observations toward the Galactic bulge using three identical $1.6~\rm m$ telescopes distributed across three continents in the Southern Hemisphere: the Cerro Tololo Inter-American Observatory (CTIO) in Chile, the South African Astronomical Observatory (SAAO) in South Africa, and the Siding Spring Observatory (SSO) in Australia. Each telescope is referred to individually as KMTC, KMTS, and KMTA.
The survey has been in full operation since 2016. 
Observations are primarily conducted in the Cousins $I$ band, with occasional exposures taken in the $V$ band to provide color information and to aid in the characterization of source stars.

The PRIME telescope is a 1.8~m near-infrared microlensing survey telescope located at SAAO. 
It conducts observations toward the highly extincted central Galactic bulge regions in the $H$ band (1.49–1.78~$\mu$m), with additional observations in the $J$ band (1.17–1.33~$\mu$m) to provide color information.
The survey commenced regular observations in 2024, while the event detection system became fully operational in 2025. 
Consequently, the microlensing events observed in 2024 were identified in 2025.
For the events analyzed in this work, the KMTNet $I$-band and PRIME $H$-band observations have typical cadences of approximately 1~hr and 1.5~hr, respectively. Additional KMTNet $V$-band observations are obtained about once per night, while PRIME $J$-band observations are obtained approximately once every three nights to provide color information.

The photometric data used for the light-curve analysis were reduced independently by each survey using their own difference image analysis (DIA; \citealt{tom96, ala98}) pipelines.
The KMTNet data were processed with the \texttt{pySIS} \citep{alb09,yan24} pipeline, while the PRIME data were reduced using a PRIME-adapted version of the DIA pipeline originally developed by \citet{bon01} for the Microlensing Observations in Astrophysics survey (MOA; \citealt{bon01,sum03}) project.

The KMTC data were calibrated to the OGLE-III photometric system \citep{szy11} and used as the reference for this analysis.
The PRIME $H$- and $J$-band data were calibrated using the VVV Infrared Astrometric Catalogue version 2 (VIRAC2; \citealt{smi25}) by matching stars between the PRIME and VIRAC2 catalogs.
The photometric zero-point offsets were determined separately for each event and band by adopting the median magnitude difference between the matched stars. For KB240211, we obtained ${\rm mag_{0,H}} = 23.74$ and ${\rm mag_{0,J}} = 23.66$, while for KB241522, we derived ${\rm mag_{0,H}} = 23.84$ and ${\rm mag_{0,J}} = 23.81$.

\subsection{KB240211}
KB240211 was first discovered by the KMTNet Alert-Finder system \citep{kim18} on 2024 March 20. 
Independently, the same microlensing event was also detected by PRIME in a post-season review of the 2024 data and designated PRIME-2024-BLG-0021.
The event is located at equatorial coordinates $\rm{(RA, Dec)_{J2000}} = (17^{\rm h}45^{\rm m}16.16^{\rm s}, -27^\circ21\arcmin31.64\arcsec)$ and Galactic coordinates $(l, b) = (1.31, 0.89)$, lying in the KMTNet BLG18 field and the PRIME GB76 field.

As will be shown in Section~\ref{sec:cmd}, this field suffers from severe extinction, with an $I$-band extinction of $A_I = 5.33$ and a reddening of $E(I-H) = 4.27$.
Because of this extremely high extinction, the $V$-band data are too faint to be useful. 
In addition, PRIME had not yet begun $J$-band observations at the beginning of the 2024 season, so no $J$-band data are available for this event. 
The KMTNet \texttt{HighMagFinder} system \citep{yan22} issued an alert on 2024 March 20 at 12:03 UT ($\text{HJD}' = 10390.0035$) that KMT-2024-BLG-0211 could peak at high magnifications. Following this alert, $I$-band follow-up observations were conducted during the peak using a 1.0\,m telescope of the LCOGT at SAAO (LCOS).
Consequently, the light-curve analysis is based on five datasets: KMTA-$I$, KMTC-$I$, KMTS-$I$, LCOS-$I$, and PRIME-$H$.

The light curve of KB240211 is shown in Figure~\ref{fig:lc_kb240211}.  
The event exhibits a clear finite-source point-lens (FSPL) signature, 
peaking at $\mathrm{HJD}' \equiv \mathrm{HJD} - 2460000 \simeq 390.7$.
Although the event duration is relatively short, with a magnification lasting less than $\sim10$~days, 
the three KMTNet observatories provide continuous coverage of the brightening phase.  
In addition, several PRIME $H$-band data points were obtained around the peak, 
which are sufficient to constrain the source brightness.

\subsection{KB241522}
KB241522 was first alerted by KMTNet on 2024 June 25.  
It was later detected independently in a post-season review of the 2024 data and designated PRIME-2024-BLG-0160.
The event is located at equatorial coordinates $\rm(RA, Dec)_{J2000} = (17^{\rm h}41^{\rm m}44.90^{\rm s}, -26^\circ17\arcmin58.31\arcsec)$
and Galactic coordinates $(l, b) = (1.80, 2.11)$, lying in the KMTNet BLG18 field and the PRIME GB75 field. 

As will be shown in Section~\ref{sec:cmd}, the extinction toward this field is moderate, with an $I$-band extinction of $A_I = 3.34$ and a reddening of $E(V-I) = 2.35$.
Because the extinction is not as severe as in the KB240211 field, reliable $V$-band data are available.
By this time, PRIME had also begun $J$-band observations, enabling both $H$- and $J$-band photometry for this event. 
Consequently, the light-curve analysis is based on eight datasets: KMTA-$I$, KMTA-$V$, KMTC-$I$, KMTC-$V$, KMTS-$I$, KMTS-$V$, PRIME-$H$,  PRIME-$J$.

The light curve of KB241522 is presented in Figure~\ref{fig:lc_KB241522}.  
This event exhibits a clear binary-lens signature, characterized by a sharp peak at 
$\mathrm{HJD}' \simeq 574$ and an additional bump around $\mathrm{HJD}' \simeq 584$.  
The overall magnification lasts for more than a month, 
providing excellent coverage from all datasets, including the KMTNet optical and PRIME near-infrared observations.

% \begin{figure}[t]
%     \includegraphics[width=1.05\textwidth]{cmd_comb.pdf}
%     \caption{
%     Color--magnitude diagrams used to determine the reddening and extinction toward KB240211 and KB241522.
%     The left panel shows the $(I-H, I)$ CMD for KB240211, while the three right panels show the $(V-I, I)$, $(I-H, I)$, and $(J-H, H)$ CMDs for KB241522.
%     Black points are stars in the matched KMTNet--PRIME catalogs, and green points are stars in the VVV--KMTNet catalog used as an external check.
%     Red and blue points indicate the measured RCG centroid and source position, respectively, and the cyan cross marks the RCG centroid estimated from the VVV--KMTNet CMD.
%     }
%     \label{fig:cmd}
% \end{figure}

\begin{figure}[t]
    \includegraphics[width=0.3\textwidth]
    {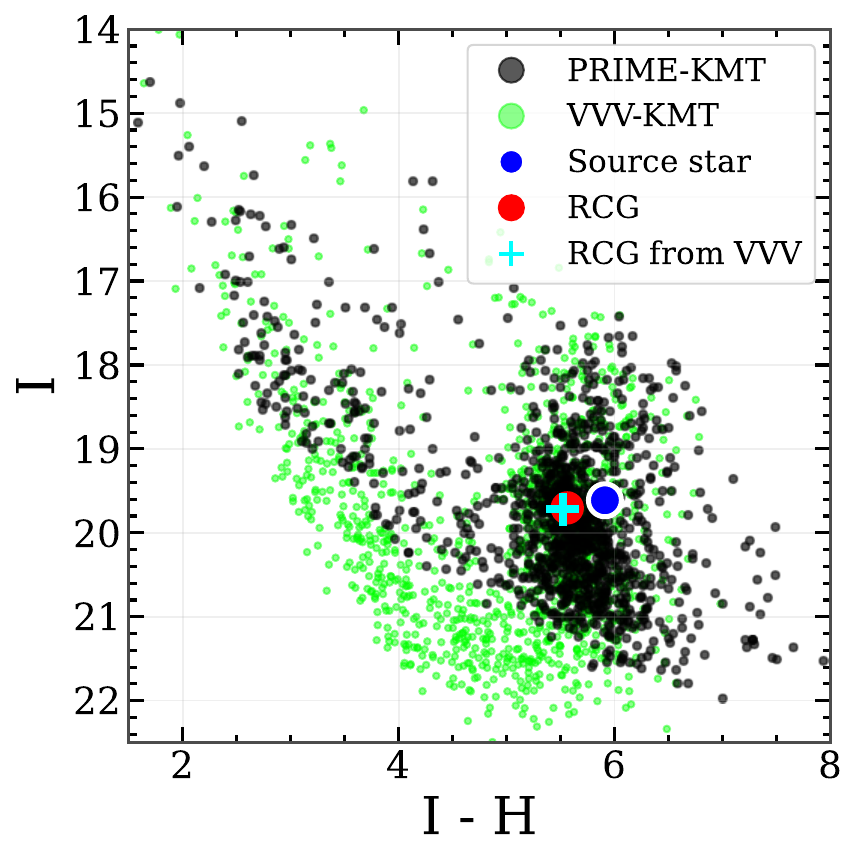}
    \centering
    \caption{
    Color--magnitude diagram used to determine the reddening and extinction toward KB240211.
    Black points are stars in the matched KMTNet--PRIME catalog, and green points are stars in the VVV--KMTNet catalog used as an external check.
    The red and blue points indicate the measured RCG centroid and source position, respectively, and the cyan cross marks the RCG centroid estimated from the VVV--KMTNet CMD.
    }
    \label{fig:cmd_pb240021}
\end{figure}

\begin{figure}[t]
    \includegraphics[width=1.0\textwidth]{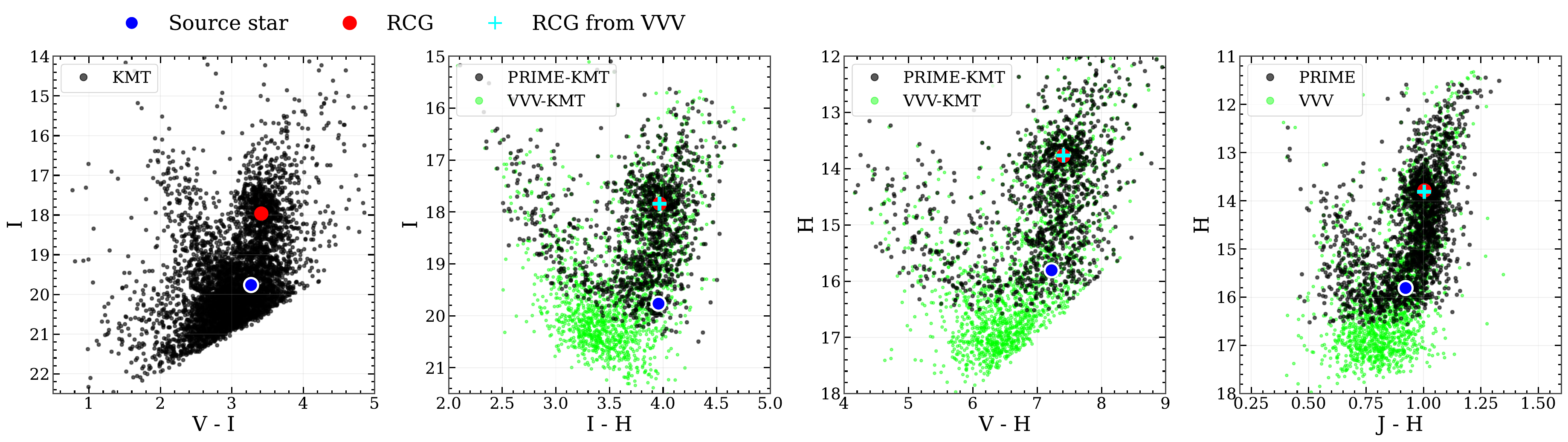}
    \caption{
    Color--magnitude diagrams used to determine the reddening and extinction toward KB241522.
    The panels show the $(V-I, I)$, $(I-H, I)$, $(V-H, H)$, and $(J-H, H)$ CMDs from left to right.
    Black points are stars in the matched KMTNet--PRIME catalogs, and green points are stars in the VVV--KMTNet catalogs used as an external check.
    The red and blue points indicate the measured RCG centroid and source position, respectively, and the cyan cross marks the RCG centroid estimated from the VVV--KMTNet CMD.
    }
    \label{fig:cmd_pb24160}
\end{figure}

\begin{table*}[t]
\centering
\caption{Observed and intrinsic RC centroids, and derived reddening and extinction for KB240211 and KB241522.}
\label{tab:RC_summary}
\begin{tabular}{llccccc}
\hline\hline
Event & Color--Mag & $(\mathrm{Color},\, \mathrm{Mag})_{\mathrm{RC}}$ & $(\mathrm{Color},\, \mathrm{Mag})_{\mathrm{RC},0}$ & $E(\mathrm{Color})$ & $A_{\mathrm{Mag}}$ \\
\hline
KB240211 & $(I-H, I)$ & $(5.561 \pm 0.009,\, 19.703 \pm 0.082)$ & $(1.29 \pm 0.06,\, 14.37 \pm 0.04)$ & $4.27 \pm 0.06$ & $5.33 \pm 0.09$ \\
\hline
KB241522 & $(V-I, I)$ & $(3.413 \pm 0.007,\, 17.710 \pm 0.032)$ & $(1.06 \pm 0.07,\, 14.37 \pm 0.04)$ & $2.35 \pm 0.07$ & $3.34 \pm 0.05$ \\
         & $(I-H, I)$ & $(3.964 \pm 0.008,\, 17.732 \pm 0.033)$ & $(1.29 \pm 0.06,\, 14.37 \pm 0.04)$ & $2.67 \pm 0.06$ & $3.36 \pm 0.05$ \\
         & $(V-H, H)$ & $(7.406 \pm 0.013,\, 13.772 \pm 0.029)$ & $(2.35 \pm 0.10,\, 13.08 \pm 0.08)$ & $5.06 \pm 0.10$ & $0.69 \pm 0.08$ \\
         & $(J-H, H)$ & $(1.004 \pm 0.002,\, 13.784 \pm 0.025)$ & $(0.55 \pm 0.03,\, 13.08 \pm 0.08)$ & $0.45 \pm 0.03$ & $0.70 \pm 0.08$ \\
\hline
\end{tabular}
\raggedright
\tablecomments{
The intrinsic RC colors and magnitudes are adopted from \citet{nat13}, \citet{ben13}, and \citet{nat21}. See Appendix~\ref{sec:rc0} for details.
}
\end{table*}

\section{Color--Magnitude Diagram} \label{sec:cmd}
To characterize the source stars of the two microlensing events, we construct color–magnitude diagrams (CMDs) of the stars in the corresponding fields.
The CMDs are used to estimate the extinction and reddening toward each event and to determine the intrinsic color and brightness of the source stars relative to the red clump giant (RCG) centroid.

The CMDs were constructed using stellar catalogs from KMTC and PRIME within a radius of $100\arcsec$ around each event.
For KB240211, we produced an $(I, I-H)$ CMD, while for KB241522, additional $(V-I)$, $(V-H)$ and $(J-H)$ CMDs were also constructed.
Because the $I$- and $H$-band catalogs originate from different instruments, astrometric alignment between the two systems was required. 

The KMTNet catalog provides stellar positions in pixel coordinates $(x, y)$, without equatorial coordinates, whereas the PRIME catalog is based on $(\mathrm{RA}, \mathrm{Dec})$.
To establish the transformation, we first visually identified 15 common stars between the two catalogs and used them as reference points to derive an initial affine transformation from the KMTNet pixel frame to the PRIME equatorial frame.
We then iteratively refined the transformation by matching stars within $0\arcsec36$, recalculating the affine parameters at each iteration, and repeating this process five times.
Finally, stars separated by less than $0\farcs36$ between the two catalogs were regarded as matches.

For the CMDs involving near-infrared bands, we also constructed comparison CMDs using stars matched between KMTNet and VVV from VIRAC2 \citep{smi25}.
These VVV--KMTNet CMDs were used as an external check on the location of the RCG sequence inferred from the KMTNet--PRIME matched catalogs.
The resulting CMDs for KB240211 and KB241522 are shown in Figures~\ref{fig:cmd_pb240021} and~\ref{fig:cmd_pb24160}, respectively.

To determine the RCG centroid in each CMD, we first fitted the luminosity function around the red clump using the model given in Equation~(4) of \citet{nat13}.
The luminosity function was modeled as the sum of a Gaussian component for the RCGs and a power-law component for the underlying background population. 
This fit provided the RC magnitude and its uncertainty.

We then measured the RC color using stars selected around the fitted RC magnitude. 
The centroid of the color distribution of these stars was adopted as the observed RC color, with the uncertainty estimated from the width of the selected RCG sequence. 
The resulting RC color and magnitude centroids were used to derive the reddening and extinction toward each event. 

To derive the extinction and reddening toward the events, we adopt the intrinsic RC magnitudes and colors from \citet{nat13}, \citet{ben13}, and \citet{nat21}.
The detailed derivations of these intrinsic values and their associated uncertainties are presented in Appendix~\ref{sec:rc0}.

The extinction and reddening toward each event were derived by comparing the observed RC centroids in the CMDs with the intrinsic RC values. 
The color excesses and extinction values were calculated as the differences between the observed and intrinsic quantities, e.g.,
\begin{align}
    E(\mathrm{Color}) &= (\mathrm{Color})_{\mathrm{RC}} - (\mathrm{Color})_{\mathrm{RC},0}, \quad 
    A_{\mathrm{Mag}} = \mathrm{Mag}_{\mathrm{RC}} - \mathrm{Mag}_{\mathrm{RC},0}, \nonumber
\end{align}
where $(\mathrm{Color})_{\mathrm{RC}}$ and $\mathrm{Mag}_{\mathrm{RC}}$ denote the observed RC centroid values in the CMDs, 
and $(\mathrm{Color})_{\mathrm{RC},0}$ and $\mathrm{Mag}_{\mathrm{RC},0}$ represent their intrinsic (dereddened) counterparts.
The resulting observed and intrinsic RC centroids, together with the derived reddening and extinction values for both events, 
are summarized in Table~\ref{tab:RC_summary}.

\section{Light Curve Analysis} \label{sec:lc}
\subsection{1L1S model for KB240211}
For KB240211, we adopt a single-lens single-source (1L1S) model including finite-source and annual-parallax effects. 
The underlying point-source point-lens (PSPL) model is characterized by three parameters \citep{pac86}: 
the time of the closest approach, $t_0$; 
the impact parameter, $u_0$, expressed in units of the angular Einstein radius $\theta_{\rm E}$; 
and the Einstein radius ($R_{\rm E}$) crossing time, $t_{\rm E}$.

In addition to the standard PSPL parameters, 
two major higher-order effects can be considered to account for subtle deviations from the simple model. 
The first is the finite-source effect \citep{wit94}. In single-lens events, this effect becomes important when the angular size of the source is comparable to the impact parameter, such that the source is no longer well approximated as a point source. The finite-source effect introduces an additional parameter, $\rho \equiv \theta_*/\theta_{\rm E}$.
The second is the microlens parallax effect, which arises from the orbital motion of the Earth around the Sun during the event.
This effect is described by the microlens parallax vector $\bm{\pi}_{\mathrm{E}} = (\pi_{\mathrm{E},N}, \pi_{\mathrm{E},E})$, \citep{gou92,gou00,gou04}
whose components represent the north and east projections of the parallax vector in the equatorial coordinate system.

Once these parameters are specified, the magnification of the source star, $A(t)$,  
is uniquely determined as a function of time.  
For convenience, we collectively denote the set of microlensing parameters as
\begin{align}
    \bm{x}_{1L1S} \equiv (t_0,\, t_{\rm E},\, u_0,\, \rho,\, \pi_{{\rm E},N},\, \pi_{{\rm E},E}).
\end{align}
The observed flux at time $t$, $F(t)$, is then given by
\begin{align}
    F(t) = F_{\rm s}\,A(t) + F_{\rm b},
\end{align}
where $F_{\rm s}$ and $F_{\rm b}$ denote the source and blended fluxes, respectively.  
Each dataset is assigned its own $(F_{\rm s}, F_{\rm b})$ parameters.

Using the parameters described above, together with several additional quantities,  
one can derive the physical properties of the lens and source systems.  
The relative proper motion between the lens and the source is given by
\begin{align}
\bm{\mu}_{\rm rel} 
= (\mu_{\rm rel,N},\mu_{\rm rel,E})
= \frac{\theta_{\rm E}}{t_{\rm E}} \cdot \frac{\bm{\pi}_{\rm E}}{|\bm{\pi}_{\rm E}|}, \label{eq:murel}
\end{align}
where the angular Einstein radius is expressed as
\begin{align}
\theta_{\rm E} = \frac{\theta_*}{\rho}. \label{eq:thE}
\end{align}
Here, $\theta_*$ denotes the angular radius of the source star,  
which is introduced as an additional parameter in our analysis.  
The physical mass and distance of the lens can then be derived from $\theta_{\rm E}$ and $\bm{\pi}_{\rm E}$ as
\begin{align}
    M_{\rm L} &= \frac{\theta_{\rm E}}{\kappa\,|\bm{\pi}_{\rm E}|}, \label{eq:ML} \\
    D_{\rm L} &= 1~\mathrm{AU} \left( |\bm{\pi}_{\rm E}|\,\theta_{\rm E} + \frac{1~\mathrm{AU}}{D_{\rm S}} \right)^{-1}, \label{eq:DL}
\end{align}
where $\kappa = 8.144~\mathrm{mas}~M_\odot^{-1}$ and $D_{\rm S}$ is the distance to the source.  
For single-lens events, $D_{\rm S}$ cannot be determined by the other model parameters  
and is therefore treated as an additional fitting parameter in our analysis.

Accordingly, the complete set of fitting parameters for the single-lens model consists of  
\begin{align}
    \bm{x}_{1L1S} = (t_0,\, t_{\rm E},\, u_0,\, \rho,\, \pi_{{\rm E},N},\, \pi_{{\rm E},E}), \quad \theta_*,~\, D_{\rm S},\,~ \{F_{\rm s}\},\,~ \{F_{\rm b}\},\,~ \{k\},
\end{align}
where $\{F_{\rm s}\}$ and $\{F_{\rm b}\}$ denote the sets of source and blend fluxes over all datasets, respectively, and $\{k\}$ denotes the set of error normalization factors introduced in Section~\ref{sec:bayes}.
From these parameters, we derive the physical quantities,
\begin{align}
    \bm{y}_{1L1S} \equiv (M_{\rm L},\, D_{\rm L},\, \mu_{{\rm rel},N},\, \mu_{{\rm rel},E}).
\end{align}
which characterize the physical nature of the lens system.

\subsection{2L1S Model for KB241522}
For KB241522, we adopt a binary-lens single-source (2L1S) model including finite-source and annual-parallax effects. 
In the 2L1S formalism, three additional parameters are introduced to describe the lens geometry:
the mass ratio between the lens components, $q$;  
the projected separation between them, $s$, expressed in units of the angular Einstein radius $\theta_{\rm E}$;  
and the angle between the projected binary axis and the source trajectory, $\alpha$.

Another higher-order effect that can be incorporated in binary-lens modeling is the lens orbital motion (LOM) \citep{dom98,iok99},  
which accounts for the relative motion of the binary components during the microlensing event.  
Assuming a circular orbit, the orbital motion is parameterized by the instantaneous velocity vector of the companion at the reference time,  
$(\gamma_1, \gamma_2, \gamma_3)$ \citep{sko11}, expressed in units normalized to $sR_{\rm E}$.  
In our analysis, we assume a circular orbit for the binary lens system.
If an eccentric orbit is adopted instead, two additional parameters are introduced to describe the initial orbital configuration:  
the line-of-sight displacement, $r_{\rm s}$, and the ratio of the orbital semi-major axis to the initial three-dimensional separation, $a_{\rm s}$, 
both normalized to the Einstein radius, $R_{\rm E}$.

Thus, the complete set of parameters required to compute the magnification curve includes  
all single-lens parameters together with the binary and orbital motion parameters,  
\begin{align}
    \bm{x}_{2L1S} = (t_0,\, t_{\rm E},\, u_0,\, q,\, s,\, \alpha,\, \rho,\, 
    \pi_{{\rm E},N},\, \pi_{{\rm E},E},\, \gamma_1,\, \gamma_2,\, \gamma_3). 
\end{align}

From the above parameters, one can further derive the physical properties of the lens and source systems.  
In contrast to single-lens events, for binary-lens events $D_{\rm S}$ can be determined from the LOM parameters.  
Under the assumption of a circular orbit, $D_{\rm S}$ can be expressed as \citep{nun26}
\begin{align}
D_{\rm S} = \frac{1~{\rm AU}}{\theta_{\rm E}} 
\left(
\left[
\frac{(1~{\rm AU})^3}{G M_{\rm L}}\,s^3
\sqrt{1 + \left(\frac{\gamma_1}{\gamma_3}\right)^2}
\left( \gamma_1^2 + \gamma_2^2 + \gamma_3^2 \right)
\right]^{1/3}
- \pi_{\rm E}
\right)^{-1}. 
\label{eq:DS_circ}
\end{align}

Furthermore, the circular LOM parameters also enable the reconstruction of the orbital configuration of the lens system  
\citep{sko11, boz21}, including the orbital semi-major axis $a$, the inclination $i$,  
the longitude of the ascending node in the North–East coordinate frame, $\Omega_{\rm NE}$ \citep{mas26},  
and the orbital phase angle with respect to the line of nodes, $\phi_0$.

Accordingly, the complete set of fitting parameters for the binary-lens model consists of
\begin{align}
    \bm{x}_{2L1S} = (t_0,\, t_{\rm E},\, u_0,\, q,\, s,\, \alpha,\, \rho,\,
    \pi_{{\rm E},N},\, \pi_{{\rm E},E},\, \gamma_1,\, \gamma_2,\, \gamma_3),
    \quad \theta_*,\, \{F_{\rm s}\},\, \{F_{\rm b}\},\, \{k\}.
\end{align}
From these fitting parameters, we derive the corresponding physical quantities,
\begin{align}
    \bm{y}_{2L1S} \equiv (M_{\rm L},\, D_{\rm L},\, D_{\rm S},\,
    \mu_{{\rm rel},N},\, \mu_{{\rm rel},E},\, a,\, i,\, \Omega_{\rm NE},\, \phi_0).
\end{align}

\subsection{Limb Darkening}\label{sec:limb_dark}
When finite-source effects are significant, we take into account the surface brightness profile of the source star, 
which is described by a linear limb-darkening law,
\begin{align}  
    S_\lambda(\vartheta) = S_\lambda(0) \left[ 1 - u_\lambda (1 - \cos\vartheta) \right],  
\end{align}
where $\vartheta$ is the angle between the normal to the stellar surface and the line of sight,
and $u_\lambda$ is the linear limb-darkening coefficient for the corresponding passband.

For the determination of the limb-darkening coefficients, we first performed preliminary light-curve fits without including limb darkening to estimate the source fluxes.  
Because the inferred source flux is only weakly dependent on the limb-darkening assumption, this approach provides sufficiently accurate initial estimates for the source brightness.  
From these fits, the observed source magnitudes were estimated to be
$I_{\rm s} \simeq 19.61$ and $H_{\rm s} \simeq 13.66$ for KB240211, and
$I_{\rm s} \simeq 19.81$ and $H_{\rm s} \simeq 15.80$ for KB241522.
For KB240211, the RC-based corrections in Table~\ref{tab:RC_summary} give
$I_{\rm s,0} \simeq 14.28$ and $(I-H)_{\rm s,0} \simeq 1.68$,
while for KB241522 they give
$I_{\rm s,0} \simeq 16.45$ and $(I-H)_{\rm s,0} \simeq 1.33$,
using the $(I-H,I)$ CMD.
Using these colors and magnitudes, we inferred the effective temperature and surface gravity of each source 
by applying the color–temperature relations from \citet{hou00}, assuming a metallicity of $[{\rm M/H}] = 0$ and a microturbulent velocity of 
$\xi = 2~{\rm km\,s^{-1}}$.  
This yielded $(T_{\rm eff}, \log g) = (4000~{\rm K},~3.0)$ for KB240211 and $(T_{\rm eff}, \log g) = (4250~{\rm K},~4.5)$ for KB241522.  
Based on these atmospheric parameters, we adopted the linear limb-darkening coefficients 
from the tabulations of \citet{cla11} corresponding to the $I$, $V$, $J$, and $H$ bands.  
For KB240211, we used $u_I = 0.652$ and $u_H = 0.431$, 
while for KB241522 we adopted $u_I = 0.610$, $u_V = 0.792$, $u_H = 0.401$, and $u_J = 0.478$.

\subsection{Source Flux and Angular Radius}
The angular radius of the source star, $\theta_*$, is empirically related to its flux and color.  
In microlensing analyses, $\theta_*$ is typically inferred from the source flux $F_{\rm s}$  
using an empirical color--surface brightness relation of the form \citep{boy14}:
\begin{align}
    \log_{10}(2\hat{\theta}_*) = a_0 + a_1 C - 0.2\,m, \label{eq:thS_mag_color}
\end{align}
where $C$ represents the intrinsic color index and $m$ is the intrinsic magnitude used in the corresponding color--surface brightness relation.

In our modeling framework, both $\{F_{\rm s}\}$ and $\theta_*$ are independently parameterized,  
as described in Section~\ref{sec:lc}.  
Rather than being used to derive $\theta_*$ directly from $F_{\rm s}$,  
the empirical color–surface brightness relation is incorporated as a prior constraint  
linking these parameters (see Section~\ref{sec:bayes}). 

We apply this relation using the following color--magnitude combinations.
For KB240211, we use $(m, C) = (I_{\rm KMTC}, I_{\rm KMTC}-H_{\rm PRIME})$.
For KB241522, we perform separate estimations using four different combinations:
$(I_{\rm KMTC}, I_{\rm KMTC}-H_{\rm PRIME})$,
$(I_{\rm KMTC}, V_{\rm KMTC}-I_{\rm KMTC})$,
$(H_{\rm PRIME}, V_{\rm KMTC}-H_{\rm PRIME})$, and
$(H_{\rm PRIME}, J_{\rm PRIME}-H_{\rm PRIME})$,
and compare the results.
The corresponding magnitudes are calculated from the source flux parameters $\{F_{\rm s}\}$
and corrected for extinction and reddening using the color excesses and extinction values
listed in Table~\ref{tab:RC_summary}, yielding the de-reddened colors and magnitudes of the source star.

The adopted coefficients are summarized as follows:
\begin{align}
    &\text{for } (I-H): \quad a_0 = 0.53026 \pm 0.00077, \quad a_1 = 0.36595 \pm 0.00079, \nonumber \\
    &\text{for } (V-I): \quad a_0 = 0.5014 \pm 0.0251, \quad a_1 = 0.4197 \pm 0.0210, \nonumber \\
    &\text{for } (V-H): \quad a_0 = 0.5145 \pm 0.0019, \quad a_1 = 0.0892 \pm 0.0009, \nonumber \\
    &\text{for } (J-H): \quad a_0 = 0.5013 \pm 0.044, \quad a_1 = 0.4312 \pm 0.12, \nonumber
\end{align}
where the $(I-H)$ and $(V-I)$ relations are taken from \citet{boy14},
and the $(V-H)$ and $(J-H)$ relations are adopted from \citet{ker04}.
We additionally include the intrinsic scatter of each color--surface brightness relation,
which we take to be $7.4\%$ for $(I-H)$, $5.1\%$ for $(V-I)$, $1.12\%$ for $(V-H)$, and $10.31\%$ for $(J-H)$.
For the $(J-H)$ relation, note that it exhibits a pronounced nonlinearity,
which largely accounts for the relatively large uncertainties in the fitted coefficients.

\subsection{Bayesian Analysis} \label{sec:bayes}

To estimate the model parameters introduced in Section~\ref{sec:lc} from the light-curve data,  
we perform a Bayesian analysis.  
Our goal is to infer the posterior probability distribution of the parameters given the observations,  
which can be written as
\begin{align}
    P(\bm{x}_{\rm 1L1S}, \theta_*, D_{\rm S}, \{F_{\rm s}\}, \{F_{\rm b}\}, \{k\} \mid \mathrm{data})
    &\propto 
    \mathcal{L}(\bm{x}_{\rm 1L1S}, \{F_{\rm s}\}, \{F_{\rm b}\}, \{k\}) \,
    \pi(\bm{x}_{\rm 1L1S}, \theta_*, D_{\rm S}, \{F_{\rm s}\}, \{F_{\rm b}\}, \{k\})\quad\text{and}
    \label{eq:bayes_1l1s}\\
    P(\bm{x}_{\rm 2L1S}, \theta_*, \{F_{\rm s}\}, \{F_{\rm b}\}, \{k\} \mid \mathrm{data})
    &\propto 
    \mathcal{L}(\bm{x}_{\rm 2L1S}, \{F_{\rm s}\}, \{F_{\rm b}\}, \{k\}) \,
    \pi(\bm{x}_{\rm 2L1S}, \theta_*, \{F_{\rm s}\}, \{F_{\rm b}\}, \{k\}),
    \label{eq:bayes_2l1s}
\end{align}
where $\mathcal{L}$ denotes the likelihood function and $\pi$ represents the prior probability distribution of the parameters.
From these posterior distributions, we also obtain the posterior distributions  
of the physical parameters of interest, $\bm{y}_{\rm 1L1S}$ and $\bm{y}_{\rm 2L1S}$.

The likelihood function, $\mathcal{L}$, quantifies the agreement between the observed fluxes 
and the model predictions.  
Assuming Gaussian uncertainties, the likelihood can be expressed as
\begin{align}
    \mathcal{L}(\bm{x}, \{F_{\rm s}\}, \{F_{\rm b}\}, \{k\})
    &= \prod_j \prod_i
    \frac{1}{\sqrt{2\pi}\,k_j\,\sigma_{ij}}
    \exp\left[
        -\frac{1}{2}
        \left(
            \frac{
                F_{ij} - \left(F_{{\rm s},j}\,A(t_{ij};\bm{x}) + F_{{\rm b},j}\right)
            }{
                k_j\,\sigma_{ij}
            }
        \right)^2
    \right] \label{eq:likelihood}
\end{align}
Here, $F_{ij}$ and $\sigma_{ij}$ are the observed flux and its reported uncertainty at epoch $t_{ij}$,  
$F_{{\rm s},j}$ and $F_{{\rm b},j}$ denote the source and blend fluxes for each dataset,  
and $k_j$ represents the error–renormalization factor that rescales the uncertainties \citep{yee12}.

We adopt weakly informative priors for the nuisance parameters.  
Uniform priors are assumed for the source and blend fluxes (i.e., $\pi_{F_{\rm s}} = \pi_{F_{\rm b}} = 1$), and scale-invariant priors ($\pi_{k_j}\propto 1/k_j$)
for the error–renormalization factors.  

For the angular source radius, $\theta_*$, we impose a conditional prior based on the color--surface brightness relation (Equation~\ref{eq:thS_mag_color}), which predicts $\log(2\theta_*)$ from the source flux parameters $\{F_{\rm s}\}$. From $\{F_{\rm s}\}$, we derive the predicted value $\widehat{\log(2\theta_*)}$ and its uncertainty $\sigma_{\log(2\theta_*)}$ by propagating the uncertainties in the color--surface brightness coefficients and in the extinction and reddening values listed in Table~\ref{tab:RC_summary}, and by including the intrinsic scatter of the adopted color--surface brightness relation.
We then adopt a Gaussian prior on $\log(2\theta_*)$:
\begin{align}
    P(\log(2\theta_*) \mid \{F_{\rm s}\})
    =
    \frac{1}{\sqrt{2\pi}\,\sigma_{\log(2\theta_*)}(\{F_{\rm s}\})}
    \exp\left[
        -\frac{1}{2}
        \left(
            \frac{
            \log(2\theta_*)-\widehat{\log(2\theta_*)}(\{F_{\rm s}\})
            }{
            \sigma_{\log(2\theta_*)}(\{F_{\rm s}\})
            }
        \right)^2
    \right].
    \label{eq:log2thS_prior}
\end{align}
Equivalently, expressed as a prior on $\theta_*$, this becomes
\begin{align}
    \pi_{\theta_*}(\theta_* \mid \{F_{\rm s}\})
    \propto
    \frac{1}{\theta_*}\,
    P(\log(2\theta_*) \mid \{F_{\rm s}\}),
    \qquad \theta_* > 0.
    \label{eq:thS_prior}
\end{align}

To simplify notation, we introduce a unified notation
\begin{align}
    \bm{X} =
    \begin{cases}
        (\bm{x}_{\rm 1L1S}, D_{\rm S}) & \\
        \bm{x}_{\rm 2L1S}              &
    \end{cases}
    \quad    \bm{Y} =
    \begin{cases}
        (\bm{y}_{\rm 1L1S}, D_{\rm S}) & \text{for 1L1S}, \\
        \bm{y}_{\rm 2L1S}              & \text{for 2L1S}.
    \end{cases}
\end{align}
For these parameters we adopt physically motivated priors implemented with the
\texttt{gapmoe} framework introduced by \citet{nun26}.
In this approach, a Galactic prior based on the Koshimoto Galactic Model \citep{kos21}
is applied to the corresponding physical quantities
$(M_{\rm L}, D_{\rm L}, D_{\rm S}, \bm{\mu}_{\rm rel})$, 
while the parameters that are independent of the Galactic model
(e.g., $t_0$, $u_0$, $q$, orbital parameters) are rescaled so that they are effectively sampled
uniformly over the allowed ranges.
The Galactic prior can then be written in the compact form
\begin{align}
    \pi_{\rm Gal}(\bm{X} \mid \theta_*)
    = \pi_{\rm Gal}\!\bigl( \bm{Y}(\bm{X}^{Gal}, \theta_*) \bigr)
      \left|
        \frac{\partial (\bm{Y}, \theta_*)}{\partial (\bm{X}^{Gal}, \theta_*)}
      \right|_{\bm{Y}={\bm Y}(\bm{X^{Gal}},\theta_*)}.
    \label{eq:pigal_X}
\end{align}
Here, $\bm{X}^{\rm Gal}$ denotes the subset of $\bm{X}$ excluding ($t_0$, $u_0$,$q$)
that are not directly tied to the Galactic model.  
The first term on the right-hand side represents the probability density of the corresponding physical parameters 
$(M_{\rm L}, D_{\rm L}, D_{\rm S}, \bm{\mu}_{\rm rel})$ evaluated from the Galactic model,  
while the second term is the Jacobian determinant that accounts for the transformation 
from the microlensing parameters to these physical quantities \citep{sko11}.

Combining the likelihood function (Equation~\ref{eq:likelihood}) with the priors described above,
the full posterior distribution for both models can be expressed compactly as
\begin{align}
    P(\bm{X}, \theta_*, \{F_{\rm s}\}, \{F_{\rm b}\}, \{k\} \mid \mathrm{data})
    &\propto 
    \mathcal{L}(\bm{x}, \{F_{\rm s}\}, \{F_{\rm b}\}, \{k\}) \,
    \pi_{\rm Gal}(\bm{X} \mid \theta_*) \,
    \pi_{\theta_*}(\theta_* \mid \{F_{\rm s}\}) \,
    \pi_k(\{k\})\,\pi_{F_{\rm s}}(\{F_{\rm s}\})\,\pi_{F_{\rm b}}(\{F_{\rm b}\}).
    \label{eq:bayes_general}
\end{align}

\subsection{Construction of the Marginalized Posterior}\label{sec:marg}

Direct sampling of all parameters is computationally expensive, because the photometric parameters
$\{F_{\rm s}\}$, $\{F_{\rm b}\}$, and $\{k\}$ each have one component for every dataset.
A common approximation is to fix the error-normalization factors $\{k\}$ using a reference model,
and then determine $\{F_{\rm s}\}$ and $\{F_{\rm b}\}$ by linear regression for each trial model.
Formally, this amounts to replacing the full likelihood with the profiled likelihood
\begin{align}
\tilde{\mathcal L}(\bm X)
=
\max_{\{F_{\rm s}\},\{F_{\rm b}\}}
\mathcal L(\bm X,\{F_{\rm s}\},\{F_{\rm b}\},\{k\}^{\rm fix}).
\label{eq:profile_likelihood}
\end{align}
Although efficient, this approximation ignores the uncertainties in
$\{F_{\rm s}\}$, $\{F_{\rm b}\}$, and $\{k\}$, and therefore tends to underestimate
the posterior uncertainty of the physical parameters.

To avoid this, we explicitly marginalize over these nuisance parameters.
The marginalized posterior for $\bm X$ is
\begin{align}
P(\bm X\mid \mathrm{data})
&\propto
\iiint
\mathcal L(\bm X,\{F_{\rm s}\},\{F_{\rm b}\},\{k\})\,
\pi_k(\{k\})\,
\pi_{F_{\rm s}}(\{F_{\rm s}\})\,
\pi_{F_{\rm b}}(\{F_{\rm b}\})
\notag\\
&\qquad\times
\left[
\int
\pi_{\rm Gal}(\bm X\mid \theta_*)\,
\pi_{\theta_*}(\theta_*\mid \{F_{\rm s}\})\,
d\theta_*
\right]
d\{F_{\rm s}\}\,d\{F_{\rm b}\}\,d\{k\}.
\label{eq:marginal_post}
\end{align}
Analytic marginalization over the linear source and blend flux parameters has also been employed in a recent microlensing inference framework \citep{miy25}.

We first define the nuisance-marginalized data term
\begin{align}
\mathcal L_{\rm marg}(\bm X)
\equiv
\int
\mathcal L(\bm X,\{F_{\rm s}\},\{F_{\rm b}\},\{k\})\,
\pi_k(\{k\})\,
\pi_{F_{\rm s}}(\{F_{\rm s}\})\,
\pi_{F_{\rm b}}(\{F_{\rm b}\})\,
d\{F_{\rm s}\}\,d\{F_{\rm b}\}\,d\{k\},
\label{eq:Lmarg}
\end{align}
and the conditional probability density for $\{F_{\rm s}\}$ at fixed $\bm X$,
\begin{align}
p(\{F_{\rm s}\}\mid \bm X,\mathrm{data})
\equiv
\frac{
\displaystyle
\int
\mathcal L(\bm X,\{F_{\rm s}\},\{F_{\rm b}\},\{k\})\,
\pi_k(\{k\})\,
\pi_{F_{\rm s}}(\{F_{\rm s}\})\,
\pi_{F_{\rm b}}(\{F_{\rm b}\})\,
d\{F_{\rm b}\}\,d\{k\}
}{
\mathcal L_{\rm marg}(\bm X)
}.
\label{eq:Fs_conditional}
\end{align}
Here and below, the conditioning on the data is suppressed when no confusion arises.

Using these definitions, Equation~\eqref{eq:marginal_post} becomes
\begin{align}
P(\bm X\mid \mathrm{data})
\propto
\mathcal L_{\rm marg}(\bm X)\,Z_{\rm Gal}(\bm X),
\label{eq:marginal_post_factorized}
\end{align}
where
\begin{align}
Z_{\rm Gal}(\bm X)
&=
\int
\left[
\int
\pi_{\rm Gal}(\bm X\mid \theta_*)\,
\pi_{\theta_*}(\theta_*\mid \{F_{\rm s}\})\,
d\theta_*
\right]
\,p(\{F_{\rm s}\}\mid \bm X,\mathrm{data})\,
d\{F_{\rm s}\}.
\label{eq:ZGal}
\end{align}
The factor $\mathcal L_{\rm marg}(\bm X)$ contains the information from the light-curve fit after marginalizing over the nuisance photometric parameters, while $Z_{\rm Gal}(\bm X)$ represents the effective Galactic prior for a given $\bm X$, obtained by marginalizing over $\theta_*$ and $\{F_{\rm s}\}$.

The term $\mathcal L_{\rm marg}(\bm X)$ can be evaluated analytically.
For each dataset, after introducing the precision parameter
$\tau_i \equiv k_i^{-2}$, the conditional posterior of
$(F_{s,i},F_{b,i},\tau_i)$ has a normal-gamma kernel under the priors
$\pi(F_{s,i})\propto 1$, $\pi(F_{b,i})\propto 1$, and
$\pi(k_i)\propto 1/k_i$.
Therefore, marginalization over $(F_{s,i},F_{b,i},\tau_i)$ yields a
closed-form expression for $\mathcal L_{\rm marg}(\bm X)$.
Likewise, marginalization over $(F_{b,i},\tau_i)$ yields a Student's
$t$ distribution for $F_{s,i}$, and thus
$p(\{F_s\}\mid \bm X,\mathrm{data})$ is a product of univariate
Student's $t$ distributions (Appendix~\ref{app:int_like}).

The remaining factor $Z_{\rm Gal}(\bm X)$ is evaluated numerically.
It can be written as the nested expectation
% \begin{align}
% Z_{\rm Gal}(\bm X)
% =
% \mathbb{E}_{\{F_{\rm s}\}\sim p(\cdot\mid \bm X,\mathrm{data})}
% \left[
% \mathbb{E}_{\theta_*\sim \pi_{\theta_*}(\cdot\mid \{F_{\rm s}\})}
% \left[
% \pi_{\rm Gal}(\bm X\mid \theta_*)
% \right]
% \right].
% \label{eq:ZGal_expectation}
% \end{align}
\begin{align}
Z_{\rm Gal}(\bm X)
&=
\mathbb{E}_{p(\{F_{\rm s}\}\mid \bm X,\mathrm{data})}
\!\left[
\mathbb{E}_{\pi_{\theta_*}(\theta_*\mid \{F_{\rm s}\})}
\!\left[
\pi_{\rm Gal}(\bm X\mid \theta_*)
\right]
\right].
\label{eq:ZGal_expectation}
\end{align}
We approximate this expectation by nested Monte Carlo sampling:
first, we draw $N$ samples $\{F_{\rm s}^{(i)}\}$ from
$p(\{F_{\rm s}\}\mid \bm X,\mathrm{data})$;
then, for each draw $\{F_{\rm s}^{(i)}\}$, we draw
$M$ samples $\theta_*^{(i,j)}$ from
$\pi_{\theta_*}(\theta_*\mid \{F_{\rm s}^{(i)}\})$.
This gives
\begin{align}
Z_{\rm Gal}(\bm X)
\simeq
\frac{1}{NM}
\sum_{i=1}^{N}\sum_{j=1}^{M}
\pi_{\rm Gal}(\bm X\mid \theta_*^{(i,j)}).
\label{eq:ZGal_MC}
\end{align}
We confirmed that $N=100$ and $M=100$ are sufficient for the present analysis,
yielding better than $10^{-2}$ fractional precision.
Because the evaluation of $\pi_{\rm Gal}(\bm X\mid \theta_*)$ in the \texttt{gapmoe} framework
is implemented in \texttt{jax} \citep{jax18} and supports \texttt{jit} compilation,
this numerical step remains computationally efficient.

\subsection{Construction of the Posterior of Physical Parameters}

Given posterior samples of $\bm X$ from Equation~\eqref{eq:marginal_post_factorized},
the posterior distribution of the physical parameters $\bm Y$ is obtained by marginalizing over $\theta_*$,
\begin{align}
P(\bm Y\mid \mathrm{data})
=
\int
\delta\!\left[\bm Y-\bm Y(\bm X,\theta_*)\right]\,
P(\theta_*\mid \bm X,\mathrm{data})\,
P(\bm X\mid \mathrm{data})\,
d\bm X\,d\theta_*.
\label{eq:Y_posterior}
\end{align}
Thus, for each posterior draw $\bm X^{(k)}\sim P(\bm X\mid \mathrm{data})$,
we only need the conditional posterior of $\theta_*$ at fixed $\bm X^{(k)}$.

For fixed $\bm X$, the integrand in Equation~\eqref{eq:ZGal} induces the joint conditional distribution
\begin{align}
p(\theta_*,\{F_{\rm s}\}\mid \bm X,\mathrm{data})
=
\frac{
\pi_{\rm Gal}(\bm X\mid \theta_*)\,
\pi_{\theta_*}(\theta_*\mid \{F_{\rm s}\})\,
p(\{F_{\rm s}\}\mid \bm X,\mathrm{data})
}{
Z_{\rm Gal}(\bm X)
}.
\label{eq:joint_theta_Fs}
\end{align}
Marginalizing over $\{F_{\rm s}\}$ gives
\begin{align}
P(\theta_*\mid \bm X,\mathrm{data})
=
\frac{
\pi_{\rm Gal}(\bm X\mid \theta_*)
\displaystyle\int
\pi_{\theta_*}(\theta_*\mid \{F_{\rm s}\})\,
p(\{F_{\rm s}\}\mid \bm X,\mathrm{data})\,
d\{F_{\rm s}\}
}{
Z_{\rm Gal}(\bm X)
}.
\label{eq:theta_conditional}
\end{align}

For each posterior sample $\bm X^{(k)}$, we reuse the same nested Monte Carlo samples
introduced for the evaluation of $Z_{\rm Gal}(\bm X^{(k)})$.
Specifically, we draw
$\{F_{\rm s}^{(k,n)}\}\sim p(\cdot\mid \bm X^{(k)},\mathrm{data})$
for $n=1,\ldots,N$, and then draw
$\theta_*^{(k,n,m)}\sim \pi_{\theta_*}(\cdot\mid \{F_{\rm s}^{(k,n)}\})$
for $m=1,\ldots,M$.
These samples are assigned importance weights
\begin{align}
\tilde w^{(k,n,m)}
\propto
\pi_{\rm Gal}\!\left(\bm X^{(k)}\mid \theta_*^{(k,n,m)}\right),
\qquad
\sum_{n,m}\tilde w^{(k,n,m)}=1.
\label{eq:norm_weight}
\end{align}
Resampling one $\theta_*^{(k,n,m)}$ according to these normalized weights yields
an approximate draw
\begin{align}
\theta_*^{(k)} \sim P(\theta_*\mid \bm X^{(k)},\mathrm{data}).
\label{eq:theta_resample}
\end{align}
We then compute
\begin{align}
\bm Y^{(k)}=\bm Y\!\left(\bm X^{(k)},\theta_*^{(k)}\right).
\label{eq:Y_draw}
\end{align}
Repeating this procedure for all posterior samples $\bm X^{(k)}$ yields an equally weighted
Monte Carlo sample $\{\bm Y^{(k)}\}$ from $P(\bm Y\mid \mathrm{data})$.

\subsection{MCMC Sampling} \label{sec:mcmc}

We sample the marginalized posterior distribution in Equation~\eqref{eq:marginal_post_factorized}
using the \texttt{emcee} ensemble sampler \citep{emcee}.
For KB240211 and KB241522, we adopt $n_{\rm walker}=28$ and $38$ walkers, respectively,
with a maximum of $3\times10^6$ iterations.

Sampling is terminated automatically when the following two conditions are both satisfied:
(i) the estimated integrated autocorrelation time $\tau$ obeys
$\tau \times 100 < N_{\rm iter}$, ensuring that the chain is sufficiently long; and
(ii) the relative change in $\tau$ between successive evaluations is smaller than 1\%,
indicating that the chains have stabilized.

\paragraph{KB240211 (1L1S model).}
For KB240211, the initial parameter values were obtained from a static (non-parallax) single-lens fit.
In this preliminary fit, the parameters $(t_0, t_{\rm E}, u_0, \rho)$ were optimized using the Nelder–Mead algorithm
under uniform priors, with the error renormalization factor fixed at $k=1$.
The source and blend fluxes ($F_{\rm s}$, $F_{\rm b}$) were determined by linear regression.
The resulting best-fit parameters were then used as the initial positions for the MCMC,
with the parallax parameters initialized at $(\pi_{{\rm E},N}, \pi_{{\rm E},E}) = (0.01, 0.01)$
and the source distance set to $D_{\rm S} = 8.1~\mathrm{kpc}$.
Because the static 1L1S model without parallax is symmetric with respect to the sign of $u_0$,  
the initial fit was performed assuming $u_0 > 0$ only.  
However, inclusion of the annual parallax effect breaks this degeneracy,  
so we ran two independent MCMC chains initialized at $u_0$ and $-u_0$,  
each restricted to its respective domain ($u_0 > 0$ or $u_0 < 0$)  
to prevent transitions between the two branches.

\paragraph{KB241522 (2L1S model).}
For KB241522, we first conducted a coarse grid search for mass ratio, $q$, projected separation in units of Einstein radius, $s$ , and the angle between binary axis and the source trajectory, $\alpha$.
The parameter space of $(q, s, \alpha)$ was divided into $11\times22\times40 = 9680$ uniformly spaced grids:
$\log q \in [-4, 0]$, $\log s \in [-0.5, 0.55]$, and $\alpha \in [0, 2\pi)$.  
At each grid point, $(q, s, \alpha)$ were fixed and the remaining parameters were optimized via the Nelder–Mead algorithm.
We then selected the best 300 grid points with the lowest $\chi^2$ values and performed refined fits allowing $(q, s, \alpha)$ to vary freely.
Among these, the best-fitting static solution improved the $\chi^2$ by more than $\Delta\chi^2 > 100$ relative to all other local minima.  
We then refitted this model including the annual parallax effect,  
and used the resulting optimized $\bm{\pi}_{\rm E}$ values as the initial conditions for the subsequent MCMC sampling.  
The orbital motion parameters $\bm{\gamma}$ were initialized at $10^{-4}$.

For the $(V-H)$ analysis, we also initialized additional MCMC runs to search for discrete parallax-related branches.
These trials were motivated by the $u_0$-flip/ecliptic parallax degeneracy and by the jerk-parallax degeneracy, which can be modified in binary-lens events when the parallax parameters are correlated with LOM parameters \citep{gou04,sko11}.
Starting from the main above parallax solution, we constructed trial seeds by reversing the sign of $u_0$, reflecting the source-trajectory angle $\alpha$, reversing the orbital-rotation parameter $\gamma_2$, and testing alternative sign domains for the components of $\bm{\pi}_{\rm E}$.
These candidate branches were sampled independently, with walkers initialized within the corresponding sign domains, to avoid missing isolated posterior modes.

\newcommand{\tzeropospmpl}{0.6974^{+0.0006}_{-0.0006}}
\newcommand{\tzeronegpmpl}{0.6974^{+0.0006}_{-0.0006}}
\newcommand{\tEpospmpl}{4.99^{+0.11}_{-0.10}}
\newcommand{\tEnegpmpl}{4.99^{+0.11}_{-0.11}}
\newcommand{\uzeropospmpl}{1.43^{+0.05}_{-0.05}}
\newcommand{\uzeronegpmpl}{-1.43^{+0.05}_{-0.05}}
\newcommand{\rhopospmpl}{4.77^{+0.11}_{-0.11}}
\newcommand{\rhonegpmpl}{4.77^{+0.11}_{-0.11}}
\newcommand{\piENpospmpl}{0.04^{+0.37}_{-0.08}}
\newcommand{\piENnegpmpl}{0.04^{+0.36}_{-0.08}}
\newcommand{\piEEpospmpl}{0.03^{+0.41}_{-0.07}}
\newcommand{\piEEnegpmpl}{0.04^{+0.42}_{-0.07}}
\newcommand{\MLpospmpl}{0.22^{+0.39}_{-0.18}}
\newcommand{\MLnegpmpl}{0.21^{+0.40}_{-0.18}}
\newcommand{\DLpospmpl}{7.0^{+0.8}_{-3.9}}
\newcommand{\DLnegpmpl}{7.0^{+0.9}_{-3.8}}
\newcommand{\DSpospmpl}{8.1^{+0.5}_{-0.6}}
\newcommand{\DSnegpmpl}{8.1^{+0.5}_{-0.6}}
\newcommand{\muNpospmpl}{7.5^{+6.0}_{-16.3}}
\newcommand{\muNnegpmpl}{7.4^{+6.1}_{-16.3}}
\newcommand{\muEpospmpl}{6.6^{+6.6}_{-14.8}}
\newcommand{\muEnegpmpl}{6.7^{+6.5}_{-14.7}}
\newcommand{\muRelpospmpl}{14.4^{+1.4}_{-1.2}}
\newcommand{\muRelnegpmpl}{14.4^{+1.4}_{-1.3}}
\newcommand{\thetaSpospmpl}{9.4^{+0.9}_{-0.8}}
\newcommand{\thetaSnegpmpl}{9.4^{+0.9}_{-0.8}}
\newcommand{\Magpospmpl}{14.28^{+0.09}_{-0.09}}
\newcommand{\Magnegpmpl}{14.28^{+0.09}_{-0.09}}
\newcommand{\Colorpospmpl}{1.64^{+0.06}_{-0.06}}
\newcommand{\Colornegpmpl}{1.64^{+0.06}_{-0.06}}
\newcommand{\thetaEpospmpl}{0.197^{+0.019}_{-0.017}}
\newcommand{\thetaEnegpmpl}{0.198^{+0.020}_{-0.017}}

\newcommand{\tzeroposlowpl}{0.6968}
\newcommand{\tzeroposmidpl}{0.6974}
\newcommand{\tzeroposhighpl}{0.6980}
\newcommand{\tzeroneglowpl}{0.6968}
\newcommand{\tzeronegmidpl}{0.6974}
\newcommand{\tzeroneghighpl}{0.6980}
\newcommand{\tEposlowpl}{4.88}
\newcommand{\tEposmidpl}{4.99}
\newcommand{\tEposhighpl}{5.09}
\newcommand{\tEneglowpl}{4.88}
\newcommand{\tEnegmidpl}{4.99}
\newcommand{\tEneghighpl}{5.10}
\newcommand{\uzeroposlowpl}{1.38}
\newcommand{\uzeroposmidpl}{1.43}
\newcommand{\uzeroposhighpl}{1.48}
\newcommand{\uzeroneglowpl}{-1.48}
\newcommand{\uzeronegmidpl}{-1.43}
\newcommand{\uzeroneghighpl}{-1.38}
\newcommand{\rhoposlowpl}{4.66}
\newcommand{\rhoposmidpl}{4.77}
\newcommand{\rhoposhighpl}{4.88}
\newcommand{\rhoneglowpl}{4.66}
\newcommand{\rhonegmidpl}{4.77}
\newcommand{\rhoneghighpl}{4.88}
\newcommand{\piENposlowpl}{-0.04}
\newcommand{\piENposmidpl}{0.04}
\newcommand{\piENposhighpl}{0.41}
\newcommand{\piENneglowpl}{-0.05}
\newcommand{\piENnegmidpl}{0.04}
\newcommand{\piENneghighpl}{0.39}
\newcommand{\piEEposlowpl}{-0.04}
\newcommand{\piEEposmidpl}{0.03}
\newcommand{\piEEposhighpl}{0.45}
\newcommand{\piEEneglowpl}{-0.04}
\newcommand{\piEEnegmidpl}{0.04}
\newcommand{\piEEneghighpl}{0.45}
\newcommand{\MLposlowpl}{0.03}
\newcommand{\MLposmidpl}{0.22}
\newcommand{\MLposhighpl}{0.61}
\newcommand{\MLneglowpl}{0.03}
\newcommand{\MLnegmidpl}{0.21}
\newcommand{\MLneghighpl}{0.61}
\newcommand{\DLposlowpl}{3.2}
\newcommand{\DLposmidpl}{7.0}
\newcommand{\DLposhighpl}{7.8}
\newcommand{\DLneglowpl}{3.2}
\newcommand{\DLnegmidpl}{7.0}
\newcommand{\DLneghighpl}{7.8}
\newcommand{\DSposlowpl}{7.5}
\newcommand{\DSposmidpl}{8.1}
\newcommand{\DSposhighpl}{8.6}
\newcommand{\DSneglowpl}{7.5}
\newcommand{\DSnegmidpl}{8.1}
\newcommand{\DSneghighpl}{8.6}
\newcommand{\muNposlowpl}{-8.8}
\newcommand{\muNposmidpl}{7.5}
\newcommand{\muNposhighpl}{13.5}
\newcommand{\muNneglowpl}{-8.9}
\newcommand{\muNnegmidpl}{7.4}
\newcommand{\muNneghighpl}{13.5}
\newcommand{\muEposlowpl}{-8.2}
\newcommand{\muEposmidpl}{6.6}
\newcommand{\muEposhighpl}{13.2}
\newcommand{\muEneglowpl}{-8.0}
\newcommand{\muEnegmidpl}{6.7}
\newcommand{\muEneghighpl}{13.2}
\newcommand{\muRelposlowpl}{13.2}
\newcommand{\muRelposmidpl}{14.4}
\newcommand{\muRelposhighpl}{15.9}
\newcommand{\muRelneglowpl}{13.2}
\newcommand{\muRelnegmidpl}{14.4}
\newcommand{\muRelneghighpl}{15.9}
\newcommand{\thetaSposlowpl}{8.6}
\newcommand{\thetaSposmidpl}{9.4}
\newcommand{\thetaSposhighpl}{10.3}
\newcommand{\thetaSneglowpl}{8.6}
\newcommand{\thetaSnegmidpl}{9.4}
\newcommand{\thetaSneghighpl}{10.4}
\newcommand{\Magposlowpl}{14.19}
\newcommand{\Magposmidpl}{14.28}
\newcommand{\Magposhighpl}{14.37}
\newcommand{\Magneglowpl}{14.19}
\newcommand{\Magnegmidpl}{14.28}
\newcommand{\Magneghighpl}{14.38}
\newcommand{\Colorposlowpl}{1.58}
\newcommand{\Colorposmidpl}{1.64}
\newcommand{\Colorposhighpl}{1.70}
\newcommand{\Colorneglowpl}{1.58}
\newcommand{\Colornegmidpl}{1.64}
\newcommand{\Colorneghighpl}{1.70}
\newcommand{\thetaEposlowpl}{0.180}
\newcommand{\thetaEposmidpl}{0.197}
\newcommand{\thetaEposhighpl}{0.217}
\newcommand{\thetaEneglowpl}{0.180}
\newcommand{\thetaEnegmidpl}{0.198}
\newcommand{\thetaEneghighpl}{0.217}

\begin{deluxetable*}{lccc}
\tabletypesize{\scriptsize}
\tablecaption{Posterior summary of the light-curve and physical parameters for KB240211.
The two columns correspond to the annual-parallax solutions with $u_0>0$ and $u_0<0$, respectively.
Values are reported as the median and the 16th--84th percentile range of the posterior distribution.}
\tablehead{
\colhead{Parameter} & \colhead{Unit} &
\colhead{$u_0 > 0$} &
\colhead{$u_0 < 0$}
}\label{tab:posterior_kb240211}
\startdata
\cutinhead{Light-curve parameters}
$t_0 - 10390$ & $\mathrm{day}$ & $\tzeropospmpl$ & $\tzeronegpmpl$ \\
$t_{\rm E}$ & $\mathrm{day}$ & $\tEpospmpl$ & $\tEnegpmpl$ \\
$u_0$ & $10^{-2}$ & $\uzeropospmpl$ & $\uzeronegpmpl$ \\
$\rho$ & $10^{-2}$ & $\rhopospmpl$ & $\rhonegpmpl$ \\
$\pi_{\rm E,N}$ & --- & $\piENpospmpl$ & $\piENnegpmpl$ \\
$\pi_{\rm E,E}$ & --- & $\piEEpospmpl$ & $\piEEnegpmpl$ \\
\cutinhead{Physical parameters}
$M_{\rm L}$ & $M_\odot$ & $\MLpospmpl$ & $\MLnegpmpl$ \\
$D_{\rm L}$ & $\mathrm{kpc}$ & $\DLpospmpl$ & $\DLnegpmpl$ \\
$D_{\rm S}$ & $\mathrm{kpc}$ & $\DSpospmpl$ & $\DSnegpmpl$ \\
$\mu_{\rm rel,N}$ & $\mathrm{mas}\,\mathrm{yr}^{-1}$ & $\muNpospmpl$ & $\muNnegpmpl$ \\
$\mu_{\rm rel,E}$ & $\mathrm{mas}\,\mathrm{yr}^{-1}$ & $\muEpospmpl$ & $\muEnegpmpl$ \\
$\mu_{\rm rel}$ & $\mathrm{mas}\,\mathrm{yr}^{-1}$ & $\muRelpospmpl$ & $\muRelnegpmpl$ \\
$\theta_*$ & $\mu\mathrm{as}$ & $\thetaSpospmpl$ & $\thetaSnegpmpl$ \\
$I_{s,0}$ & mag & $\Magpospmpl$ & $\Magnegpmpl$ \\
$(I-H)_{s,0}$ & mag & $\Colorpospmpl$ & $\Colornegpmpl$ \\
$\theta_{\rm E}$ & $\mathrm{mas}$ & $\thetaEpospmpl$ & $\thetaEnegpmpl$ \\
\enddata
\tablecomments{
The marginalized intervals of $\mu_{\rm rel,N}$ and $\mu_{\rm rel,E}$ should be interpreted with caution: Figure~\ref{fig:corner_y_kb240211} shows that the posterior in this plane is nearly ring-like, so $\mu_{\rm rel}$ is much better constrained than the direction of $\boldsymbol{\mu}_{\rm rel}$.
}
\end{deluxetable*}
\newcommand{\tzeroIHpmKB}{0.186^{+0.065}_{-0.057}}
\newcommand{\tzeroVIpmKB}{0.197^{+0.063}_{-0.061}}
\newcommand{\tzeroJHpmKB}{0.185^{+0.063}_{-0.056}}
\newcommand{\tzeroVHpmKB}{0.189^{+0.063}_{-0.053}}
\newcommand{\tzeroVHnegpmKB}{0.147^{+0.050}_{-0.044}}
\newcommand{\tEIHpmKB}{148.01^{+5.94}_{-5.01}}
\newcommand{\tEVIpmKB}{147.42^{+5.72}_{-4.94}}
\newcommand{\tEJHpmKB}{147.62^{+5.78}_{-4.89}}
\newcommand{\tEVHpmKB}{148.08^{+5.52}_{-4.71}}
\newcommand{\tEVHnegpmKB}{143.39^{+4.55}_{-4.05}}
\newcommand{\uzeroIHpmKB}{3.70^{+0.14}_{-0.15}}
\newcommand{\uzeroVIpmKB}{3.71^{+0.14}_{-0.15}}
\newcommand{\uzeroJHpmKB}{3.71^{+0.14}_{-0.15}}
\newcommand{\uzeroVHpmKB}{3.69^{+0.13}_{-0.14}}
\newcommand{\uzeroVHnegpmKB}{-3.77^{+0.13}_{-0.12}}
\newcommand{\rhoIHpmKB}{0.54^{+0.03}_{-0.03}}
\newcommand{\rhoVIpmKB}{0.54^{+0.03}_{-0.03}}
\newcommand{\rhoJHpmKB}{0.54^{+0.03}_{-0.03}}
\newcommand{\rhoVHpmKB}{0.54^{+0.03}_{-0.03}}
\newcommand{\rhoVHnegpmKB}{0.56^{+0.02}_{-0.02}}
\newcommand{\qIHpmKB}{0.995^{+0.004}_{-0.008}}
\newcommand{\qVIpmKB}{0.995^{+0.004}_{-0.009}}
\newcommand{\qJHpmKB}{0.995^{+0.004}_{-0.008}}
\newcommand{\qVHpmKB}{0.995^{+0.004}_{-0.009}}
\newcommand{\qVHnegpmKB}{0.995^{+0.004}_{-0.008}}
\newcommand{\sIHpmKB}{0.255^{+0.004}_{-0.005}}
\newcommand{\sVIpmKB}{0.256^{+0.004}_{-0.005}}
\newcommand{\sJHpmKB}{0.256^{+0.004}_{-0.005}}
\newcommand{\sVHpmKB}{0.255^{+0.004}_{-0.004}}
\newcommand{\sVHnegpmKB}{0.257^{+0.004}_{-0.004}}
\newcommand{\alphaIHpmKB}{1.027^{+0.006}_{-0.007}}
\newcommand{\alphaVIpmKB}{1.025^{+0.007}_{-0.007}}
\newcommand{\alphaJHpmKB}{1.027^{+0.007}_{-0.007}}
\newcommand{\alphaVHpmKB}{1.027^{+0.006}_{-0.006}}
\newcommand{\alphaVHnegpmKB}{-1.030^{+0.005}_{-0.005}}
\newcommand{\piENIHpmKB}{0.154^{+0.023}_{-0.024}}
\newcommand{\piENVIpmKB}{0.149^{+0.023}_{-0.026}}
\newcommand{\piENJHpmKB}{0.154^{+0.022}_{-0.026}}
\newcommand{\piENVHpmKB}{0.155^{+0.022}_{-0.026}}
\newcommand{\piENVHnegpmKB}{-0.070^{+0.033}_{-0.035}}
\newcommand{\piEEIHpmKB}{-0.057^{+0.006}_{-0.005}}
\newcommand{\piEEVIpmKB}{-0.058^{+0.005}_{-0.006}}
\newcommand{\piEEJHpmKB}{-0.057^{+0.005}_{-0.006}}
\newcommand{\piEEVHpmKB}{-0.057^{+0.006}_{-0.006}}
\newcommand{\piEEVHnegpmKB}{-0.075^{+0.003}_{-0.003}}
\newcommand{\goneIHpmKB}{-11.52^{+0.47}_{-0.49}}
\newcommand{\goneVIpmKB}{-11.59^{+0.48}_{-0.49}}
\newcommand{\goneJHpmKB}{-11.52^{+0.47}_{-0.49}}
\newcommand{\goneVHpmKB}{-11.69^{+0.53}_{-0.48}}
\newcommand{\goneVHnegpmKB}{-11.12^{+0.44}_{-0.42}}
\newcommand{\gtwoIHpmKB}{4.54^{+1.09}_{-1.08}}
\newcommand{\gtwoVIpmKB}{4.81^{+1.16}_{-1.13}}
\newcommand{\gtwoJHpmKB}{4.54^{+1.08}_{-1.16}}
\newcommand{\gtwoVHpmKB}{4.53^{+1.00}_{-1.00}}
\newcommand{\gtwoVHnegpmKB}{-6.18^{+1.12}_{-1.08}}
\newcommand{\gthreeIHpmKB}{17.18^{+0.91}_{-0.88}}
\newcommand{\gthreeVIpmKB}{17.62^{+1.01}_{-0.97}}
\newcommand{\gthreeJHpmKB}{17.23^{+1.01}_{-1.00}}
\newcommand{\gthreeVHpmKB}{17.40^{+0.85}_{-0.92}}
\newcommand{\gthreeVHnegpmKB}{18.27^{+0.91}_{-0.92}}
\newcommand{\piEIHpmKB}{0.164^{+0.020}_{-0.021}}
\newcommand{\piEVIpmKB}{0.160^{+0.020}_{-0.022}}
\newcommand{\piEJHpmKB}{0.164^{+0.020}_{-0.022}}
\newcommand{\piEVHpmKB}{0.165^{+0.019}_{-0.022}}
\newcommand{\piEVHnegpmKB}{0.103^{+0.025}_{-0.019}}

\newcommand{\MLIHpmKB}{0.365^{+0.058}_{-0.044}}
\newcommand{\MLVIpmKB}{0.309^{+0.055}_{-0.047}}
\newcommand{\MLJHpmKB}{0.365^{+0.071}_{-0.056}}
\newcommand{\MLVHpmKB}{0.335^{+0.050}_{-0.037}}
\newcommand{\MLVHnegpmKB}{0.523^{+0.116}_{-0.104}}
\newcommand{\DLIHpmKB}{4.70^{+0.38}_{-0.32}}
\newcommand{\DLVIpmKB}{5.30^{+0.62}_{-0.42}}
\newcommand{\DLJHpmKB}{4.75^{+0.50}_{-0.47}}
\newcommand{\DLVHpmKB}{4.93^{+0.32}_{-0.27}}
\newcommand{\DLVHnegpmKB}{5.68^{+0.40}_{-0.40}}
\newcommand{\DSIHpmKB}{7.58^{+0.32}_{-0.37}}
\newcommand{\DSVIpmKB}{8.00^{+0.76}_{-0.38}}
\newcommand{\DSJHpmKB}{7.60^{+0.40}_{-0.45}}
\newcommand{\DSVHpmKB}{7.77^{+0.33}_{-0.30}}
\newcommand{\DSVHnegpmKB}{7.71^{+0.27}_{-0.29}}
\newcommand{\muNIHpmKB}{1.14^{+0.11}_{-0.10}}
\newcommand{\muNVIpmKB}{0.94^{+0.11}_{-0.13}}
\newcommand{\muNJHpmKB}{1.13^{+0.18}_{-0.15}}
\newcommand{\muNVHpmKB}{1.05^{+0.06}_{-0.06}}
\newcommand{\muNVHnegpmKB}{-0.75^{+0.26}_{-0.14}}
\newcommand{\muEIHpmKB}{-0.35^{+0.09}_{-0.11}}
\newcommand{\muEVIpmKB}{-0.31^{+0.08}_{-0.10}}
\newcommand{\muEJHpmKB}{-0.36^{+0.09}_{-0.12}}
\newcommand{\muEVHpmKB}{-0.33^{+0.08}_{-0.11}}
\newcommand{\muEVHnegpmKB}{-0.78^{+0.19}_{-0.20}}
\newcommand{\aIHpmKB}{0.70^{+0.04}_{-0.04}}
\newcommand{\aVIpmKB}{0.65^{+0.04}_{-0.04}}
\newcommand{\aJHpmKB}{0.70^{+0.05}_{-0.05}}
\newcommand{\aVHpmKB}{0.68^{+0.03}_{-0.03}}
\newcommand{\aVHnegpmKB}{0.75^{+0.05}_{-0.04}}
\newcommand{\cosiIHpmKB}{0.178^{+0.039}_{-0.040}}
\newcommand{\cosiVIpmKB}{0.186^{+0.039}_{-0.042}}
\newcommand{\cosiJHpmKB}{0.178^{+0.038}_{-0.043}}
\newcommand{\cosiVHpmKB}{0.175^{+0.036}_{-0.037}}
\newcommand{\cosiVHnegpmKB}{-0.237^{+0.040}_{-0.037}}
\newcommand{\OmegaIHpmKB}{-1.501^{+0.077}_{-0.095}}
\newcommand{\OmegaVIpmKB}{-1.519^{+0.082}_{-0.104}}
\newcommand{\OmegaJHpmKB}{-1.501^{+0.077}_{-0.103}}
\newcommand{\OmegaVHpmKB}{-1.497^{+0.077}_{-0.101}}
\newcommand{\OmegaVHnegpmKB}{-1.149^{+0.301}_{-0.213}}
\newcommand{\phizeroIHpmKB}{0.601^{+0.033}_{-0.031}}
\newcommand{\phizeroVIpmKB}{0.593^{+0.033}_{-0.031}}
\newcommand{\phizeroJHpmKB}{0.600^{+0.034}_{-0.031}}
\newcommand{\phizeroVHpmKB}{0.601^{+0.036}_{-0.032}}
\newcommand{\phizeroVHnegpmKB}{0.565^{+0.025}_{-0.024}}
\newcommand{\thetaSIHpmKB}{2.62^{+0.25}_{-0.21}}
\newcommand{\thetaSVIpmKB}{2.19^{+0.22}_{-0.28}}
\newcommand{\thetaSJHpmKB}{2.61^{+0.41}_{-0.33}}
\newcommand{\thetaSVHpmKB}{2.43^{+0.11}_{-0.11}}
\newcommand{\thetaSVHnegpmKB}{2.45^{+0.11}_{-0.11}}
\newcommand{\MagIHpmKB}{16.410^{+0.070}_{-0.066}}
\newcommand{\MagVIpmKB}{16.429^{+0.068}_{-0.064}}
\newcommand{\MagJHpmKB}{15.107^{+0.092}_{-0.092}}
\newcommand{\MagVHpmKB}{15.123^{+0.086}_{-0.089}}
\newcommand{\MagVHnegpmKB}{15.100^{+0.088}_{-0.086}}
\newcommand{\ColorIHpmKB}{1.283^{+0.066}_{-0.065}}
\newcommand{\ColorVIpmKB}{0.917^{+0.070}_{-0.071}}
\newcommand{\ColorJHpmKB}{0.469^{+0.027}_{-0.027}}
\newcommand{\ColorVHpmKB}{2.171^{+0.099}_{-0.096}}
\newcommand{\ColorVHnegpmKB}{2.171^{+0.094}_{-0.100}}
\newcommand{\thetaEIHpmKB}{0.49^{+0.05}_{-0.04}}
\newcommand{\thetaEVIpmKB}{0.40^{+0.04}_{-0.05}}
\newcommand{\thetaEJHpmKB}{0.48^{+0.08}_{-0.06}}
\newcommand{\thetaEVHpmKB}{0.45^{+0.02}_{-0.02}}
\newcommand{\thetaEVHnegpmKB}{0.44^{+0.02}_{-0.02}}

\newcommand{\tzeroIHlowKB}{0.129}
\newcommand{\tzeroIHmidKB}{0.186}
\newcommand{\tzeroIHhighKB}{0.252}
\newcommand{\tzeroVIlowKB}{0.136}
\newcommand{\tzeroVImidKB}{0.197}
\newcommand{\tzeroVIhighKB}{0.261}
\newcommand{\tzeroJHlowKB}{0.129}
\newcommand{\tzeroJHmidKB}{0.185}
\newcommand{\tzeroJHhighKB}{0.248}
\newcommand{\tzeroVHlowKB}{0.136}
\newcommand{\tzeroVHmidKB}{0.189}
\newcommand{\tzeroVHhighKB}{0.252}
\newcommand{\tzeroVHneglowKB}{0.102}
\newcommand{\tzeroVHnegmidKB}{0.147}
\newcommand{\tzeroVHneghighKB}{0.196}
\newcommand{\tEIHlowKB}{142.99}
\newcommand{\tEIHmidKB}{148.01}
\newcommand{\tEIHhighKB}{153.94}
\newcommand{\tEVIlowKB}{142.48}
\newcommand{\tEVImidKB}{147.42}
\newcommand{\tEVIhighKB}{153.13}
\newcommand{\tEJHlowKB}{142.73}
\newcommand{\tEJHmidKB}{147.62}
\newcommand{\tEJHhighKB}{153.39}
\newcommand{\tEVHlowKB}{143.36}
\newcommand{\tEVHmidKB}{148.08}
\newcommand{\tEVHhighKB}{153.59}
\newcommand{\tEVHneglowKB}{139.35}
\newcommand{\tEVHnegmidKB}{143.39}
\newcommand{\tEVHneghighKB}{147.95}
\newcommand{\uzeroIHlowKB}{3.54}
\newcommand{\uzeroIHmidKB}{3.70}
\newcommand{\uzeroIHhighKB}{3.84}
\newcommand{\uzeroVIlowKB}{3.56}
\newcommand{\uzeroVImidKB}{3.71}
\newcommand{\uzeroVIhighKB}{3.85}
\newcommand{\uzeroJHlowKB}{3.56}
\newcommand{\uzeroJHmidKB}{3.71}
\newcommand{\uzeroJHhighKB}{3.84}
\newcommand{\uzeroVHlowKB}{3.56}
\newcommand{\uzeroVHmidKB}{3.69}
\newcommand{\uzeroVHhighKB}{3.83}
\newcommand{\uzeroVHneglowKB}{-3.89}
\newcommand{\uzeroVHnegmidKB}{-3.77}
\newcommand{\uzeroVHneghighKB}{-3.64}
\newcommand{\rhoIHlowKB}{0.51}
\newcommand{\rhoIHmidKB}{0.54}
\newcommand{\rhoIHhighKB}{0.57}
\newcommand{\rhoVIlowKB}{0.51}
\newcommand{\rhoVImidKB}{0.54}
\newcommand{\rhoVIhighKB}{0.57}
\newcommand{\rhoJHlowKB}{0.52}
\newcommand{\rhoJHmidKB}{0.54}
\newcommand{\rhoJHhighKB}{0.57}
\newcommand{\rhoVHlowKB}{0.52}
\newcommand{\rhoVHmidKB}{0.54}
\newcommand{\rhoVHhighKB}{0.57}
\newcommand{\rhoVHneglowKB}{0.53}
\newcommand{\rhoVHnegmidKB}{0.56}
\newcommand{\rhoVHneghighKB}{0.58}
\newcommand{\qIHlowKB}{0.987}
\newcommand{\qIHmidKB}{0.995}
\newcommand{\qIHhighKB}{0.999}
\newcommand{\qVIlowKB}{0.986}
\newcommand{\qVImidKB}{0.995}
\newcommand{\qVIhighKB}{0.999}
\newcommand{\qJHlowKB}{0.987}
\newcommand{\qJHmidKB}{0.995}
\newcommand{\qJHhighKB}{0.999}
\newcommand{\qVHlowKB}{0.986}
\newcommand{\qVHmidKB}{0.995}
\newcommand{\qVHhighKB}{0.999}
\newcommand{\qVHneglowKB}{0.988}
\newcommand{\qVHnegmidKB}{0.995}
\newcommand{\qVHneghighKB}{0.999}
\newcommand{\sIHlowKB}{0.251}
\newcommand{\sIHmidKB}{0.255}
\newcommand{\sIHhighKB}{0.260}
\newcommand{\sVIlowKB}{0.251}
\newcommand{\sVImidKB}{0.256}
\newcommand{\sVIhighKB}{0.260}
\newcommand{\sJHlowKB}{0.251}
\newcommand{\sJHmidKB}{0.256}
\newcommand{\sJHhighKB}{0.260}
\newcommand{\sVHlowKB}{0.250}
\newcommand{\sVHmidKB}{0.255}
\newcommand{\sVHhighKB}{0.259}
\newcommand{\sVHneglowKB}{0.254}
\newcommand{\sVHnegmidKB}{0.257}
\newcommand{\sVHneghighKB}{0.261}
\newcommand{\alphaIHlowKB}{1.019}
\newcommand{\alphaIHmidKB}{1.027}
\newcommand{\alphaIHhighKB}{1.033}
\newcommand{\alphaVIlowKB}{1.018}
\newcommand{\alphaVImidKB}{1.025}
\newcommand{\alphaVIhighKB}{1.032}
\newcommand{\alphaJHlowKB}{1.020}
\newcommand{\alphaJHmidKB}{1.027}
\newcommand{\alphaJHhighKB}{1.033}
\newcommand{\alphaVHlowKB}{1.021}
\newcommand{\alphaVHmidKB}{1.027}
\newcommand{\alphaVHhighKB}{1.033}
\newcommand{\alphaVHneglowKB}{-1.035}
\newcommand{\alphaVHnegmidKB}{-1.030}
\newcommand{\alphaVHneghighKB}{-1.025}
\newcommand{\piENIHlowKB}{0.130}
\newcommand{\piENIHmidKB}{0.154}
\newcommand{\piENIHhighKB}{0.177}
\newcommand{\piENVIlowKB}{0.123}
\newcommand{\piENVImidKB}{0.149}
\newcommand{\piENVIhighKB}{0.172}
\newcommand{\piENJHlowKB}{0.128}
\newcommand{\piENJHmidKB}{0.154}
\newcommand{\piENJHhighKB}{0.176}
\newcommand{\piENVHlowKB}{0.129}
\newcommand{\piENVHmidKB}{0.155}
\newcommand{\piENVHhighKB}{0.177}
\newcommand{\piENVHneglowKB}{-0.106}
\newcommand{\piENVHnegmidKB}{-0.070}
\newcommand{\piENVHneghighKB}{-0.038}
\newcommand{\piEEIHlowKB}{-0.062}
\newcommand{\piEEIHmidKB}{-0.057}
\newcommand{\piEEIHhighKB}{-0.051}
\newcommand{\piEEVIlowKB}{-0.063}
\newcommand{\piEEVImidKB}{-0.058}
\newcommand{\piEEVIhighKB}{-0.052}
\newcommand{\piEEJHlowKB}{-0.062}
\newcommand{\piEEJHmidKB}{-0.057}
\newcommand{\piEEJHhighKB}{-0.051}
\newcommand{\piEEVHlowKB}{-0.062}
\newcommand{\piEEVHmidKB}{-0.057}
\newcommand{\piEEVHhighKB}{-0.051}
\newcommand{\piEEVHneglowKB}{-0.078}
\newcommand{\piEEVHnegmidKB}{-0.075}
\newcommand{\piEEVHneghighKB}{-0.072}
\newcommand{\goneIHlowKB}{-12.01}
\newcommand{\goneIHmidKB}{-11.52}
\newcommand{\goneIHhighKB}{-11.05}
\newcommand{\goneVIlowKB}{-12.08}
\newcommand{\goneVImidKB}{-11.59}
\newcommand{\goneVIhighKB}{-11.11}
\newcommand{\goneJHlowKB}{-12.01}
\newcommand{\goneJHmidKB}{-11.52}
\newcommand{\goneJHhighKB}{-11.05}
\newcommand{\goneVHlowKB}{-12.17}
\newcommand{\goneVHmidKB}{-11.69}
\newcommand{\goneVHhighKB}{-11.16}
\newcommand{\goneVHneglowKB}{-11.54}
\newcommand{\goneVHnegmidKB}{-11.12}
\newcommand{\goneVHneghighKB}{-10.67}
\newcommand{\gtwoIHlowKB}{3.46}
\newcommand{\gtwoIHmidKB}{4.54}
\newcommand{\gtwoIHhighKB}{5.63}
\newcommand{\gtwoVIlowKB}{3.68}
\newcommand{\gtwoVImidKB}{4.81}
\newcommand{\gtwoVIhighKB}{5.97}
\newcommand{\gtwoJHlowKB}{3.38}
\newcommand{\gtwoJHmidKB}{4.54}
\newcommand{\gtwoJHhighKB}{5.62}
\newcommand{\gtwoVHlowKB}{3.53}
\newcommand{\gtwoVHmidKB}{4.53}
\newcommand{\gtwoVHhighKB}{5.52}
\newcommand{\gtwoVHneglowKB}{-7.25}
\newcommand{\gtwoVHnegmidKB}{-6.18}
\newcommand{\gtwoVHneghighKB}{-5.05}
\newcommand{\gthreeIHlowKB}{16.30}
\newcommand{\gthreeIHmidKB}{17.18}
\newcommand{\gthreeIHhighKB}{18.09}
\newcommand{\gthreeVIlowKB}{16.66}
\newcommand{\gthreeVImidKB}{17.62}
\newcommand{\gthreeVIhighKB}{18.63}
\newcommand{\gthreeJHlowKB}{16.23}
\newcommand{\gthreeJHmidKB}{17.23}
\newcommand{\gthreeJHhighKB}{18.24}
\newcommand{\gthreeVHlowKB}{16.48}
\newcommand{\gthreeVHmidKB}{17.40}
\newcommand{\gthreeVHhighKB}{18.25}
\newcommand{\gthreeVHneglowKB}{17.35}
\newcommand{\gthreeVHnegmidKB}{18.27}
\newcommand{\gthreeVHneghighKB}{19.17}
\newcommand{\piEIHlowKB}{0.143}
\newcommand{\piEIHmidKB}{0.164}
\newcommand{\piEIHhighKB}{0.184}
\newcommand{\piEVIlowKB}{0.138}
\newcommand{\piEVImidKB}{0.160}
\newcommand{\piEVIhighKB}{0.180}
\newcommand{\piEJHlowKB}{0.142}
\newcommand{\piEJHmidKB}{0.164}
\newcommand{\piEJHhighKB}{0.184}
\newcommand{\piEVHlowKB}{0.143}
\newcommand{\piEVHmidKB}{0.165}
\newcommand{\piEVHhighKB}{0.185}
\newcommand{\piEVHneglowKB}{0.084}
\newcommand{\piEVHnegmidKB}{0.103}
\newcommand{\piEVHneghighKB}{0.129}

\newcommand{\MLIHlowKB}{0.321}
\newcommand{\MLIHmidKB}{0.365}
\newcommand{\MLIHhighKB}{0.423}
\newcommand{\MLVIlowKB}{0.262}
\newcommand{\MLVImidKB}{0.309}
\newcommand{\MLVIhighKB}{0.364}
\newcommand{\MLJHlowKB}{0.309}
\newcommand{\MLJHmidKB}{0.365}
\newcommand{\MLJHhighKB}{0.435}
\newcommand{\MLVHlowKB}{0.298}
\newcommand{\MLVHmidKB}{0.335}
\newcommand{\MLVHhighKB}{0.385}
\newcommand{\MLVHneglowKB}{0.418}
\newcommand{\MLVHnegmidKB}{0.523}
\newcommand{\MLVHneghighKB}{0.638}
\newcommand{\DLIHlowKB}{4.39}
\newcommand{\DLIHmidKB}{4.70}
\newcommand{\DLIHhighKB}{5.08}
\newcommand{\DLVIlowKB}{4.88}
\newcommand{\DLVImidKB}{5.30}
\newcommand{\DLVIhighKB}{5.92}
\newcommand{\DLJHlowKB}{4.28}
\newcommand{\DLJHmidKB}{4.75}
\newcommand{\DLJHhighKB}{5.25}
\newcommand{\DLVHlowKB}{4.66}
\newcommand{\DLVHmidKB}{4.93}
\newcommand{\DLVHhighKB}{5.25}
\newcommand{\DLVHneglowKB}{5.29}
\newcommand{\DLVHnegmidKB}{5.68}
\newcommand{\DLVHneghighKB}{6.09}
\newcommand{\DSIHlowKB}{7.21}
\newcommand{\DSIHmidKB}{7.58}
\newcommand{\DSIHhighKB}{7.90}
\newcommand{\DSVIlowKB}{7.63}
\newcommand{\DSVImidKB}{8.00}
\newcommand{\DSVIhighKB}{8.76}
\newcommand{\DSJHlowKB}{7.15}
\newcommand{\DSJHmidKB}{7.60}
\newcommand{\DSJHhighKB}{8.00}
\newcommand{\DSVHlowKB}{7.47}
\newcommand{\DSVHmidKB}{7.77}
\newcommand{\DSVHhighKB}{8.10}
\newcommand{\DSVHneglowKB}{7.42}
\newcommand{\DSVHnegmidKB}{7.71}
\newcommand{\DSVHneghighKB}{7.98}
\newcommand{\muNIHlowKB}{1.04}
\newcommand{\muNIHmidKB}{1.14}
\newcommand{\muNIHhighKB}{1.25}
\newcommand{\muNVIlowKB}{0.81}
\newcommand{\muNVImidKB}{0.94}
\newcommand{\muNVIhighKB}{1.05}
\newcommand{\muNJHlowKB}{0.98}
\newcommand{\muNJHmidKB}{1.13}
\newcommand{\muNJHhighKB}{1.30}
\newcommand{\muNVHlowKB}{0.99}
\newcommand{\muNVHmidKB}{1.05}
\newcommand{\muNVHhighKB}{1.11}
\newcommand{\muNVHneglowKB}{-0.89}
\newcommand{\muNVHnegmidKB}{-0.75}
\newcommand{\muNVHneghighKB}{-0.49}
\newcommand{\muEIHlowKB}{-0.46}
\newcommand{\muEIHmidKB}{-0.35}
\newcommand{\muEIHhighKB}{-0.27}
\newcommand{\muEVIlowKB}{-0.41}
\newcommand{\muEVImidKB}{-0.31}
\newcommand{\muEVIhighKB}{-0.23}
\newcommand{\muEJHlowKB}{-0.47}
\newcommand{\muEJHmidKB}{-0.36}
\newcommand{\muEJHhighKB}{-0.26}
\newcommand{\muEVHlowKB}{-0.43}
\newcommand{\muEVHmidKB}{-0.33}
\newcommand{\muEVHhighKB}{-0.24}
\newcommand{\muEVHneglowKB}{-0.98}
\newcommand{\muEVHnegmidKB}{-0.78}
\newcommand{\muEVHneghighKB}{-0.59}
\newcommand{\aIHlowKB}{0.67}
\newcommand{\aIHmidKB}{0.70}
\newcommand{\aIHhighKB}{0.75}
\newcommand{\aVIlowKB}{0.62}
\newcommand{\aVImidKB}{0.65}
\newcommand{\aVIhighKB}{0.69}
\newcommand{\aJHlowKB}{0.66}
\newcommand{\aJHmidKB}{0.70}
\newcommand{\aJHhighKB}{0.76}
\newcommand{\aVHlowKB}{0.65}
\newcommand{\aVHmidKB}{0.68}
\newcommand{\aVHhighKB}{0.71}
\newcommand{\aVHneglowKB}{0.71}
\newcommand{\aVHnegmidKB}{0.75}
\newcommand{\aVHneghighKB}{0.80}
\newcommand{\cosiIHlowKB}{0.137}
\newcommand{\cosiIHmidKB}{0.178}
\newcommand{\cosiIHhighKB}{0.217}
\newcommand{\cosiVIlowKB}{0.144}
\newcommand{\cosiVImidKB}{0.186}
\newcommand{\cosiVIhighKB}{0.225}
\newcommand{\cosiJHlowKB}{0.135}
\newcommand{\cosiJHmidKB}{0.178}
\newcommand{\cosiJHhighKB}{0.216}
\newcommand{\cosiVHlowKB}{0.138}
\newcommand{\cosiVHmidKB}{0.175}
\newcommand{\cosiVHhighKB}{0.212}
\newcommand{\cosiVHneglowKB}{-0.274}
\newcommand{\cosiVHnegmidKB}{-0.237}
\newcommand{\cosiVHneghighKB}{-0.197}
\newcommand{\OmegaIHlowKB}{-1.596}
\newcommand{\OmegaIHmidKB}{-1.501}
\newcommand{\OmegaIHhighKB}{-1.424}
\newcommand{\OmegaVIlowKB}{-1.623}
\newcommand{\OmegaVImidKB}{-1.519}
\newcommand{\OmegaVIhighKB}{-1.437}
\newcommand{\OmegaJHlowKB}{-1.605}
\newcommand{\OmegaJHmidKB}{-1.501}
\newcommand{\OmegaJHhighKB}{-1.424}
\newcommand{\OmegaVHlowKB}{-1.597}
\newcommand{\OmegaVHmidKB}{-1.497}
\newcommand{\OmegaVHhighKB}{-1.420}
\newcommand{\OmegaVHneglowKB}{-1.362}
\newcommand{\OmegaVHnegmidKB}{-1.149}
\newcommand{\OmegaVHneghighKB}{-0.848}
\newcommand{\phizeroIHlowKB}{0.570}
\newcommand{\phizeroIHmidKB}{0.601}
\newcommand{\phizeroIHhighKB}{0.634}
\newcommand{\phizeroVIlowKB}{0.563}
\newcommand{\phizeroVImidKB}{0.593}
\newcommand{\phizeroVIhighKB}{0.626}
\newcommand{\phizeroJHlowKB}{0.569}
\newcommand{\phizeroJHmidKB}{0.600}
\newcommand{\phizeroJHhighKB}{0.634}
\newcommand{\phizeroVHlowKB}{0.569}
\newcommand{\phizeroVHmidKB}{0.601}
\newcommand{\phizeroVHhighKB}{0.637}
\newcommand{\phizeroVHneglowKB}{0.540}
\newcommand{\phizeroVHnegmidKB}{0.565}
\newcommand{\phizeroVHneghighKB}{0.590}
\newcommand{\thetaSIHlowKB}{2.42}
\newcommand{\thetaSIHmidKB}{2.62}
\newcommand{\thetaSIHhighKB}{2.87}
\newcommand{\thetaSVIlowKB}{1.90}
\newcommand{\thetaSVImidKB}{2.19}
\newcommand{\thetaSVIhighKB}{2.40}
\newcommand{\thetaSJHlowKB}{2.28}
\newcommand{\thetaSJHmidKB}{2.61}
\newcommand{\thetaSJHhighKB}{3.01}
\newcommand{\thetaSVHlowKB}{2.32}
\newcommand{\thetaSVHmidKB}{2.43}
\newcommand{\thetaSVHhighKB}{2.54}
\newcommand{\thetaSVHneglowKB}{2.34}
\newcommand{\thetaSVHnegmidKB}{2.45}
\newcommand{\thetaSVHneghighKB}{2.56}
\newcommand{\MagIHlowKB}{16.344}
\newcommand{\MagIHmidKB}{16.410}
\newcommand{\MagIHhighKB}{16.480}
\newcommand{\MagVIlowKB}{16.365}
\newcommand{\MagVImidKB}{16.429}
\newcommand{\MagVIhighKB}{16.496}
\newcommand{\MagJHlowKB}{15.015}
\newcommand{\MagJHmidKB}{15.107}
\newcommand{\MagJHhighKB}{15.198}
\newcommand{\MagVHlowKB}{15.033}
\newcommand{\MagVHmidKB}{15.123}
\newcommand{\MagVHhighKB}{15.209}
\newcommand{\MagVHneglowKB}{15.014}
\newcommand{\MagVHnegmidKB}{15.100}
\newcommand{\MagVHneghighKB}{15.188}
\newcommand{\ColorIHlowKB}{1.218}
\newcommand{\ColorIHmidKB}{1.283}
\newcommand{\ColorIHhighKB}{1.349}
\newcommand{\ColorVIlowKB}{0.847}
\newcommand{\ColorVImidKB}{0.917}
\newcommand{\ColorVIhighKB}{0.987}
\newcommand{\ColorJHlowKB}{0.442}
\newcommand{\ColorJHmidKB}{0.469}
\newcommand{\ColorJHhighKB}{0.495}
\newcommand{\ColorVHlowKB}{2.075}
\newcommand{\ColorVHmidKB}{2.171}
\newcommand{\ColorVHhighKB}{2.269}
\newcommand{\ColorVHneglowKB}{2.071}
\newcommand{\ColorVHnegmidKB}{2.171}
\newcommand{\ColorVHneghighKB}{2.265}
\newcommand{\thetaEIHlowKB}{0.44}
\newcommand{\thetaEIHmidKB}{0.49}
\newcommand{\thetaEIHhighKB}{0.53}
\newcommand{\thetaEVIlowKB}{0.35}
\newcommand{\thetaEVImidKB}{0.40}
\newcommand{\thetaEVIhighKB}{0.45}
\newcommand{\thetaEJHlowKB}{0.42}
\newcommand{\thetaEJHmidKB}{0.48}
\newcommand{\thetaEJHhighKB}{0.55}
\newcommand{\thetaEVHlowKB}{0.43}
\newcommand{\thetaEVHmidKB}{0.45}
\newcommand{\thetaEVHhighKB}{0.47}
\newcommand{\thetaEVHneglowKB}{0.42}
\newcommand{\thetaEVHnegmidKB}{0.44}
\newcommand{\thetaEVHneghighKB}{0.46}
\begin{deluxetable*}{lcccccc}
\tabletypesize{\scriptsize}
\tablecaption{Posterior summary of the light-curve and physical parameters for KB241522.
The first four solution columns correspond to the analyses using the
$(I-H)$, $(V-I)$, $(J-H)$, and $(V-H)$ color constraints, respectively.
The final column gives the secondary $(V-H)$ solution with $u_0<0$.
Values are reported as the median and the 16th--84th percentile range of the posterior distribution.}
\tablehead{
\colhead{Parameter} & \colhead{Unit} & \colhead{$I,\ I-H$} & \colhead{$I,\ V-I$} & \colhead{$H,\ J-H$} & \colhead{$H,\ V-H$} & \colhead{$H,\ V-H\ (u_0<0)$}
}\label{tab:posterior_kb241522}
\startdata
\cutinhead{Light-curve parameters}
$t_0 - 10578$ & $\mathrm{day}$ & $\tzeroIHpmKB$ & $\tzeroVIpmKB$ & $\tzeroJHpmKB$ & $\tzeroVHpmKB$ & $\tzeroVHnegpmKB$ \\
$t_{\rm E}$ & $\mathrm{day}$ & $\tEIHpmKB$ & $\tEVIpmKB$ & $\tEJHpmKB$ & $\tEVHpmKB$ & $\tEVHnegpmKB$ \\
$u_0$ & $10^{-2}$ & $\uzeroIHpmKB$ & $\uzeroVIpmKB$ & $\uzeroJHpmKB$ & $\uzeroVHpmKB$ & $\uzeroVHnegpmKB$ \\
$\rho$ & $10^{-2}$ & $\rhoIHpmKB$ & $\rhoVIpmKB$ & $\rhoJHpmKB$ & $\rhoVHpmKB$ & $\rhoVHnegpmKB$ \\
$q$ & --- & $\qIHpmKB$ & $\qVIpmKB$ & $\qJHpmKB$ & $\qVHpmKB$ & $\qVHnegpmKB$ \\
$s$ & --- & $\sIHpmKB$ & $\sVIpmKB$ & $\sJHpmKB$ & $\sVHpmKB$ & $\sVHnegpmKB$ \\
$\alpha$ & $\mathrm{rad}$ & $\alphaIHpmKB$ & $\alphaVIpmKB$ & $\alphaJHpmKB$ & $\alphaVHpmKB$ & $\alphaVHnegpmKB$ \\
$\pi_{\rm E,N}$ & --- & $\piENIHpmKB$ & $\piENVIpmKB$ & $\piENJHpmKB$ & $\piENVHpmKB$ & $\piENVHnegpmKB$ \\
$\pi_{\rm E,E}$ & --- & $\piEEIHpmKB$ & $\piEEVIpmKB$ & $\piEEJHpmKB$ & $\piEEVHpmKB$ & $\piEEVHnegpmKB$ \\
$\gamma_1$ & $10^{-3}\,\mathrm{day}^{-1}$ & $\goneIHpmKB$ & $\goneVIpmKB$ & $\goneJHpmKB$ & $\goneVHpmKB$ & $\goneVHnegpmKB$ \\
$\gamma_2$ & $10^{-3}\,\mathrm{day}^{-1}$ & $\gtwoIHpmKB$ & $\gtwoVIpmKB$ & $\gtwoJHpmKB$ & $\gtwoVHpmKB$ & $\gtwoVHnegpmKB$ \\
$\gamma_3$ & $10^{-3}\,\mathrm{day}^{-1}$ & $\gthreeIHpmKB$ & $\gthreeVIpmKB$ & $\gthreeJHpmKB$ & $\gthreeVHpmKB$ & $\gthreeVHnegpmKB$ \\
$\pi_{\rm E}$ & --- & $\piEIHpmKB$ & $\piEVIpmKB$ & $\piEJHpmKB$ & $\piEVHpmKB$ & $\piEVHnegpmKB$ \\
\cutinhead{Physical parameters}
$M_{\rm L}$ & $M_\odot$ & $\MLIHpmKB$ & $\MLVIpmKB$ & $\MLJHpmKB$ & $\MLVHpmKB$ & $\MLVHnegpmKB$ \\
$D_{\rm L}$ & $\mathrm{kpc}$ & $\DLIHpmKB$ & $\DLVIpmKB$ & $\DLJHpmKB$ & $\DLVHpmKB$ & $\DLVHnegpmKB$ \\
$D_{\rm S}$ & $\mathrm{kpc}$ & $\DSIHpmKB$ & $\DSVIpmKB$ & $\DSJHpmKB$ & $\DSVHpmKB$ & $\DSVHnegpmKB$ \\
$\mu_{\rm rel,N}$ & $\mathrm{mas}\,\mathrm{yr}^{-1}$ & $\muNIHpmKB$ & $\muNVIpmKB$ & $\muNJHpmKB$ & $\muNVHpmKB$ & $\muNVHnegpmKB$ \\
$\mu_{\rm rel,E}$ & $\mathrm{mas}\,\mathrm{yr}^{-1}$ & $\muEIHpmKB$ & $\muEVIpmKB$ & $\muEJHpmKB$ & $\muEVHpmKB$ & $\muEVHnegpmKB$ \\
$a$ & $\mathrm{au}$ & $\aIHpmKB$ & $\aVIpmKB$ & $\aJHpmKB$ & $\aVHpmKB$ & $\aVHnegpmKB$ \\
$\cos i$ & --- & $\cosiIHpmKB$ & $\cosiVIpmKB$ & $\cosiJHpmKB$ & $\cosiVHpmKB$ & $\cosiVHnegpmKB$ \\
$\Omega_{\rm NE}$ & $\mathrm{rad}$ & $\OmegaIHpmKB$ & $\OmegaVIpmKB$ & $\OmegaJHpmKB$ & $\OmegaVHpmKB$ & $\OmegaVHnegpmKB$ \\
$\phi_0$ & $\mathrm{rad}$ & $\phizeroIHpmKB$ & $\phizeroVIpmKB$ & $\phizeroJHpmKB$ & $\phizeroVHpmKB$ & $\phizeroVHnegpmKB$ \\
$\theta_*$ & $\mu\mathrm{as}$ & $\thetaSIHpmKB$ & $\thetaSVIpmKB$ & $\thetaSJHpmKB$ & $\thetaSVHpmKB$ & $\thetaSVHnegpmKB$ \\
$m_{s,0}^{a}$ & mag & $\MagIHpmKB$ & $\MagVIpmKB$ & $\MagJHpmKB$ & $\MagVHpmKB$ & $\MagVHnegpmKB$ \\
$\mathrm{Color}_{s,0}^{a}$ & mag & $\ColorIHpmKB$ & $\ColorVIpmKB$ & $\ColorJHpmKB$ & $\ColorVHpmKB$ & $\ColorVHnegpmKB$ \\
$\theta_{\rm E}$ & $\mathrm{mas}$ & $\thetaEIHpmKB$ & $\thetaEVIpmKB$ & $\thetaEJHpmKB$ & $\thetaEVHpmKB$ & $\thetaEVHnegpmKB$ \\
\enddata
\tablenotetext{a}{
For each column, $m_{s,0}$ and $\mathrm{Color}_{s,0}$ denote the
dereddened source magnitude and color specified by the column header.
}
\end{deluxetable*}
\begin{figure}[t]
    \centering
    \includegraphics[width=0.7\textwidth]{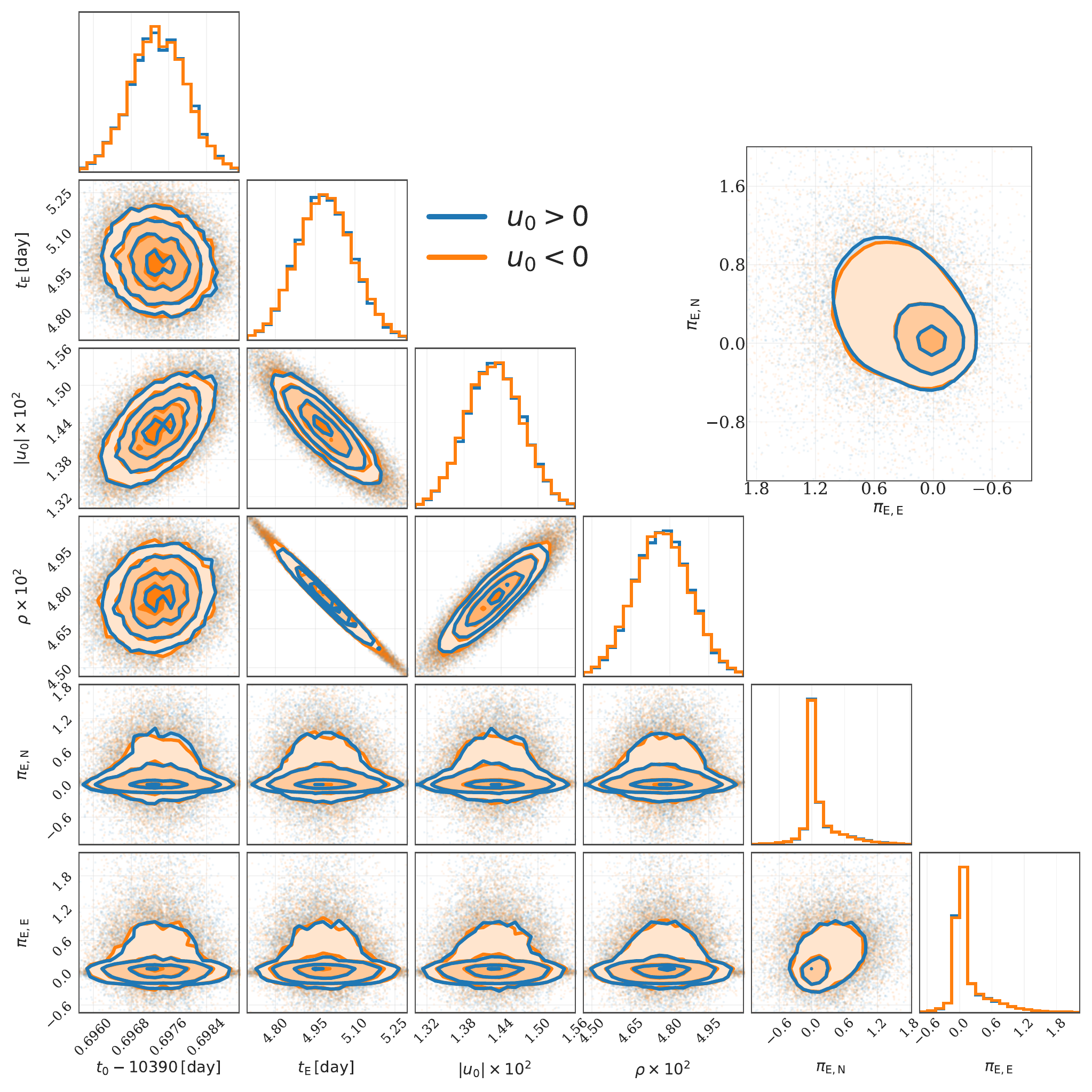}
    \caption{
        Posterior distributions of the microlensing parameters for KB240211.
        The blue and orange contours correspond to the $u_0>0$ and $u_0<0$ parallax solutions, respectively.
        Contours correspond to the 0.5$\sigma$, 1$\sigma$, 1.5$\sigma$, and 2$\sigma$ confidence levels.
    }
    \label{fig:corner_x_kb240211}
\end{figure}

\begin{figure}[t]
    \centering
    \includegraphics[width=0.7\textwidth]{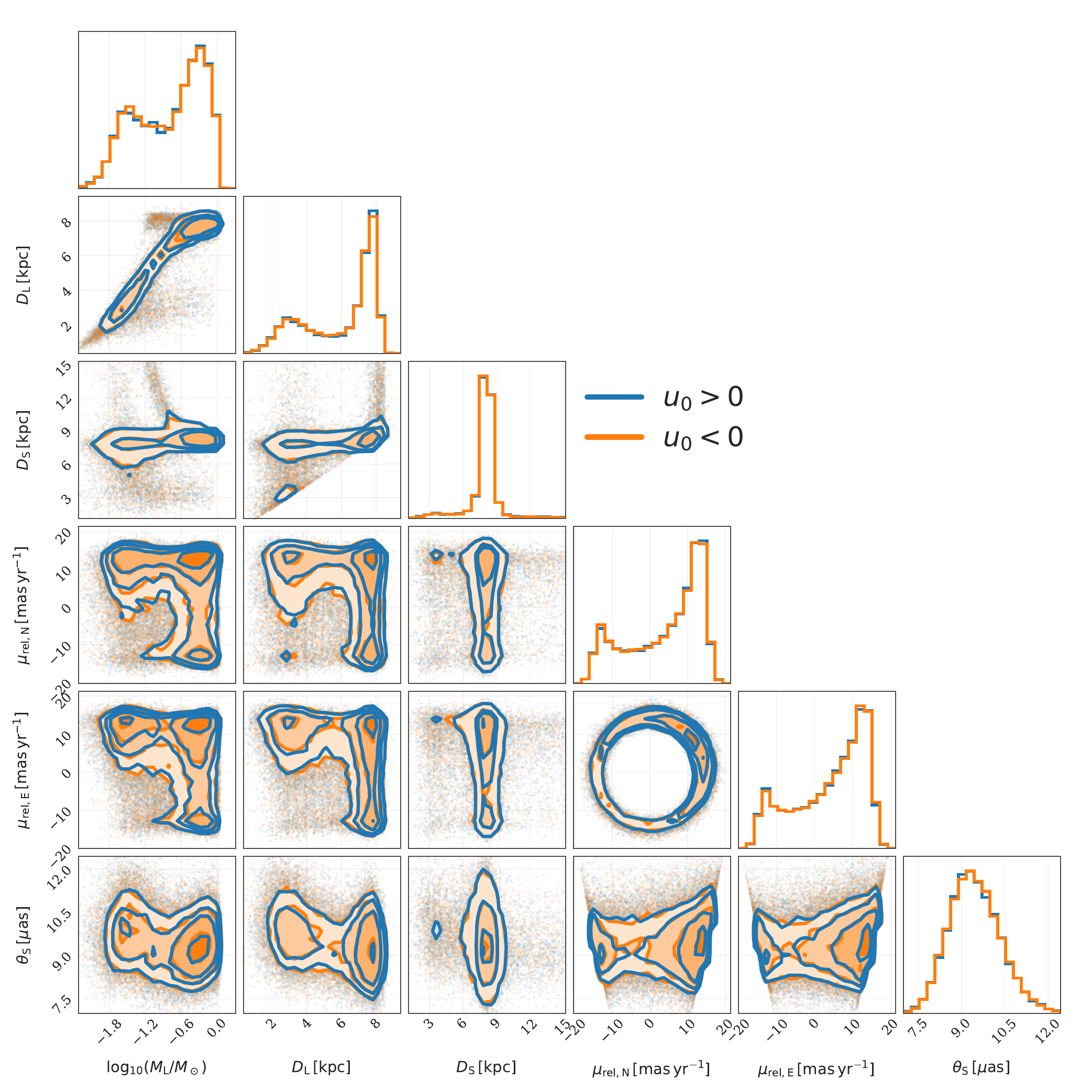}
    \caption{
        Posterior distributions of the physical parameters for KB240211.
        The color coding is the same as in Figure~\ref{fig:corner_x_kb240211}. Contours correspond to the 0.5$\sigma$, 1$\sigma$, 1.5$\sigma$, and 2$\sigma$ confidence levels.
    }
    \label{fig:corner_y_kb240211}
\end{figure}

\begin{figure}[t]
    \centering
    \includegraphics[width=0.9\textwidth]{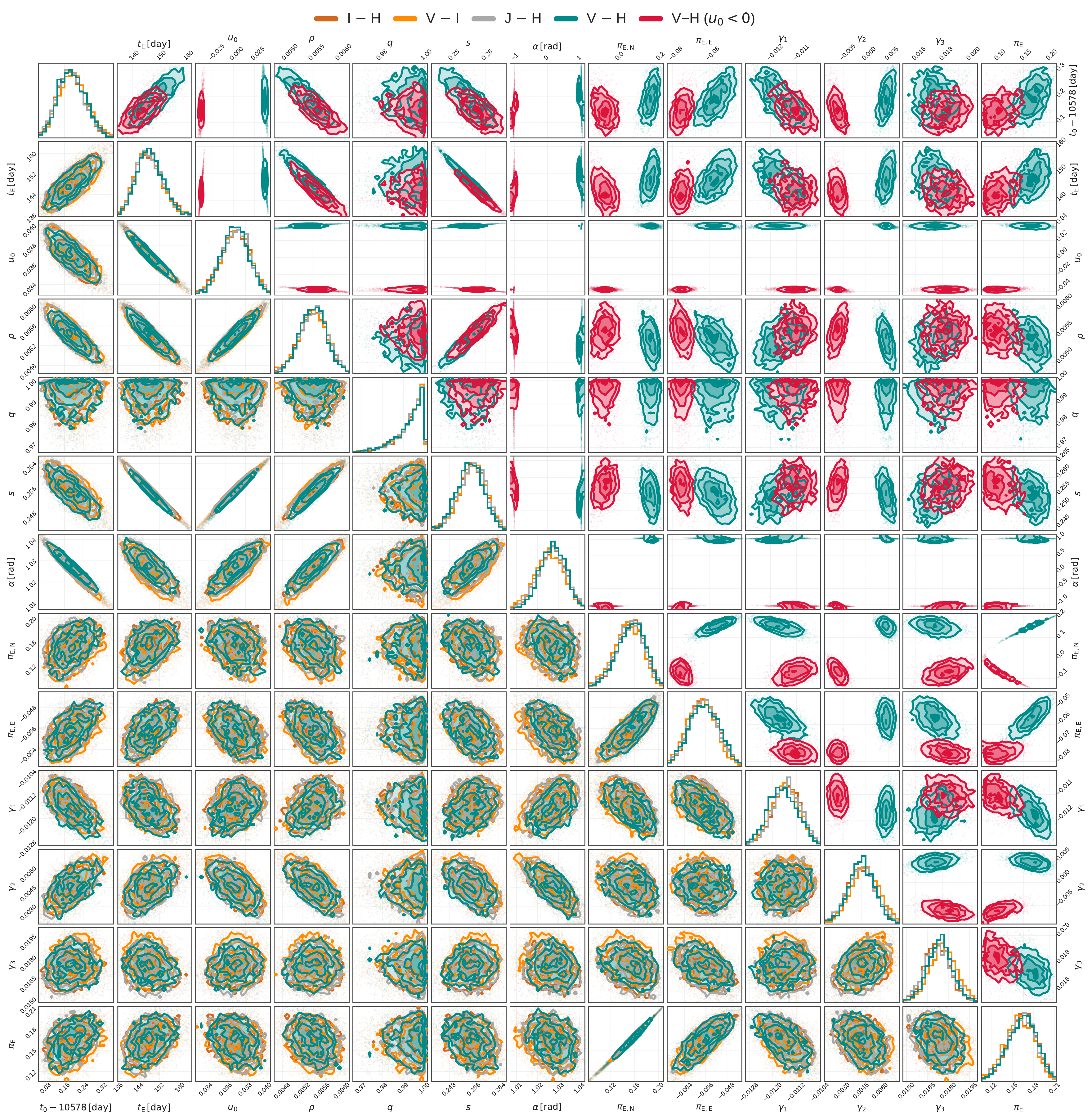}
    \caption{
    Posterior distributions of the microlensing parameters for KB241522.
    The contours show the results obtained with the $(I-H)$, $(V-I)$,
    $(J-H)$, and $(V-H)$ source-color constraints, together with the
    additional $V-H$ solution on the $u_0<0$ parallax branch.
    The $u_0<0$ branch is shown separately because its parameter ranges,
    in particular those of $u_0$, $\alpha$, and $\bm{\pi}_{\rm E}$,
    differ from those of the main $u_0>0$ branch.
    Contours correspond to the 0.5$\sigma$, 1$\sigma$, 1.5$\sigma$,
    and 2$\sigma$ confidence levels.
    }
    \label{fig:corner_x_kb241522}
\end{figure}

\begin{figure}[t]
    \centering
    \includegraphics[width=0.9\textwidth]{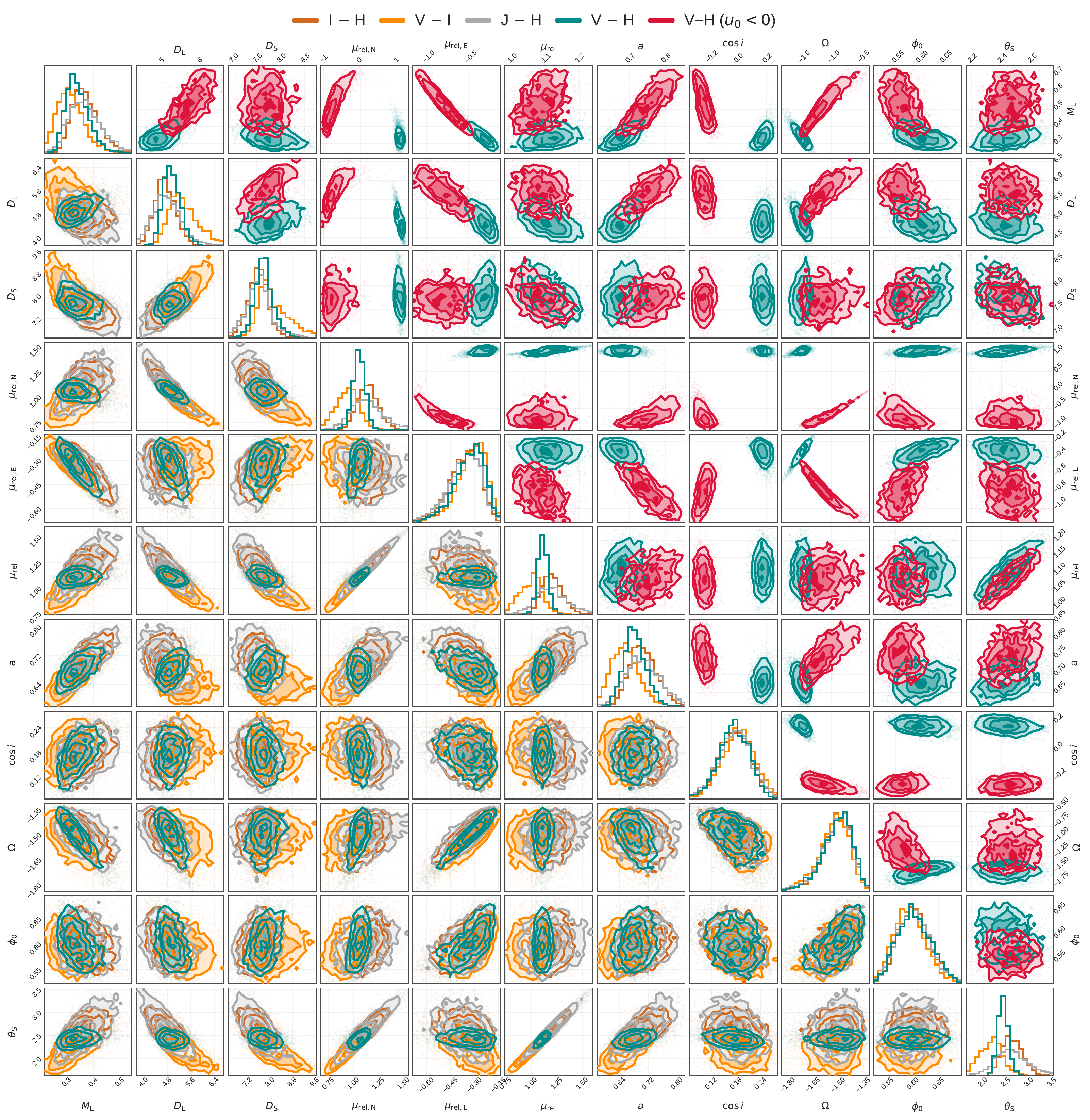}
    \caption{
    Posterior distributions of the physical parameters for KB241522.
    The color coding is the same as in Figure~\ref{fig:corner_x_kb241522},
    including the additional $V-H$ solution on the $u_0<0$ parallax branch.
    Contours correspond to the 0.5$\sigma$, 1$\sigma$, 1.5$\sigma$,
    and 2$\sigma$ confidence levels.
    }
    \label{fig:corner_y_kb241522}
\end{figure}

\begin{figure}[t]
    \centering
    \includegraphics[width=0.5\textwidth]{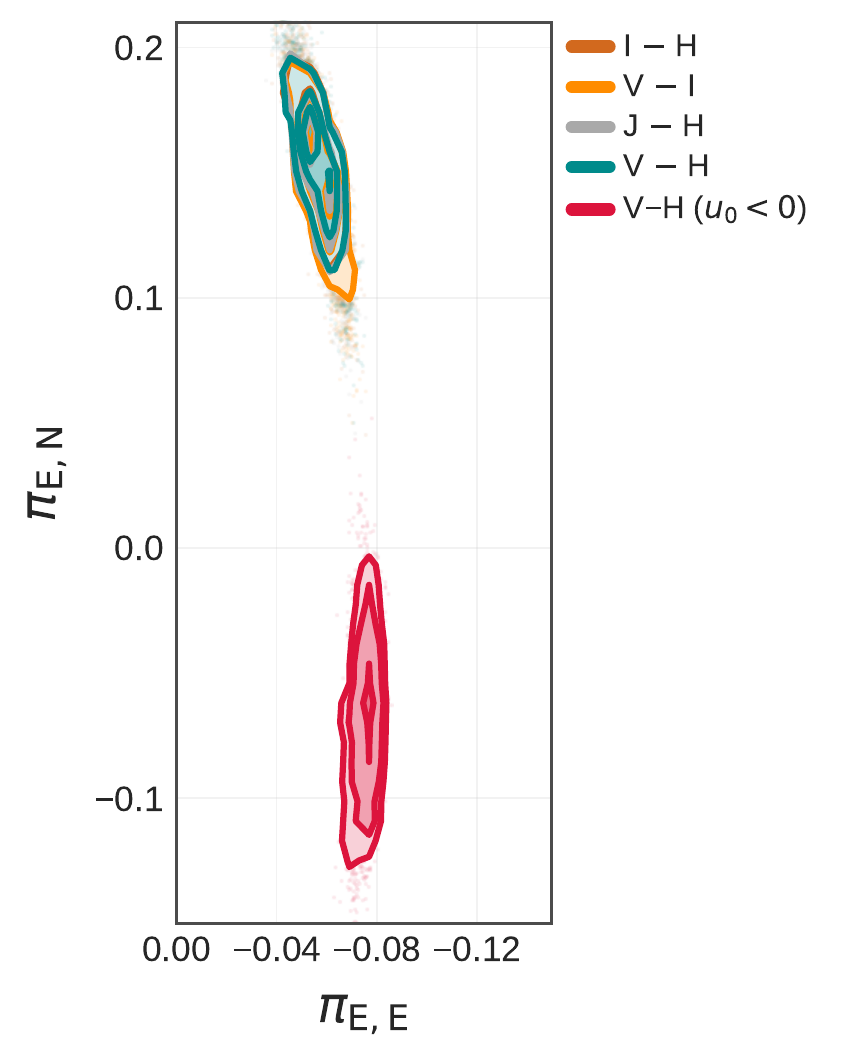}
    \caption{
    Comparison of the posterior distributions of the microlens parallax vector components for KB241522.
    The contours are identical to those shown in
    Figure~\ref{fig:corner_x_kb241522} and are replotted in the
    $(\pi_{E,E},\pi_{E,N})$ plane with a common aspect ratio for clarity.
    }
    \label{fig:parallax_comparison_kb241522}
\end{figure}

\begin{figure}[t]
    \centering
    \includegraphics[width=1\textwidth]{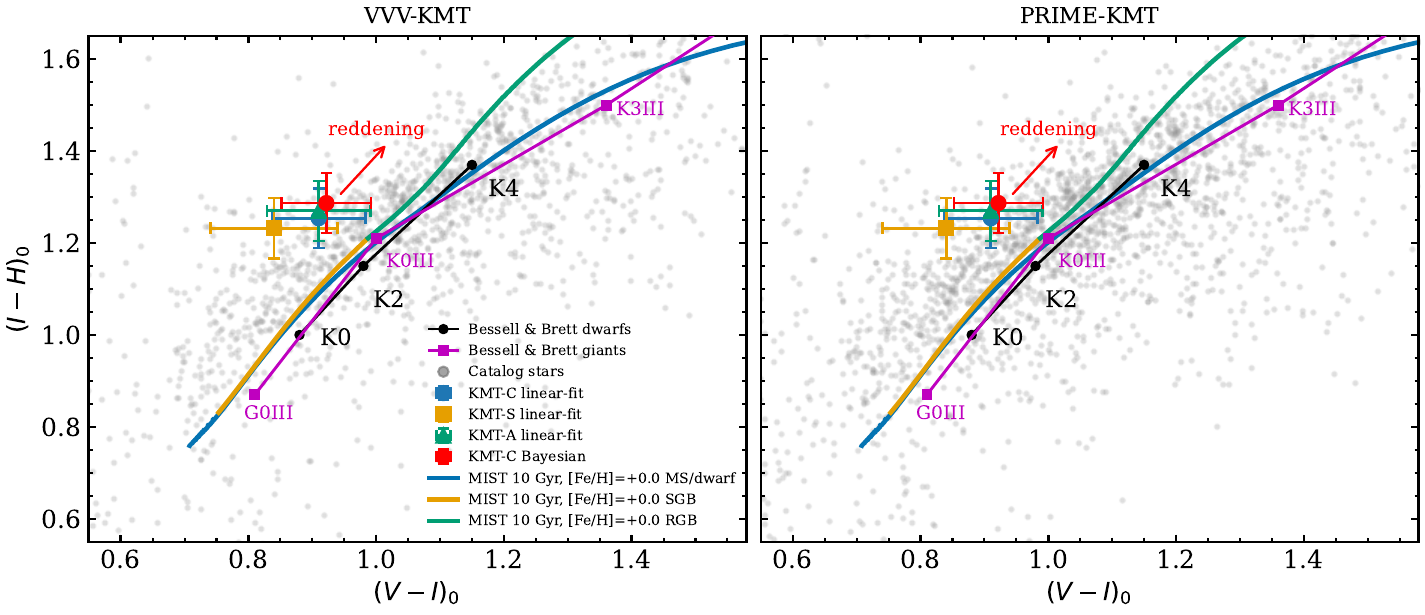}
    \caption{
    Comparison of the inferred source colors of KB241522 with the empirical stellar color--color relations of \citet{bes98}, solar-metallicity 10-Gyr MIST isochrones \citep{cho16,dot16}, and catalog stars in the same field.
    The colored symbols show the source colors obtained from the KMTC Bayesian analysis and from conventional linear fits to the KMTC, KMTA, and KMTS data.
    The arrow indicates the reddening direction.
    }
    \label{fig:bessell_brett_vi_ih_relation}
\end{figure}

\begin{deluxetable*}{llcc}
\tablecaption{Dereddened source-color estimates for KB241522.\label{tab:source_color_method_comparison}}
\tablehead{
\colhead{Method} &
\colhead{Observatory} &
\colhead{$(V-I)_0$} &
\colhead{$(I-H)_0$}
}
\startdata
Bayesian analysis & KMTC & $0.922 \pm 0.070$ & $1.287 \pm 0.065$ \\
Linear regression              & KMTC & $0.910 \pm 0.073$ & $1.254 \pm 0.065$ \\
Linear regression              & KMTA & $0.910 \pm 0.081$ & $1.270 \pm 0.065$ \\
Linear regression              & KMTS & $0.840 \pm 0.099$ & $1.232 \pm 0.065$
\enddata
\end{deluxetable*}

\section{Results} \label{sec:result}
\subsection{KB240211}\label{sec:result_KB240211}

For KB240211, we find two parallax solutions corresponding to $u_0>0$ and $u_0<0$. 
The posterior summaries of the microlensing and physical parameters are given in Table~\ref{tab:posterior_kb240211},
and the corresponding posterior distributions are shown in Figures~\ref{fig:corner_x_kb240211} and \ref{fig:corner_y_kb240211}.
The two branches give nearly identical values of the main light-curve parameters, including $t_{\rm E}$, $|u_0|$, and $\rho$, and therefore the sign degeneracy in $u_0$ does not affect the basic interpretation of the event.

The source position in the $(I-H,I)$ color--magnitude diagram
(Figure~\ref{fig:cmd_pb240021}) is located near the red giant branch, indicating that the source is a giant star.
The inferred angular source radius is correspondingly large, $\theta_* \simeq 9.4~\mu\mathrm{as}$, and the finite-source effect is clearly detected. 
Combined with the measured normalized source radius, this gives an angular Einstein radius of $\theta_{\rm E}\simeq 0.20~\mathrm{mas}$. Although this value is not unusual for a Galactic microlensing event, the short event timescale of $t_{\rm E}\simeq 5~\mathrm{days}$ implies a relatively large lens--source relative proper motion.

The physical-parameter posterior favors a low-mass lens, with a median mass of about $0.2~M_\odot$ for both $u_0$ branches. However, the microlens parallax is only weakly constrained, resulting in bimodal posterior distributions for the lens mass and distance. 
Thus, KB240211 is best interpreted as a short-timescale finite-source event caused by a likely low-mass stellar lens, but the current data do not uniquely determine the lens mass or distance.

\subsection{KB241522}\label{sec:result_KB241522}
For KB241522, we derived posterior distributions for the main $u_0>0$ branch using four source-color constraints, namely $(I-H)$, $(V-I)$, $(J-H)$, and $(V-H)$.
The posterior summaries are listed in Table~\ref{tab:posterior_kb241522}, while the corresponding microlensing-parameter, physical-parameter, and parallax-vector distributions are shown in Figures~\ref{fig:corner_x_kb241522}, \ref{fig:corner_y_kb241522}, and \ref{fig:parallax_comparison_kb241522}.

For the $(V-H)$ constraint, we also tested three additional parallax-related branch seeds: one ecliptic-like $u_0$-flip seed and two jerk-parallax-motivated seeds, allowing for their possible coupling to LOM \citep{gou04,sko11}.
The two jerk-parallax-motivated seeds converged to the boundary at $\pi_{\rm E,N}\simeq0$ and were not retained as independent solutions.
The ecliptic-like seed survived as a secondary branch with $u_0<0$, with the corresponding reversal of $\alpha$ and $\gamma_2$, and with $\pi_{\rm E,N}$ on the opposite-sign side relative to the main branch.
It is disfavored relative to the main branch by $\Delta\chi^2\simeq25$, corresponding to a relative posterior weight of $\exp(-\Delta\chi^2/2)\simeq4\times10^{-6}$ for comparable prior volumes.
This secondary branch is included in Table~\ref{tab:posterior_kb241522} and Figures~\ref{fig:corner_x_kb241522} and \ref{fig:corner_y_kb241522}; its physical implications are discussed in Section~\ref{sec:disc_degeneracy}.

Focusing on the main $u_0>0$ branch, the light-curve parameters are nearly identical among the four color analyses.
Thus, within the same branch, the microlensing solution itself is insensitive to the adopted source-color constraint.

However, while the light-curve parameters are nearly identical among the different color analyses, the inferred physical parameters show a systematic dependence on the adopted color constraint.
As shown in Table~\ref{tab:posterior_kb241522}, the $(I-H)$ and $(J-H)$ analyses give similar values of $\theta_*$, $\theta_{\rm E}$, $M_{\rm L}$, and $D_{\rm L}$, whereas the $(V-I)$ analysis gives a smaller angular source radius, a smaller angular Einstein radius, a smaller lens mass, and a larger lens distance.
The $(V-H)$ analysis gives intermediate values between these two groups and generally provides the tightest constraints on the source properties.
Although all four solutions indicate a compact binary composed of low-mass stars, the physical parameters are therefore not fully independent of the color used to estimate the source properties.

This difference arises from the inferred source colors.
Figure~\ref{fig:bessell_brett_vi_ih_relation} compares the measured $(V-I)_0$ and $(I-H)_0$ source colors with the empirical dwarf and giant color--color sequences of \citet{bes98}, as well as with a solar-metallicity, 10-Gyr MIST isochrone \citep{cho16,dot16}.
The measured source position is offset from both the empirical sequences and the theoretical isochrone tracks by more than the quoted uncertainties.
In the CMDs shown in Figure~\ref{fig:cmd_pb24160}, the source is not cleanly located on the main-sequence locus and may be slightly evolved.
However, this possibility does not remove the color discrepancy: in the color--color plane, the inferred source colors are inconsistent not only with the main-sequence locus but also with the subgiant and red-giant branches of the isochrone. Thus, the $(V-I)$ and $(I-H)$ color constraints cannot be simultaneously explained by a single consistent stellar color under the adopted calibration.

We therefore conclude that the light-curve solution for KB241522 is robust, but the inferred source properties and lens physical parameters are affected by a systematic discrepancy between the optical and optical--near-infrared source-color estimates, which we discuss further below.

\section{Discussion}\label{sec:disc}
We analyzed two microlensing events, KB240211 and KB241522, which have different physical characters.
KB240211 is a short-timescale finite-source event with $t_{\rm E}\simeq5~{\rm days}$, but its short duration does not imply an unusually small Einstein radius.
The inferred value, $\theta_{\rm E}\simeq0.20~{\rm mas}$, is within the ordinary range for Galactic microlensing events.
Instead, the short timescale is mainly caused by the relatively large lens--source relative proper motion, $\mu_{\rm rel}\simeq14~{\rm mas~yr^{-1}}$.
The source is a red giant, and the lens posterior favors an ordinary low-mass stellar lens, most likely an M dwarf, although the mass and distance remain broad because the microlens parallax is weakly constrained.

KB241522 is a long-timescale binary-lens event produced by a compact, nearly equal-mass, low-mass stellar binary.
For the main $u_0>0$ branch, the mass ratio is $q\simeq0.995$, and the total lens mass inferred from the different source-color constraints is $M_{\rm L}\simeq0.3$--$0.37~M_\odot$, corresponding to two M-dwarf components with masses of roughly $0.15$--$0.18~M_\odot$ each.
The inferred distance, $D_{\rm L}\simeq5~{\rm kpc}$, places the lens in the foreground disk or inner disk.
The semimajor axis is $a\simeq0.65$--$0.70~{\rm au}$, implying an orbital period of order one year.
Thus, the system is close in the microlensing sense, with $s\simeq0.255<1$, and physically compact, but it is better described as a detached sub-au M-dwarf binary than as an interacting close binary.
The main interest of this event is therefore not that the binary itself is exotic, but that the long timescale allows both annual parallax and LOM to be constrained, leading to meaningful estimates of the mass, distance, and orbital geometry.

\subsection{Role of joint optical--near-infrared observations in constraining the angular source radius}
Constraining the angular source radius $\theta_*$ is a key step in deriving the physical properties of microlensing events.
For finite-source events, the light curve measures the normalized source radius $\rho$, but converting this measurement into the angular Einstein radius, $\theta_{\rm E}=\theta_*/\rho$, requires an estimate of $\theta_*$ from the dereddened source color and magnitude.
The availability of suitable source-color information therefore directly determines whether the finite-source signal can be translated into physical constraints on the lens.

KB240211 illustrates this point.
Because the event lies in a highly extincted field and has a short timescale, the optical data alone do not provide a reliable conventional color estimate for the source.
The KMTNet--PRIME $(I-H)$ color, however, makes it possible to estimate $\theta_*$ and hence to infer $\theta_{\rm E}$ and the lens physical parameters from the measured finite-source effect.
In this case, the near-infrared data are not merely supplementary photometry, but an essential component of the physical interpretation.

This role is especially important for events toward highly extincted regions near the Galactic center, where optical source characterization is often limited.
Joint optical--near-infrared observations can therefore expand the range of analyzable microlensing events by preserving the link between finite-source measurements and physical parameter inference.
At the same time, as demonstrated by KB241522, combining optical and near-infrared source-color information may also introduce calibration and consistency issues, which we discuss in the next subsection.

\subsection{Source-color inconsistency and sanity checks}
\label{sec:disc_source_color}
The physical parameters inferred for KB241522 depend on the adopted source-color constraint.
This dependence is primarily driven by differences in the inferred angular source radius, $\theta_*$, rather than by differences in the microlensing light-curve parameters.

Figure~\ref{fig:bessell_brett_vi_ih_relation} illustrates the origin of this behavior.
The inferred $(V-I)_0$ and $(I-H)_0$ source colors are offset from the empirical dwarf and giant color--color relations of \citet{bes98} and from a solar-metallicity, 10-Gyr MIST isochrone \citep{cho16,dot16}.
Thus, the different color constraints do not correspond to a single position on the standard stellar color--color loci, leading to different estimates of $\theta_*$.

However, the catalog stars in the same field exhibit substantial scatter around the empirical and theoretical color--color relations.
The source lies within the broader distribution of these field stars and is therefore not exceptionally unusual relative to the observed stellar population.
This suggests that there may be additional uncertainty in the source size inferred from the colors that is not represented by the adopted color--surface-brightness relations and photometric uncertainties alone.

Because the present analysis uses the marginalized Bayesian framework introduced in Section~\ref{sec:lc}, it is important to verify that the apparent color inconsistency is not produced by the new methodology.
It is also necessary to test whether the result could arise from a problem in one of the individual KMTNet datasets.
We therefore performed a series of sanity checks using conventional source-color measurements.

For each of the KMTC, KMTA, and KMTS datasets, we measured the intrinsic source colors using linear regression.
The resulting values are listed in Table~\ref{tab:source_color_method_comparison} and shown in Figure~\ref{fig:bessell_brett_vi_ih_relation}.
KMTC provides the most precise $(V-I)_0$ measurement of the three observatories.
KMTA yields a more precise $(V-I)_0$ measurement than KMTS because it obtained more $V$-band measurements while the event was bright.
The regression measurements from the three observatories are mutually consistent within their uncertainties, providing no evidence for a problem in the KMTNet photometric reductions.

For KMTC, the regression-based source colors also agree closely with the posterior medians obtained from the marginalized Bayesian analysis.
This agreement demonstrates that the source-color offset is not introduced by the new inference framework.
Although the measurements show modest site-to-site differences, all three observatories place the source on the same side of the standard color--color relations and reproduce the same qualitative offset.
Thus, the color inconsistency is independently present in the KMTC, KMTA, and KMTS datasets.

Closer inspection of Figure~\ref{fig:bessell_brett_vi_ih_relation} shows that the field-star color--color sequence exhibits scatter comparable to the offset of the source from the standard color--color relations.
The discrepancy therefore appears to be a genuine feature of the measured source colors rather than a problem with the underlying data.
The field-star distribution further suggests that real stars in this field exhibit intrinsic color scatter that is not fully captured by the idealized empirical relations or by the current uncertainty model.

Taken together, these tests provide no evidence that the offset is caused by either the marginalized Bayesian methodology or a reduction problem specific to one KMTNet observatory.
Instead, the discrepancy appears to reflect genuine scatter in the measured stellar colors.

A larger sample of jointly observed KMTNet--PRIME events will be required to quantify this additional scatter and determine how it should be propagated into estimates of $\theta_*$ and the lens physical parameters.
Future high-angular-resolution imaging can provide an independent test by measuring the lens--source relative proper motion and lens flux.

\subsection{Secondary parallax branch in KB241522}
\label{sec:disc_degeneracy}
The branch search for the $(V-H)$ analysis yielded a secondary solution with $u_0<0$.
This branch is connected to the ecliptic-like $u_0$-flip parallax degeneracy, including the corresponding changes in $\alpha$ and $\gamma_2$, but it is not an equally likely degenerate solution.
As shown in Figure~\ref{fig:parallax_comparison_kb241522}, its parallax vector, $(\pi_{\rm E,N},\pi_{\rm E,E})=(-0.070,-0.075)$, is not a simple mirror image of the main $(V-H)$ solution, $(+0.155,-0.057)$.
The explicitly jerk-parallax-motivated seeds did not form independent posterior modes, so we regard the surviving branch as a disfavored ecliptic-like local minimum modified by correlations with LOM, rather than as a distinct jerk-parallax solution.

The secondary branch is worse than the main $(V-H)$ solution by $\Delta\chi^2\simeq25$.
We therefore do not base the interpretation of KB241522 on this branch, but we retain it to indicate the possible scale of the remaining parallax-branch systematic.
Because the two branches use the same $(V-H)$ source-color constraint, their angular source and Einstein radii are nearly unchanged: $\theta_*=2.43~\mu{\rm as}$ and $\theta_{\rm E}=0.45~{\rm mas}$ for the main branch, compared with $\theta_*=2.45~\mu{\rm as}$ and $\theta_{\rm E}=0.44~{\rm mas}$ for the secondary branch.
The main physical difference comes from the smaller parallax amplitude of the secondary branch, $|\bm{\pi}_{\rm E}|=0.103$ instead of $0.165$, which shifts the posterior lens mass and distance from $M_{\rm L}=0.335^{+0.050}_{-0.037}~M_\odot$ and $D_{\rm L}=4.93^{+0.32}_{-0.27}~{\rm kpc}$ to $M_{\rm L}=0.523^{+0.116}_{-0.104}~M_\odot$ and $D_{\rm L}=5.68^{+0.40}_{-0.40}~{\rm kpc}$.

The observational prediction that differs the most is the direction of the lens--source relative proper motion.
The main branch gives $(\mu_{\rm rel,N},\mu_{\rm rel,E})=(+1.05,-0.33)~{\rm mas~yr^{-1}}$, whereas the secondary branch gives $(-0.75,-0.78)~{\rm mas~yr^{-1}}$.
Although the secondary branch is disfavored, future high-angular-resolution imaging could test it directly by measuring the lens--source relative proper-motion vector once the lens and source are resolved.

\section{Summary and Conclusions}\label{sec:concl}
We have analyzed two 2024 microlensing events, KMT-2024-BLG-0211 and KMT-2024-BLG-1522, using optical data from KMTNet and near-infrared $H$- and $J$-band data from PRIME. The main goal of this work was to investigate how optical--near-infrared source-color information affects the determination of the angular source radius and the inferred physical properties of the lens system.

For KMT-2024-BLG-0211, we find that the event is a short-timescale finite-source event with a giant source. The two degenerate solutions with opposite signs of $u_0$ give nearly identical microlensing and physical parameters, so the sign degeneracy does not affect the interpretation of the event. The KMTNet--PRIME $I-H$ color provides the source angular radius, which, together with the measured finite-source parameter, yields the angular Einstein radius. The event is consistent with a low-mass stellar lens, although the weak microlens-parallax constraint leaves the lens mass and distance broadly distributed.

For KMT-2024-BLG-1522, we compared analyses based on four different source-color constraints: $(I-H)$, $(V-I)$, $(V-H)$, and $(J-H)$. 
The microlensing parameters are nearly identical among the four cases, indicating that the light-curve solution is robust and consistently favors a nearly equal-mass binary lens. 
Among the four constraints, the $(V-H)$ analysis generally provides the tightest posterior constraints.
However, the derived source properties and lens physical parameters depend on the adopted color constraint. This dependence is driven by the angular source radius inferred from the measured source color.

A representative comparison in the $(V-I)_{\rm KMT}$ versus $(I-H)_{\rm KMT,PRIME}$ color--color plane shows that the inferred source colors do not correspond to a single consistent stellar type under the adopted calibration. This indicates an offset from the standard color--color relations, rather than a difference in the microlensing geometry. However, the source lies within the scatter of observed field stars, suggesting that such deviations may reflect additional scatter in the stellar colors not captured by the empirical relations or the current uncertainty model.

This result motivates a broader investigation of source-color measurements in events observed by both KMTNet and PRIME. In particular, it will be important to determine whether similar inconsistencies appear in other events, or whether KB241522 is an isolated case. Future adaptive-optics follow-up observations will provide an independent test by resolving the lens and source, measuring the lens--source relative proper motion and lens flux, and thereby constraining $\theta_{\rm E}$, $\theta_*$, the lens mass, and the lens distance independently of the color-based estimates.

\begin{acknowledgments}
K.N. is supported by the Japan Society for the Promotion of Science (JSPS)
Research Fellowship for Young Scientists (DC) and by the Next-Generation
Global Leaders Overseas Training Program of The University of Osaka.

The PRIME project is supported by JSPS KAKENHI Grant Number
JP16H06287, JP22H00153, JP25H00668, JP19KK0082, JP20H04754,
JP24H01811 and JPJSCCA20210003.
We acknowledge financial support from the Astrobiology Center.

This research has made use of the KMTNet system operated by the Korea
Astronomy and Space Science Institute (KASI) at three host sites of CTIO
in Chile, SAAO in South Africa, and SSO in Australia. Data transfer from
the host site to KASI was supported by the Korea Research Environment
Open NETwork (KREONET).

This work makes use of observations from the Las Cumbres Observatory
global telescope network.

The authors used ChatGPT (OpenAI) to assist with language editing and improving the clarity of the manuscript.
The authors reviewed and edited all AI-assisted text and take full responsibility for the content of the manuscript.
\end{acknowledgments}

\appendix

\section{Derivation of the Intrinsic RC Parameters} \label{sec:rc0}
We derive the intrinsic magnitudes and colors of the RCGs adopted in this work from the literature. 
Following \citet{nat13}, the intrinsic $I$-band magnitude and colors are taken to be
\begin{align}
    I_{\mathrm{RC},0} = 14.373, \quad
    (V-K_s)_{\mathrm{RC},0} = 2.44, \quad
    (H-K_s)_{\mathrm{RC},0} = 0.09. \nonumber
\end{align}
From \citet{ben13}, we adopt
\begin{align}
    (V-I)_{\mathrm{RC},0} = 1.06. \nonumber
\end{align}

Combining these intrinsic colors, we obtain
\begin{align}
    (I-H)_{\mathrm{RC},0} 
        = -(V-I)_{\mathrm{RC},0} + (V-K_s)_{\mathrm{RC},0} - (H-K_s)_{\mathrm{RC},0} = 1.29, \nonumber
\end{align}
which yields
\begin{align}
    H_{\mathrm{RC},0} = I_{\mathrm{RC},0} - (I-H)_{\mathrm{RC},0} = 13.083. \nonumber
\end{align}

Finally, adopting $(J-K_s)_{\mathrm{RC},0} = 0.64$ from \citet{nat21}, we derive
\begin{align}
    (J-H)_{\mathrm{RC},0} 
        &= (J-K_s)_{\mathrm{RC},0} - (H-K_s)_{\mathrm{RC},0} = 0.55. \nonumber
\end{align}

The uncertainties in $I_{\mathrm{RC},0}$ and $(V-I)_{\mathrm{RC},0}$ are taken from the respective literature values, 
whereas conservative 5\% fractional uncertainties are assumed for $(I-H)_{\mathrm{RC},0}$ and $(J-H)_{\mathrm{RC},0}$. 
The uncertainty in $H_{\mathrm{RC},0}$ is obtained through standard error propagation from $I_{\mathrm{RC},0}$ and $(I-H)_{\mathrm{RC},0}$.

\section{Analytic marginalization over the photometric nuisance parameters}
\label{app:int_like}

In this appendix, we derive closed-form expressions for the
nuisance-marginalized likelihood $L_{\rm marg}(\bm X)$ and for the
conditional distribution $p(\{F_s\}\mid \bm X,\mathrm{data})$
introduced in Equations~\eqref{eq:Lmarg} and \eqref{eq:Fs_conditional}.

Because the light-curve likelihood factorizes over datasets, it is
sufficient to consider a single dataset $j$ and then take the product
over $j$. For notational simplicity, we omit the dataset index in the
intermediate steps and restore it at the end.

\subsection{Single-dataset linear model}

For a fixed microlensing parameter vector $\bm X$, let
\begin{align}
\bm F^{\rm obs}
\equiv
\begin{bmatrix}
F_1\\
\vdots\\
F_N
\end{bmatrix}
\end{align}
be the vector of observed fluxes for one dataset. We define the linear
parameter vector
\begin{align}
\bm\beta
\equiv
\begin{bmatrix}
F_s\\
F_b
\end{bmatrix},
\end{align}
and the design matrix
\begin{align}
\mathbf M
\equiv
\begin{bmatrix}
A(t_1;\bm X) & 1\\
\vdots & \vdots\\
A(t_N;\bm X) & 1
\end{bmatrix}.
\end{align}
We also define the weight matrix based on the reported photometric
uncertainties,
\begin{align}
\mathbf W_0
\equiv
\mathrm{diag}\!\left(
\sigma_1^{-2},\dots,\sigma_N^{-2}
\right),
\end{align}
and introduce the precision parameter
\begin{align}
\tau \equiv k^{-2}.
\end{align}

Then the likelihood for this dataset can be written as
\begin{align}
p(\bm F^{\rm obs}\mid \bm\beta,\tau,\bm X)
&=
(2\pi)^{-N/2}
|\mathbf W_0|^{1/2}
\tau^{N/2}
\exp\!\left[
-\frac{\tau}{2}
(\bm F^{\rm obs}-\mathbf M\bm\beta)^\top
\mathbf W_0
(\bm F^{\rm obs}-\mathbf M\bm\beta)
\right].
\label{eq:app_like_single}
\end{align}

We adopt the improper priors
\begin{align}
\pi(F_s)\propto 1,\qquad
\pi(F_b)\propto 1,\qquad
\pi(\tau)\propto \frac{1}{\tau},
\end{align}
which are equivalent to the priors
$\pi(F_s)\propto 1$, $\pi(F_b)\propto 1$, and $\pi(k)\propto 1/k$
used in Section~\ref{sec:bayes}.

For the single dataset under consideration, define
\begin{align}
\mathcal K(\bm\beta,\tau;\bm X)
&\equiv
p(\bm F^{\rm obs}\mid \bm\beta,\tau,\bm X)\,
\pi(\bm\beta)\,\pi(\tau).
\end{align}
Up to a multiplicative constant independent of $(\bm\beta,\tau)$, this
kernel is given by
\begin{align}
\mathcal K(\bm\beta,\tau;\bm X)
\propto
\tau^{N/2-1}
\exp\!\left[
-\frac{\tau}{2}
(\bm F^{\rm obs}-\mathbf M\bm\beta)^\top
\mathbf W_0
(\bm F^{\rm obs}-\mathbf M\bm\beta)
\right].
\label{eq:app_joint_kernel}
\end{align}
This has the normal-gamma form in $(\bm\beta,\tau)$, or equivalently
the normal-inverse-gamma form in $(\bm\beta,k^2)$.

\subsection{Marginal likelihood for one dataset}

We define
\begin{align}
\mathbf H &\equiv \mathbf M^\top \mathbf W_0 \mathbf M,\\
\mathbf\Sigma_0 &\equiv \mathbf H^{-1},\\
\hat{\bm\beta} &\equiv \mathbf\Sigma_0 \mathbf M^\top \mathbf W_0 \bm F^{\rm obs},\\
\mathrm{RSS} &\equiv
(\bm F^{\rm obs}-\mathbf M\hat{\bm\beta})^\top
\mathbf W_0
(\bm F^{\rm obs}-\mathbf M\hat{\bm\beta}).
\end{align}
Then the quadratic form in Equation~\eqref{eq:app_joint_kernel} can be
written as
\begin{align}
(\bm F^{\rm obs}-\mathbf M\bm\beta)^\top
\mathbf W_0
(\bm F^{\rm obs}-\mathbf M\bm\beta)
=
(\bm\beta-\hat{\bm\beta})^\top
\mathbf H
(\bm\beta-\hat{\bm\beta})
+
\mathrm{RSS},
\label{eq:app_complete_square}
\end{align}
where $\hat{\bm\beta}$ is the weighted least-squares solution.

Using Equation~\eqref{eq:app_complete_square}, we first integrate over
$\bm\beta$. For fixed $\tau$, define
\begin{align}
\mathcal K_\tau(\tau;\bm X)
\equiv
\int_{\mathbb R^2}
\mathcal K(\bm\beta,\tau;\bm X)\,d\bm\beta.
\end{align}
Then
\begin{align}
\mathcal K_\tau(\tau;\bm X)
&=
(2\pi)^{-(N-2)/2}
|\mathbf W_0|^{1/2}
|\mathbf H|^{-1/2}
\tau^{(N-2)/2-1}
\exp\!\left(
-\frac{\tau}{2}\mathrm{RSS}
\right),
\label{eq:app_marg_tau}
\end{align}
where we used
\begin{align}
\int_{\mathbb R^2}
\exp\!\left[
-\frac{\tau}{2}
(\bm\beta-\hat{\bm\beta})^\top
\mathbf H
(\bm\beta-\hat{\bm\beta})
\right]
\,d\bm\beta
=
(2\pi)\tau^{-1}|\mathbf H|^{-1/2}.
\end{align}

The nuisance-marginalized likelihood appearing in
Equation~\eqref{eq:Lmarg} for the single dataset under consideration is therefore
\begin{align}
\mathcal{L}_{\rm marg}(\bm X)
&=
\int_0^\infty
\mathcal K_\tau(\tau;\bm X)\,d\tau \\
&=
(2\pi)^{-(N-2)/2}
|\mathbf W_0|^{1/2}
|\mathbf H|^{-1/2}
\int_0^\infty
\tau^{(N-2)/2-1}
\exp\!\left(
-\frac{\tau}{2}\mathrm{RSS}
\right)
\,d\tau.
\end{align}
Using the Gamma-function identity
\begin{align}
\int_0^\infty \tau^{a-1}e^{-b\tau}\,d\tau
=
\frac{\Gamma(a)}{b^a},
\end{align}
with $a=(N-2)/2$ and $b=\mathrm{RSS}/2$, we obtain
\begin{align}
\mathcal{L}_{\rm marg}(\bm X)
&=
\pi^{-(N-2)/2}
|\mathbf W_0|^{1/2}
|\mathbf H|^{-1/2}
\Gamma\!\left(\frac{N-2}{2}\right)
\mathrm{RSS}^{-(N-2)/2}.
\label{eq:app_single_Lmarg}
\end{align}

Applying the same calculation to each dataset independently, the full
nuisance-marginalized likelihood in Equation~\eqref{eq:Lmarg} is
obtained by taking the product of the corresponding single-dataset
factors.

\subsection{Conditional posterior of \texorpdfstring{$F_s$}{Fs}}

Next, we derive the conditional distribution of $F_s$ by marginalizing
Equation~\eqref{eq:app_joint_kernel} over $F_b$ and $\tau$. Using
Equation~\eqref{eq:app_complete_square}, we may write
\begin{align}
\mathcal K(\bm\beta,\tau;\bm X)
\propto
\tau^{N/2-1}
\exp\!\left[
-\frac{\tau}{2}
\left\{
(\bm\beta-\hat{\bm\beta})^\top \mathbf H (\bm\beta-\hat{\bm\beta})
+\mathrm{RSS}
\right\}
\right].
\label{eq:app_joint_kernel_cs}
\end{align}
Let
\begin{align}
\mathbf H
=
\begin{bmatrix}
h_{ss} & h_{sb}\\
h_{sb} & h_{bb}
\end{bmatrix},
\qquad
\hat{\bm\beta}
=
\begin{bmatrix}
\hat F_s\\
\hat F_b
\end{bmatrix}.
\end{align}
Then the quadratic form can be decomposed as
\begin{align}
(\bm\beta-\hat{\bm\beta})^\top \mathbf H (\bm\beta-\hat{\bm\beta})
&=
h_{bb}
\left[
F_b-\hat F_b-\frac{h_{sb}}{h_{bb}}(F_s-\hat F_s)
\right]^2
+
\left(
h_{ss}-\frac{h_{sb}^2}{h_{bb}}
\right)
(F_s-\hat F_s)^2.
\end{align}
Using the identity
\begin{align}
h_{ss}-\frac{h_{sb}^2}{h_{bb}}
=
\frac{1}{[\mathbf\Sigma_0]_{11}},
\end{align}
we obtain
\begin{align}
(\bm\beta-\hat{\bm\beta})^\top \mathbf H (\bm\beta-\hat{\bm\beta})
&=
h_{bb}
\left[
F_b-\hat F_b-\frac{h_{sb}}{h_{bb}}(F_s-\hat F_s)
\right]^2
+
\frac{(F_s-\hat F_s)^2}{[\mathbf\Sigma_0]_{11}}.
\label{eq:app_quad_fs}
\end{align}

We first integrate over $F_b$ and define
\begin{align}
\mathcal K_s(F_s,\tau;\bm X)
\equiv
\int_{-\infty}^{\infty}
\mathcal K(\bm\beta,\tau;\bm X)\,dF_b.
\end{align}
Then
\begin{align}
\mathcal K_s(F_s,\tau;\bm X)
&\propto
\tau^{(N-1)/2-1}
\exp\!\left[
-\frac{\tau}{2}
\left\{
\mathrm{RSS}
+
\frac{(F_s-\hat F_s)^2}{[\mathbf\Sigma_0]_{11}}
\right\}
\right].
\label{eq:app_fs_tau_kernel}
\end{align}

Integrating over $\tau$, we obtain
\begin{align}
\mathcal K_s(F_s;\bm X)
&\equiv
\int_0^\infty
\mathcal K_s(F_s,\tau;\bm X)\,d\tau \\
&\propto
\left[
\mathrm{RSS}
+
\frac{(F_s-\hat F_s)^2}{[\mathbf\Sigma_0]_{11}}
\right]^{-(N-1)/2}.
\label{eq:app_fs_kernel}
\end{align}
This can be rewritten as
\begin{align}
\mathcal K_s(F_s;\bm X)
&\propto
\left[
1+
\frac{(F_s-\hat F_s)^2}
{(N-2)\,\lambda^2}
\right]^{-(N-1)/2},
\end{align}
where
\begin{align}
\lambda^2
\equiv
\frac{\mathrm{RSS}}{N-2}[\mathbf\Sigma_0]_{11}.
\end{align}
This is the kernel of a univariate Student's $t$ distribution with
$\nu=N-2$ degrees of freedom, location $\hat F_s$, and scale
$\lambda^2$. 
Therefore,
\begin{align}
p(F_s\mid \bm F^{\rm obs},\bm X)
=
t_{N-2}\!\left(
F_s;\,
\hat F_s,\,
\frac{\mathrm{RSS}}{N-2}[\mathbf\Sigma_0]_{11}
\right).
\label{eq:app_Fs_student_single}
\end{align}

Repeating the same calculation for each dataset, the full conditional
distribution in Equation~\eqref{eq:Fs_conditional} is obtained as the product
of the corresponding single-dataset Student's $t$ factors.

For clarity, we define the univariate Student's $t$ density as
\begin{align}
t_\nu(z;\mu,\lambda^2)
\equiv
\frac{\Gamma\!\left(\frac{\nu+1}{2}\right)}
{\Gamma\!\left(\frac{\nu}{2}\right)\sqrt{\pi\nu\lambda^2}}
\left[
1+\frac{(z-\mu)^2}{\nu\lambda^2}
\right]^{-(\nu+1)/2}.
\label{eq:app_student_def}
\end{align}

\subsection{Remarks}

The nuisance-parameter integral in Equation~\eqref{eq:app_single_Lmarg} is finite for $N>2$ provided that $\mathbf M$ has full column rank.
Under the same condition, the conditional distribution of $F_s$ in
Equation~\eqref{eq:app_Fs_student_single} is well defined. Therefore,
both $L_{\rm marg}(\bm X)$ and $p(\{F_s\}\mid \bm X,\mathrm{data})$
can be evaluated analytically for each dataset, and the latter can be
sampled directly as a product of univariate Student's $t$
distributions.

\bibliographystyle{aasjournalv7}
\bibliography{reference}

@ARTICLE{nat21,
       author = {{Nataf}, David M. and {Cassisi}, Santi and {Casagrande}, Luca and {Yuan}, Wenlong and {Riess}, Adam G.},
        title = "{On the Color-Metallicity Relation of the Red Clump and the Reddening toward the Magellanic Clouds}",
      journal = {\apj},
         year = 2021,
        month = apr,
       volume = {910},
       number = {2},
          eid = {121},
        pages = {121},
          doi = {10.3847/1538-4357/abe530},
archivePrefix = {arXiv},
       eprint = {2006.03603},
 primaryClass = {astro-ph.SR},
       adsurl = {https://ui.adsabs.harvard.edu/abs/2021ApJ...910..121N}
}

@ARTICLE{nat13,
       author = {{Nataf}, David M. and {Gould}, Andrew and {Fouqu{\'e}}, Pascal and {Gonzalez}, Oscar A. and {Johnson}, Jennifer A. and {Skowron}, Jan and {Udalski}, Andrzej and {Szyma{\'n}ski}, Micha{\l} K. and {Kubiak}, Marcin and {Pietrzy{\'n}ski}, Grzegorz and {Soszy{\'n}ski}, Igor and {Ulaczyk}, Krzysztof and {Wyrzykowski}, {\L}ukasz and {Poleski}, Rados{\l}aw},
        title = "{Reddening and Extinction toward the Galactic Bulge from OGLE-III: The Inner Milky Way's R$_{V}$ \raisebox{-0.5ex}\textasciitilde 2.5 Extinction Curve}",
      journal = {\apj},
         year = 2013,
        month = jun,
       volume = {769},
       number = {2},
          eid = {88},
        pages = {88},
          doi = {10.1088/0004-637X/769/2/88},
archivePrefix = {arXiv},
       eprint = {1208.1263},
 primaryClass = {astro-ph.GA},
       adsurl = {https://ui.adsabs.harvard.edu/abs/2013ApJ...769...88N}
}

@ARTICLE{cla11,
       author = {{Claret}, A. and {Bloemen}, S.},
        title = "{Gravity and limb-darkening coefficients for the Kepler, CoRoT, Spitzer, uvby, UBVRIJHK, and Sloan photometric systems}",
      journal = {\aap},
         year = 2011,
        month = may,
       volume = {529},
          eid = {A75},
        pages = {A75},
          doi = {10.1051/0004-6361/201116451},
       adsurl = {https://ui.adsabs.harvard.edu/abs/2011A&A...529A..75C}
}

@ARTICLE{hou00,
       author = {{Houdashelt}, M.~L. and {Bell}, R.~A. and {Sweigart}, A.~V.},
        title = "{Improved Color-Temperature Relations and Bolometric Corrections for Cool Stars}",
      journal = {\aj},
         year = 2000,
        month = mar,
       volume = {119},
       number = {3},
        pages = {1448-1469},
          doi = {10.1086/301243},
archivePrefix = {arXiv},
       eprint = {astro-ph/9911367},
 primaryClass = {astro-ph},
       adsurl = {https://ui.adsabs.harvard.edu/abs/2000AJ....119.1448H}
}

@ARTICLE{boy14,
       author = {{Boyajian}, Tabetha S. and {van Belle}, Gerard and {von Braun}, Kaspar},
        title = "{Stellar Diameters and Temperatures. IV. Predicting Stellar Angular Diameters}",
      journal = {\aj},
         year = 2014,
        month = mar,
       volume = {147},
       number = {3},
          eid = {47},
        pages = {47},
          doi = {10.1088/0004-6256/147/3/47},
archivePrefix = {arXiv},
       eprint = {1311.4901},
 primaryClass = {astro-ph.SR},
       adsurl = {https://ui.adsabs.harvard.edu/abs/2014AJ....147...47B}
}

@ARTICLE{ben13,
       author = {{Bensby}, T. and {Yee}, J.~C. and {Feltzing}, S. and {Johnson}, J.~A. and {Gould}, A. and {Cohen}, J.~G. and {Asplund}, M. and {Mel{\'e}ndez}, J. and {Lucatello}, S. and {Han}, C. and {Thompson}, I. and {Gal-Yam}, A. and {Udalski}, A. and {Bennett}, D.~P. and {Bond}, I.~A. and {Kohei}, W. and {Sumi}, T. and {Suzuki}, D. and {Suzuki}, K. and {Takino}, S. and {Tristram}, P. and {Yamai}, N. and {Yonehara}, A.},
        title = "{Chemical evolution of the Galactic bulge as traced by microlensed dwarf and subgiant stars. V. Evidence for a wide age distribution and a complex MDF}",
      journal = {\aap},
         year = 2013,
        month = jan,
       volume = {549},
          eid = {A147},
        pages = {A147},
          doi = {10.1051/0004-6361/201220678},
archivePrefix = {arXiv},
       eprint = {1211.6848},
 primaryClass = {astro-ph.GA},
       adsurl = {https://ui.adsabs.harvard.edu/abs/2013A&A...549A.147B}
}

@ARTICLE{ker04,
       author = {{Kervella}, P. and {Th{\'e}venin}, F. and {Di Folco}, E. and {S{\'e}gransan}, D.},
        title = "{The angular sizes of dwarf stars and subgiants. Surface brightness relations calibrated by interferometry}",
      journal = {\aap},
         year = 2004,
        month = oct,
       volume = {426},
        pages = {297-307},
          doi = {10.1051/0004-6361:20035930},
archivePrefix = {arXiv},
       eprint = {astro-ph/0404180},
 primaryClass = {astro-ph},
       adsurl = {https://ui.adsabs.harvard.edu/abs/2004A&A...426..297K}
}

@ARTICLE{kim16,
       author = {{Kim}, Seung-Lee and {Lee}, Chung-Uk and {Park}, Byeong-Gon and {Kim}, Dong-Jin and {Cha}, Sang-Mok and {Lee}, Yongseok and {Han}, Cheongho and {Chun}, Moo-Young and {Yuk}, Insoo},
        title = "{KMTNET: A Network of 1.6 m Wide-Field Optical Telescopes Installed at Three Southern Observatories}",
      journal = {Journal of Korean Astronomical Society},
         year = 2016,
        month = feb,
       volume = {49},
       number = {1},
        pages = {37-44},
          doi = {10.5303/JKAS.2016.49.1.37},
       adsurl = {https://ui.adsabs.harvard.edu/abs/2016JKAS...49...37K}
}

@ARTICLE{sum25,
       author = {{Sumi}, Takahiro and {Buckley}, David A.~H. and {Kutyrev}, Alexander S. and {Tamura}, Motohide and {Bennett}, David P. and {Bond}, Ian A. and {Cataldo}, Giuseppe and {Durbak}, Joseph M. and {Cenko}, S. Bradley and {Fixsen}, Dale and {Guiffreda}, Orion and {Hamada}, Ryusei and {Hirao}, Yuki and {Idei}, Asahi and {Kelly}, Dan and {Loose}, Markus and {Lotkin}, Gennadiy N. and {Lyness}, Eric I. and {Maher}, Stephen and {Makida}, Shuma and {Matsunaga}, Noriyuki and {Miyazaki}, Shota and {Mosby}, Gregory and {Moseley}, Samuel H. and {Nagai}, Tutumi and {Nagano}, Togo and {Nakayama}, Seiya and {Nishio}, Mayu and {Nunota}, Kansuke and {Ogawa}, Ryo and {Oishi}, Ryunosuke and {Okumoto}, Yui and {Rattenbury}, Nicholas J. and {Satoh}, Yuki K. and {Sharp}, Elmer H. and {Suzuki}, Daisuke and {Tamaoki}, Takuto and {Troja}, Eleonora and {White}, Sarah V. and {Yama}, Hibiki},
        title = "{The Prime Focus Infrared Microlensing Experiment (PRIME): First Results}",
      journal = {\aj},
         year = 2025,
        month = dec,
       volume = {170},
       number = {6},
          eid = {338},
        pages = {338},
          doi = {10.3847/1538-3881/ae14f5},
archivePrefix = {arXiv},
       eprint = {2508.14474},
 primaryClass = {astro-ph.EP},
       adsurl = {https://ui.adsabs.harvard.edu/abs/2025AJ....170..338S}
}

@ARTICLE{kim18,
       author = {{Kim}, Hyoun-Woo and {Hwang}, Kyu-Ha and {Shvartzvald}, Yossi and {Yee}, Jennifer C. and {Albrow}, Michael D. and {Cha}, Sang-Mok and {Chung}, Sun-Ju and {Gould}, Andrew and {Han}, Cheongho and {Jung}, Youn Kil and {Kim}, Dong-Jin and {Kim}, Seung-Lee and {Lee}, Chung-Uk and {Lee}, Dong-Joo and {Lee}, Yongseok and {Park}, Byeong-Gon and {Pogge}, Richard W. and {Ryu}, Yoon-Hyun and {Shin}, In-Gu and {Zang}, Weicheng},
        title = "{The Korea Microlensing Telescope Network (KMTNet) Alert Algorithm and Alert System}",
      journal = {arXiv e-prints},
         year = 2018,
        month = jun,
          eid = {arXiv:1806.07545},
        pages = {arXiv:1806.07545},
          doi = {10.48550/arXiv.1806.07545},
archivePrefix = {arXiv},
       eprint = {1806.07545},
 primaryClass = {astro-ph.IM},
       adsurl = {https://ui.adsabs.harvard.edu/abs/2018arXiv180607545K}
}

@ARTICLE{szy11,
       author = {{Szyma{\'n}ski}, M.~K. and {Udalski}, A. and {Soszy{\'n}ski}, I. and {Kubiak}, M. and {Pietrzy{\'n}ski}, G. and {Poleski}, R. and {Wyrzykowski}, {\L}. and {Ulaczyk}, K.},
        title = "{The Optical Gravitational Lensing Experiment. OGLE-III Photometric Maps of the Galactic Bulge Fields}",
      journal = {\actaa},
         year = 2011,
        month = jun,
       volume = {61},
       number = {2},
        pages = {83-102},
          doi = {10.48550/arXiv.1107.4008},
archivePrefix = {arXiv},
       eprint = {1107.4008},
 primaryClass = {astro-ph.SR},
       adsurl = {https://ui.adsabs.harvard.edu/abs/2011AcA....61...83S}
}

@ARTICLE{alb09,
       author = {{Albrow}, M.~D. and {Horne}, K. and {Bramich}, D.~M. and {Fouqu{\'e}}, P. and {Miller}, V.~R. and {Beaulieu}, J.-P. and {Coutures}, C. and {Menzies}, J. and {Williams}, A. and {Batista}, V. and {Bennett}, D.~P. and {Brillant}, S. and {Cassan}, A. and {Dieters}, S. and {Dominis Prester}, D. and {Donatowicz}, J. and {Greenhill}, J. and {Kains}, N. and {Kane}, S.~R. and {Kubas}, D. and {Marquette}, J.~B. and {Pollard}, K.~R. and {Sahu}, K.~C. and {Tsapras}, Y. and {Wambsganss}, J. and {Zub}, M.},
        title = "{Difference imaging photometry of blended gravitational microlensing events with a numerical kernel}",
      journal = {\mnras},
         year = 2009,
        month = aug,
       volume = {397},
       number = {4},
        pages = {2099-2105},
          doi = {10.1111/j.1365-2966.2009.15098.x},
archivePrefix = {arXiv},
       eprint = {0905.3003},
 primaryClass = {astro-ph.SR},
       adsurl = {https://ui.adsabs.harvard.edu/abs/2009MNRAS.397.2099A}
}

@ARTICLE{yan24,
       author = {{Yang}, Hongjing and {Yee}, Jennifer C. and {Hwang}, Kyu-Ha and {Qian}, Qiyue and {Bond}, Ian A. and {Gould}, Andrew and {Hu}, Zhecheng and {Zhang}, Jiyuan and {Mao}, Shude and {Zhu}, Wei and {Albrow}, Michael D. and {Chung}, Sun-Ju and {Kim}, Seung-Lee and {Park}, Byeong-Gon and {Han}, Cheongho and {Jung}, Youn Kil and {Ryu}, Yoon-Hyun and {Shin}, In-Gu and {Shvartzvald}, Yossi and {Cha}, Sang-Mok and {Kim}, Dong-Jin and {Kim}, Hyoun-Woo and {Lee}, Chung-Uk and {Lee}, Dong-Joo and {Lee}, Yongseok and {Pogge}, Richard W. and {Zang}, Weicheng and {Abe}, Fumio and {Barry}, Richard and {Bennett}, David P. and {Bhattacharya}, Aparna and {Donachie}, Martin and {Fujii}, Hirosane and {Fukui}, Akihiko and {Hirao}, Yuki and {Itow}, Yoshitaka and {Kirikawa}, Rintaro and {Kondo}, Iona and {Koshimoto}, Naoki and {Silva}, Stela Ishitani and {Li}, Man Cheung Alex and {Matsubara}, Yutaka and {Muraki}, Yasushi and {Suzuki}, Daisuke and {Tristram}, Paul J. and {Yonehara}, Atsunori and {Ranc}, Cl{\'e}ment and {Miyazaki}, Shota and {Olmschenk}, Greg and {Rattenbury}, Nicholas J. and {Satoh}, Yuki and {Shoji}, Hikaru and {Sumi}, Takahiro and {Tanaka}, Yuzuru and {Yamawaki}, Tsubasa},
        title = "{Systematic reanalysis of KMTNet microlensing events, paper I: Updates of the photometry pipeline and a new planet candidate}",
      journal = {\mnras},
         year = 2024,
        month = feb,
       volume = {528},
       number = {1},
        pages = {11-27},
          doi = {10.1093/mnras/stad3672},
archivePrefix = {arXiv},
       eprint = {2311.04876},
 primaryClass = {astro-ph.EP},
       adsurl = {https://ui.adsabs.harvard.edu/abs/2024MNRAS.528...11Y}
}

@ARTICLE{tom96,
       author = {{Tomaney}, Austin B. and {Crotts}, Arlin P.~S.},
        title = "{Expanding the Realm of Microlensing Surveys with Difference Image Photometry}",
      journal = {\aj},
         year = 1996,
        month = dec,
       volume = {112},
        pages = {2872},
          doi = {10.1086/118228},
archivePrefix = {arXiv},
       eprint = {astro-ph/9610066},
 primaryClass = {astro-ph},
       adsurl = {https://ui.adsabs.harvard.edu/abs/1996AJ....112.2872T}
}

@ARTICLE{ala98,
       author = {{Alard}, C. and {Lupton}, Robert H.},
        title = "{A Method for Optimal Image Subtraction}",
      journal = {\apj},
         year = 1998,
        month = aug,
       volume = {503},
       number = {1},
        pages = {325-331},
          doi = {10.1086/305984},
archivePrefix = {arXiv},
       eprint = {astro-ph/9712287},
 primaryClass = {astro-ph},
       adsurl = {https://ui.adsabs.harvard.edu/abs/1998ApJ...503..325A}
}

@ARTICLE{bon01,
       author = {{Bond}, I.~A. and {Abe}, F. and {Dodd}, R.~J. and {Hearnshaw}, J.~B. and {Honda}, M. and {Jugaku}, J. and {Kilmartin}, P.~M. and {Marles}, A. and {Masuda}, K. and {Matsubara}, Y. and {Muraki}, Y. and {Nakamura}, T. and {Nankivell}, G. and {Noda}, S. and {Noguchi}, C. and {Ohnishi}, K. and {Rattenbury}, N.~J. and {Reid}, M. and {Saito}, To. and {Sato}, H. and {Sekiguchi}, M. and {Skuljan}, J. and {Sullivan}, D.~J. and {Sumi}, T. and {Takeuti}, M. and {Watase}, Y. and {Wilkinson}, S. and {Yamada}, R. and {Yanagisawa}, T. and {Yock}, P.~C.~M.},
        title = "{Real-time difference imaging analysis of MOA Galactic bulge observations during 2000}",
      journal = {\mnras},
         year = 2001,
        month = nov,
       volume = {327},
       number = {3},
        pages = {868-880},
          doi = {10.1046/j.1365-8711.2001.04776.x},
archivePrefix = {arXiv},
       eprint = {astro-ph/0102181},
 primaryClass = {astro-ph},
       adsurl = {https://ui.adsabs.harvard.edu/abs/2001MNRAS.327..868B}
}

@ARTICLE{pac86,
       author = {{Paczynski}, B.},
        title = "{Gravitational Microlensing by the Galactic Halo}",
      journal = {\apj},
         year = 1986,
        month = may,
       volume = {304},
        pages = {1},
          doi = {10.1086/164140},
       adsurl = {https://ui.adsabs.harvard.edu/abs/1986ApJ...304....1P}
}

@ARTICLE{wit94,
       author = {{Witt}, Hans J. and {Mao}, Shude},
        title = "{Can Lensed Stars Be Regarded as Pointlike for Microlensing by MACHOs?}",
      journal = {\apj},
         year = 1994,
        month = aug,
       volume = {430},
        pages = {505},
          doi = {10.1086/174426},
       adsurl = {https://ui.adsabs.harvard.edu/abs/1994ApJ...430..505W}
}

@ARTICLE{gou92,
       author = {{Gould}, Andrew},
        title = "{Extending the MACHO Search to approximately 10 6 M sub sun}",
      journal = {\apj},
         year = 1992,
        month = jun,
       volume = {392},
        pages = {442},
          doi = {10.1086/171443},
       adsurl = {https://ui.adsabs.harvard.edu/abs/1992ApJ...392..442G}
}

@ARTICLE{dom98,
       author = {{Dominik}, M.},
        title = "{Galactic microlensing with rotating binaries}",
      journal = {\aap},
         year = 1998,
        month = jan,
       volume = {329},
        pages = {361-374},
          doi = {10.48550/arXiv.astro-ph/9702039},
archivePrefix = {arXiv},
       eprint = {astro-ph/9702039},
 primaryClass = {astro-ph},
       adsurl = {https://ui.adsabs.harvard.edu/abs/1998A&A...329..361D}
}

@ARTICLE{iok99,
       author = {{Ioka}, K. and {Nishi}, R. and {Kan-Ya}, Y.},
        title = "{Kepler Rotation Effects on the Binary-Lens Microlensing Events}",
      journal = {Progress of Theoretical Physics},
         year = 1999,
        month = nov,
       volume = {102},
       number = {5},
        pages = {983-1000},
          doi = {10.1143/PTP.102.983},
archivePrefix = {arXiv},
       eprint = {astro-ph/9910337},
 primaryClass = {astro-ph},
       adsurl = {https://ui.adsabs.harvard.edu/abs/1999PThPh.102..983I}
}

@ARTICLE{sko11,
       author = {{Skowron}, J. and {Udalski}, A. and {Gould}, A. and {Dong}, Subo and {Monard}, L.~A.~G. and {Han}, C. and {Nelson}, C.~R. and {McCormick}, J. and {Moorhouse}, D. and {Thornley}, G. and {Maury}, A. and {Bramich}, D.~M. and {Greenhill}, J. and {Koz{\l}owski}, S. and {Bond}, I. and {Poleski}, R. and {Wyrzykowski}, {\L}. and {Ulaczyk}, K. and {Kubiak}, M. and {Szyma{\'n}ski}, M.~K. and {Pietrzy{\'n}ski}, G. and {Soszy{\'n}ski}, I. and {OGLE Collaboration} and {Gaudi}, B.~S. and {Yee}, J.~C. and {Hung}, L.-W. and {Pogge}, R.~W. and {DePoy}, D.~L. and {Lee}, C.-U. and {Park}, B.-G. and {Allen}, W. and {Mallia}, F. and {Drummond}, J. and {Bolt}, G. and {{\ensuremath{\mu}}FUN Collaboration} and {Allan}, A. and {Browne}, P. and {Clay}, N. and {Dominik}, M. and {Fraser}, S. and {Horne}, K. and {Kains}, N. and {Mottram}, C. and {Snodgrass}, C. and {Steele}, I. and {Street}, R.~A. and {Tsapras}, Y. and {RoboNet Collaboration} and {Abe}, F. and {Bennett}, D.~P. and {Botzler}, C.~S. and {Douchin}, D. and {Freeman}, M. and {Fukui}, A. and {Furusawa}, K. and {Hayashi}, F. and {Hearnshaw}, J.~B. and {Hosaka}, S. and {Itow}, Y. and {Kamiya}, K. and {Kilmartin}, P.~M. and {Korpela}, A. and {Lin}, W. and {Ling}, C.~H. and {Makita}, S. and {Masuda}, K. and {Matsubara}, Y. and {Muraki}, Y. and {Nagayama}, T. and {Miyake}, N. and {Nishimoto}, K. and {Ohnishi}, K. and {Perrott}, Y.~C. and {Rattenbury}, N. and {Saito}, To. and {Skuljan}, L. and {Sullivan}, D.~J. and {Sumi}, T. and {Suzuki}, D. and {Sweatman}, W.~L. and {Tristram}, P.~J. and {Wada}, K. and {Yock}, P.~C.~M. and {MOA Collaboration} and {Beaulieu}, J.-P. and {Fouqu{\'e}}, P. and {Albrow}, M.~D. and {Batista}, V. and {Brillant}, S. and {Caldwell}, J.~A.~R. and {Cassan}, A. and {Cole}, A. and {Cook}, K.~H. and {Coutures}, Ch. and {Dieters}, S. and {Dominis Prester}, D. and {Donatowicz}, J. and {Kane}, S.~R. and {Kubas}, D. and {Marquette}, J.-B. and {Martin}, R. and {Menzies}, J. and {Sahu}, K.~C. and {Wambsganss}, J. and {Williams}, A. and {Zub}, M. and {PLANET Collaboration}},
        title = "{Binary Microlensing Event OGLE-2009-BLG-020 Gives Verifiable Mass, Distance, and Orbit Predictions}",
      journal = {\apj},
         year = 2011,
        month = sep,
       volume = {738},
       number = {1},
          eid = {87},
        pages = {87},
          doi = {10.1088/0004-637X/738/1/87},
archivePrefix = {arXiv},
       eprint = {1101.3312},
 primaryClass = {astro-ph.SR},
       adsurl = {https://ui.adsabs.harvard.edu/abs/2011ApJ...738...87S}
}

@ARTICLE{boz21,
       author = {{Bozza}, Valerio and {Khalouei}, Elahe and {Bachelet}, Etienne},
        title = "{A public code for astrometric microlensing with contour integration}",
      journal = {\mnras},
         year = 2021,
        month = jul,
       volume = {505},
       number = {1},
        pages = {126-135},
          doi = {10.1093/mnras/stab1376},
archivePrefix = {arXiv},
       eprint = {2011.04780},
 primaryClass = {astro-ph.IM},
       adsurl = {https://ui.adsabs.harvard.edu/abs/2021MNRAS.505..126B}
}

@misc{nun26,
  author = {{Nunota}, Kansuke and {Masuda}, Kento},
  title = "{A Bayesian Inference Framework for Binary Lens Events Fully Incorporating Higher-Order Effects}",
  year = {2026},
  howpublished = {Accepted for publication in The Astronomical Journal}
}

@misc{mas26,
      title={On the Reparameterization Between Cartesian Position-Velocity Vectors and Orbital Elements in the Kepler Problem}, 
      author={Kento Masuda and Kansuke Nunota},
      year={2026},
      eprint={2605.12982},
      archivePrefix={arXiv},
      primaryClass={astro-ph.EP},
      url={https://arxiv.org/abs/2605.12982}, 
}

@ARTICLE{yee12,
       author = {{Yee}, J.~C. and {Shvartzvald}, Y. and {Gal-Yam}, A. and {Bond}, I.~A. and {Udalski}, A. and {Koz{\l}owski}, S. and {Han}, C. and {Gould}, A. and {Skowron}, J. and {Suzuki}, D. and {Abe}, F. and {Bennett}, D.~P. and {Botzler}, C.~S. and {Chote}, P. and {Freeman}, M. and {Fukui}, A. and {Furusawa}, K. and {Itow}, Y. and {Kobara}, S. and {Ling}, C.~H. and {Masuda}, K. and {Matsubara}, Y. and {Miyake}, N. and {Muraki}, Y. and {Ohmori}, K. and {Ohnishi}, K. and {Rattenbury}, N.~J. and {Saito}, To. and {Sullivan}, D.~J. and {Sumi}, T. and {Suzuki}, K. and {Sweatman}, W.~L. and {Takino}, S. and {Tristram}, P.~J. and {Wada}, K. and {MOA Collaboration} and {Szyma{\'n}ski}, M.~K. and {Kubiak}, M. and {Pietrzy{\'n}ski}, G. and {Soszy{\'n}ski}, I. and {Poleski}, R. and {Ulaczyk}, K. and {Wyrzykowski}, {\L}. and {Pietrukowicz}, P. and {OGLE Collaboration} and {Allen}, W. and {Almeida}, L.~A. and {Batista}, V. and {Bos}, M. and {Christie}, G. and {DePoy}, D.~L. and {Dong}, Subo and {Drummond}, J. and {Finkelman}, I. and {Gaudi}, B.~S. and {Gorbikov}, E. and {Henderson}, C. and {Higgins}, D. and {Jablonski}, F. and {Kaspi}, S. and {Manulis}, I. and {Maoz}, D. and {McCormick}, J. and {McGregor}, D. and {Monard}, L.~A.~G. and {Moorhouse}, D. and {Mu{\~n}oz}, J.~A. and {Natusch}, T. and {Ngan}, H. and {Ofek}, E. and {Pogge}, R.~W. and {Santallo}, R. and {Tan}, T.-G. and {Thornley}, G. and {Shin}, I.-G. and {Choi}, J.-Y. and {Park}, S.-Y. and {Lee}, C.-U. and {Koo}, J.-R. and {{\ensuremath{\mu}}FUN Collaboration}},
        title = "{MOA-2011-BLG-293Lb: A Test of Pure Survey Microlensing Planet Detections}",
      journal = {\apj},
         year = 2012,
        month = aug,
       volume = {755},
       number = {2},
          eid = {102},
        pages = {102},
          doi = {10.1088/0004-637X/755/2/102},
archivePrefix = {arXiv},
       eprint = {1201.1002},
 primaryClass = {astro-ph.EP},
       adsurl = {https://ui.adsabs.harvard.edu/abs/2012ApJ...755..102Y}
}

@ARTICLE{kos21,
       author = {{Koshimoto}, Naoki and {Baba}, Junichi and {Bennett}, David P.},
        title = "{A Parametric Galactic Model toward the Galactic Bulge Based on Gaia and Microlensing Data}",
      journal = {\apj},
         year = 2021,
        month = aug,
       volume = {917},
       number = {2},
          eid = {78},
        pages = {78},
          doi = {10.3847/1538-4357/ac07a8},
archivePrefix = {arXiv},
       eprint = {2104.03306},
 primaryClass = {astro-ph.GA},
       adsurl = {https://ui.adsabs.harvard.edu/abs/2021ApJ...917...78K}
}

@software{jax18,
    author      = {James Bradbury and Roy Frostig and Peter Hawkins and Matthew James Johnson and Chris Leary and Dougal Maclaurin and George Necula and Adam Paszke and Jake Vander{P}las and Skye Wanderman-{M}ilne and Qiao Zhang},
    title       = {{JAX}: composable transformations of {P}ython+{N}um{P}y programs},
    url         = {http://github.com/google/jax},
    version     = {0.2.5},
    year        = {2018}
}

@article{emcee,
  author       = {Foreman-Mackey, D. and Hogg, D. W. and Lang, D. and Goodman, J.},
  year         = {2013},
  title        = {emcee: The MCMC Hammer},
  journal      = {PASP},
  volume       = {125},
  pages        = {306},
  doi          = {10.1086/670067}
}

@ARTICLE{sum03,
       author = {{Sumi}, T. and {Abe}, F. and {Bond}, I.~A. and {Dodd}, R.~J. and {Hearnshaw}, J.~B. and {Honda}, M. and {Honma}, M. and {Kan-ya}, Y. and {Kilmartin}, P.~M. and {Masuda}, K. and {Matsubara}, Y. and {Muraki}, Y. and {Nakamura}, T. and {Nishi}, R. and {Noda}, S. and {Ohnishi}, K. and {Petterson}, O.~K.~L. and {Rattenbury}, N.~J. and {Reid}, M. and {Saito}, To. and {Saito}, Y. and {Sato}, H. and {Sekiguchi}, M. and {Skuljan}, J. and {Sullivan}, D.~J. and {Takeuti}, M. and {Tristram}, P.~J. and {Wilkinson}, S. and {Yanagisawa}, T. and {Yock}, P.~C.~M.},
        title = "{Microlensing Optical Depth toward the Galactic Bulge from Microlensing Observations in Astrophysics Group Observations during 2000 with Difference Image Analysis}",
      journal = {\apj},
         year = 2003,
        month = jul,
       volume = {591},
       number = {1},
        pages = {204-227},
          doi = {10.1086/375212},
archivePrefix = {arXiv},
       eprint = {astro-ph/0207604},
 primaryClass = {astro-ph},
       adsurl = {https://ui.adsabs.harvard.edu/abs/2003ApJ...591..204S}
}

@ARTICLE{uda15,
       author = {{Udalski}, A. and {Szyma{\'n}ski}, M.~K. and {Szyma{\'n}ski}, G.},
        title = "{OGLE-IV: Fourth Phase of the Optical Gravitational Lensing Experiment}",
      journal = {\actaa},
         year = 2015,
        month = mar,
       volume = {65},
       number = {1},
        pages = {1-38},
          doi = {10.48550/arXiv.1504.05966},
archivePrefix = {arXiv},
       eprint = {1504.05966},
 primaryClass = {astro-ph.SR},
       adsurl = {https://ui.adsabs.harvard.edu/abs/2015AcA....65....1U}
}

@ARTICLE{mao91,
       author = {{Mao}, Shude and {Paczynski}, Bohdan},
        title = "{Gravitational Microlensing by Double Stars and Planetary Systems}",
      journal = {\apjl},
         year = 1991,
        month = jun,
       volume = {374},
        pages = {L37},
          doi = {10.1086/186066},
       adsurl = {https://ui.adsabs.harvard.edu/abs/1991ApJ...374L..37M}
}

@ARTICLE{bon04,
       author = {{Bond}, I.~A. and {Udalski}, A. and {Jaroszy{\'n}ski}, M. and {Rattenbury}, N.~J. and {Paczy{\'n}ski}, B. and {Soszy{\'n}ski}, I. and {Wyrzykowski}, L. and {Szyma{\'n}ski}, M.~K. and {Kubiak}, M. and {Szewczyk}, O. and {{\.Z}ebru{\'n}}, K. and {Pietrzy{\'n}ski}, G. and {Abe}, F. and {Bennett}, D.~P. and {Eguchi}, S. and {Furuta}, Y. and {Hearnshaw}, J.~B. and {Kamiya}, K. and {Kilmartin}, P.~M. and {Kurata}, Y. and {Masuda}, K. and {Matsubara}, Y. and {Muraki}, Y. and {Noda}, S. and {Okajima}, K. and {Sako}, T. and {Sekiguchi}, T. and {Sullivan}, D.~J. and {Sumi}, T. and {Tristram}, P.~J. and {Yanagisawa}, T. and {Yock}, P.~C.~M. and {OGLE Collaboration}},
        title = "{OGLE 2003-BLG-235/MOA 2003-BLG-53: A Planetary Microlensing Event}",
      journal = {\apjl},
         year = 2004,
        month = may,
       volume = {606},
       number = {2},
        pages = {L155-L158},
          doi = {10.1086/420928},
archivePrefix = {arXiv},
       eprint = {astro-ph/0404309},
 primaryClass = {astro-ph},
       adsurl = {https://ui.adsabs.harvard.edu/abs/2004ApJ...606L.155B}
}

@ARTICLE{gou22,
       author = {{Gould}, Andrew and {Jung}, Youn Kil and {Hwang}, Kyu-Ha and {Dong}, Subo and {Albrow}, Michael D. and {Chung}, Sun-Ju and {Han}, Cheongho and {Ryu}, Yoon-Hyun and {Shin}, In-Gu and {Shvartzvald}, Yossi and {Yang}, Hongjing and {Yee}, Jennifer C. and {Zang}, Weicheng and {Cha}, Sang-Mok and {Kim}, Dong-Jin and {Kim}, Seung-Lee and {Lee}, Chung-Uk and {Lee}, Dong-Joo and {Lee}, Yongseok and {Park}, Byeong-Gon and {Pogge}, Richard W.},
        title = "{Free-Floating Planets, the Einstein Desert, and 'OUMUAMUA}",
      journal = {Journal of Korean Astronomical Society},
         year = 2022,
        month = oct,
       volume = {55},
        pages = {173-194},
          doi = {10.5303/JKAS.2022.55.5.173},
archivePrefix = {arXiv},
       eprint = {2204.03269},
 primaryClass = {astro-ph.EP},
       adsurl = {https://ui.adsabs.harvard.edu/abs/2022JKAS...55..173G}
}

@ARTICLE{sum23,
       author = {{Sumi}, Takahiro and {Koshimoto}, Naoki and {Bennett}, David P. and {Rattenbury}, Nicholas J. and {Abe}, Fumio and {Barry}, Richard and {Bhattacharya}, Aparna and {Bond}, Ian A. and {Fujii}, Hirosane and {Fukui}, Akihiko and {Hamada}, Ryusei and {Hirao}, Yuki and {Silva}, Stela Ishitani and {Itow}, Yoshitaka and {Kirikawa}, Rintaro and {Kondo}, Iona and {Matsubara}, Yutaka and {Miyazaki}, Shota and {Muraki}, Yasushi and {Olmschenk}, Greg and {Ranc}, Cl{\'e}ment and {Satoh}, Yuki and {Suzuki}, Daisuke and {Tomoyoshi}, Mio and {Tristram}, Paul. J. and {Vandorou}, Aikaterini and {Yama}, Hibiki and {Yamashita}, Kansuke},
        title = "{Free-floating Planet Mass Function from MOA-II 9 yr Survey toward the Galactic Bulge}",
      journal = {\aj},
         year = 2023,
        month = sep,
       volume = {166},
       number = {3},
          eid = {108},
        pages = {108},
          doi = {10.3847/1538-3881/ace688},
archivePrefix = {arXiv},
       eprint = {2303.08280},
 primaryClass = {astro-ph.EP},
       adsurl = {https://ui.adsabs.harvard.edu/abs/2023AJ....166..108S}
}

@ARTICLE{kos23,
       author = {{Koshimoto}, Naoki and {Sumi}, Takahiro and {Bennett}, David P. and {Bozza}, Valerio and {Mr{\'o}z}, Przemek and {Udalski}, Andrzej and {Rattenbury}, Nicholas J. and {Abe}, Fumio and {Barry}, Richard and {Bhattacharya}, Aparna and {Bond}, Ian A. and {Fujii}, Hirosane and {Fukui}, Akihiko and {Hamada}, Ryusei and {Hirao}, Yuki and {Silva}, Stela Ishitani and {Itow}, Yoshitaka and {Kirikawa}, Rintaro and {Kondo}, Iona and {Matsubara}, Yutaka and {Miyazaki}, Shota and {Muraki}, Yasushi and {Olmschenk}, Greg and {Ranc}, Cl{\'e}ment and {Satoh}, Yuki and {Suzuki}, Daisuke and {Tomoyoshi}, Mio and {Tristram}, Paul J. and {Vandorou}, Aikaterini and {Yama}, Hibiki and {Yamashita}, Kansuke},
        title = "{Terrestrial- and Neptune-mass Free-Floating Planet Candidates from the MOA-II 9 yr Galactic Bulge Survey}",
      journal = {\aj},
         year = 2023,
        month = sep,
       volume = {166},
       number = {3},
          eid = {107},
        pages = {107},
          doi = {10.3847/1538-3881/ace689},
archivePrefix = {arXiv},
       eprint = {2303.08279},
 primaryClass = {astro-ph.EP},
       adsurl = {https://ui.adsabs.harvard.edu/abs/2023AJ....166..107K}
}

@ARTICLE{mro17,
       author = {{Mr{\'o}z}, Przemek and {Udalski}, Andrzej and {Skowron}, Jan and {Poleski}, Rados{\l}aw and {Koz{\l}owski}, Szymon and {Szyma{\'n}ski}, Micha{\l} K. and {Soszy{\'n}ski}, Igor and {Wyrzykowski}, {\L}ukasz and {Pietrukowicz}, Pawe{\l} and {Ulaczyk}, Krzysztof and {Skowron}, Dorota and {Pawlak}, Micha{\l}},
        title = "{No large population of unbound or wide-orbit Jupiter-mass planets}",
      journal = {\nat},
         year = 2017,
        month = aug,
       volume = {548},
       number = {7666},
        pages = {183-186},
          doi = {10.1038/nature23276},
archivePrefix = {arXiv},
       eprint = {1707.07634},
 primaryClass = {astro-ph.EP},
       adsurl = {https://ui.adsabs.harvard.edu/abs/2017Natur.548..183M}
}

@ARTICLE{sah22,
       author = {{Sahu}, Kailash C. and {Anderson}, Jay and {Casertano}, Stefano and {Bond}, Howard E. and {Udalski}, Andrzej and {Dominik}, Martin and {Calamida}, Annalisa and {Bellini}, Andrea and {Brown}, Thomas M. and {Rejkuba}, Marina and {Bajaj}, Varun and {Kains}, No{\'e} and {Ferguson}, Henry C. and {Fryer}, Chris L. and {Yock}, Philip and {Mr{\'o}z}, Przemek and {Koz{\l}owski}, Szymon and {Pietrukowicz}, Pawe{\l} and {Poleski}, Radek and {Skowron}, Jan and {Soszy{\'n}ski}, Igor and {Szyma{\'n}ski}, Micha{\l} K. and {Ulaczyk}, Krzysztof and {Wyrzykowski}, {\L}ukasz and {Barry}, Richard K. and {Bennett}, David P. and {Bond}, Ian A. and {Hirao}, Yuki and {Silva}, Stela Ishitani and {Kondo}, Iona and {Koshimoto}, Naoki and {Ranc}, Cl{\'e}ment and {Rattenbury}, Nicholas J. and {Sumi}, Takahiro and {Suzuki}, Daisuke and {Tristram}, Paul J. and {Vandorou}, Aikaterini and {Beaulieu}, Jean-Philippe and {Marquette}, Jean-Baptiste and {Cole}, Andrew and {Fouqu{\'e}}, Pascal and {Hill}, Kym and {Dieters}, Stefan and {Coutures}, Christian and {Dominis-Prester}, Dijana and {Bennett}, Clara and {Bachelet}, Etienne and {Menzies}, John and {Albrow}, Michael and {Pollard}, Karen and {Gould}, Andrew and {Yee}, Jennifer C. and {Allen}, William and {Almeida}, Leonardo A. and {Christie}, Grant and {Drummond}, John and {Gal-Yam}, Avishay and {Gorbikov}, Evgeny and {Jablonski}, Francisco and {Lee}, Chung-Uk and {Maoz}, Dan and {Manulis}, Ilan and {McCormick}, Jennie and {Natusch}, Tim and {Pogge}, Richard W. and {Shvartzvald}, Yossi and {J{\o}rgensen}, Uffe G. and {Alsubai}, Khalid A. and {Andersen}, Michael I. and {Bozza}, Valerio and {Novati}, Sebastiano Calchi and {Burgdorf}, Martin and {Hinse}, Tobias C. and {Hundertmark}, Markus and {Husser}, Tim-Oliver and {Kerins}, Eamonn and {Longa-Pe{\~n}a}, Penelope and {Mancini}, Luigi and {Penny}, Matthew and {Rahvar}, Sohrab and {Ricci}, Davide and {Sajadian}, Sedighe and {Skottfelt}, Jesper and {Snodgrass}, Colin and {Southworth}, John and {Tregloan-Reed}, Jeremy and {Wambsganss}, Joachim and {Wertz}, Olivier and {Tsapras}, Yiannis and {Street}, Rachel A. and {Bramich}, D.~M. and {Horne}, Keith and {Steele}, Iain A. and {RoboNet Collaboration}},
        title = "{An Isolated Stellar-mass Black Hole Detected through Astrometric Microlensing}",
      journal = {\apj},
         year = 2022,
        month = jul,
       volume = {933},
       number = {1},
          eid = {83},
        pages = {83},
          doi = {10.3847/1538-4357/ac739e},
archivePrefix = {arXiv},
       eprint = {2201.13296},
 primaryClass = {astro-ph.SR},
       adsurl = {https://ui.adsabs.harvard.edu/abs/2022ApJ...933...83S}
}

@ARTICLE{lam22,
       author = {{Lam}, Casey Y. and {Lu}, Jessica R. and {Udalski}, Andrzej and {Bond}, Ian and {Bennett}, David P. and {Skowron}, Jan and {Mr{\'o}z}, Przemek and {Poleski}, Radek and {Sumi}, Takahiro and {Szyma{\'n}ski}, Micha{\l} K. and {Koz{\l}owski}, Szymon and {Pietrukowicz}, Pawe{\l} and {Soszy{\'n}ski}, Igor and {Ulaczyk}, Krzysztof and {Wyrzykowski}, {\L}ukasz and {Miyazaki}, Shota and {Suzuki}, Daisuke and {Koshimoto}, Naoki and {Rattenbury}, Nicholas J. and {Hosek}, Matthew W. and {Abe}, Fumio and {Barry}, Richard and {Bhattacharya}, Aparna and {Fukui}, Akihiko and {Fujii}, Hirosane and {Hirao}, Yuki and {Itow}, Yoshitaka and {Kirikawa}, Rintaro and {Kondo}, Iona and {Matsubara}, Yutaka and {Matsumoto}, Sho and {Muraki}, Yasushi and {Olmschenk}, Greg and {Ranc}, Cl{\'e}ment and {Okamura}, Arisa and {Satoh}, Yuki and {Silva}, Stela Ishitani and {Toda}, Taiga and {Tristram}, Paul J. and {Vandorou}, Aikaterini and {Yama}, Hibiki and {Abrams}, Natasha S. and {Agarwal}, Shrihan and {Rose}, Sam and {Terry}, Sean K.},
        title = "{An Isolated Mass-gap Black Hole or Neutron Star Detected with Astrometric Microlensing}",
      journal = {\apjl},
         year = 2022,
        month = jul,
       volume = {933},
       number = {1},
          eid = {L23},
        pages = {L23},
          doi = {10.3847/2041-8213/ac7442},
archivePrefix = {arXiv},
       eprint = {2202.01903},
 primaryClass = {astro-ph.GA},
       adsurl = {https://ui.adsabs.harvard.edu/abs/2022ApJ...933L..23L}
}

@ARTICLE{mro19,
       author = {{Mr{\'o}z}, Przemek and {Udalski}, Andrzej and {Skowron}, Jan and {Szyma{\'n}ski}, Micha{\l} K. and {Soszy{\'n}ski}, Igor and {Wyrzykowski}, {\L}ukasz and {Pietrukowicz}, Pawe{\l} and {Koz{\l}owski}, Szymon and {Poleski}, Rados{\l}aw and {Ulaczyk}, Krzysztof and {Rybicki}, Krzysztof and {Iwanek}, Patryk},
        title = "{Microlensing Optical Depth and Event Rate toward the Galactic Bulge from 8 yr of OGLE-IV Observations}",
      journal = {\apjs},
         year = 2019,
        month = oct,
       volume = {244},
       number = {2},
          eid = {29},
        pages = {29},
          doi = {10.3847/1538-4365/ab426b},
archivePrefix = {arXiv},
       eprint = {1906.02210},
 primaryClass = {astro-ph.SR},
       adsurl = {https://ui.adsabs.harvard.edu/abs/2019ApJS..244...29M}
}

@ARTICLE{yoo04,
       author = {{Yoo}, Jaiyul and {DePoy}, D.~L. and {Gal-Yam}, A. and {Gaudi}, B.~S. and {Gould}, A. and {Han}, C. and {Lipkin}, Y. and {Maoz}, D. and {Ofek}, E.~O. and {Park}, B.-G. and {Pogge}, R.~W. and {Mu-Fun Collaboration} and {Udalski}, A. and {Soszy{\'n}ski}, I. and {Wyrzykowski}, {\L}. and {Kubiak}, M. and {Szyma{\'n}ski}, M. and {Pietrzy{\'n}ski}, G. and {Szewczyk}, O. and {{\.Z}ebru{\'n}}, K. and {OGLE Collaboration}},
        title = "{OGLE-2003-BLG-262: Finite-Source Effects from a Point-Mass Lens}",
      journal = {\apj},
         year = 2004,
        month = mar,
       volume = {603},
       number = {1},
        pages = {139-151},
          doi = {10.1086/381241},
archivePrefix = {arXiv},
       eprint = {astro-ph/0309302},
 primaryClass = {astro-ph},
       adsurl = {https://ui.adsabs.harvard.edu/abs/2004ApJ...603..139Y}
}

@ARTICLE{ben08,
       author = {{Bennett}, D.~P. and {Bond}, I.~A. and {Udalski}, A. and {Sumi}, T. and {Abe}, F. and {Fukui}, A. and {Furusawa}, K. and {Hearnshaw}, J.~B. and {Holderness}, S. and {Itow}, Y. and {Kamiya}, K. and {Korpela}, A.~V. and {Kilmartin}, P.~M. and {Lin}, W. and {Ling}, C.~H. and {Masuda}, K. and {Matsubara}, Y. and {Miyake}, N. and {Muraki}, Y. and {Nagaya}, M. and {Okumura}, T. and {Ohnishi}, K. and {Perrott}, Y.~C. and {Rattenbury}, N.~J. and {Sako}, T. and {Saito}, To. and {Sato}, S. and {Skuljan}, L. and {Sullivan}, D.~J. and {Sweatman}, W.~L. and {Tristram}, P.~J. and {Yock}, P.~C.~M. and {Kubiak}, M. and {Szyma{\'n}ski}, M.~K. and {Pietrzy{\'n}ski}, G. and {Soszy{\'n}ski}, I. and {Szewczyk}, O. and {Wyrzykowski}, {\L}. and {Ulaczyk}, K. and {Batista}, V. and {Beaulieu}, J.~P. and {Brillant}, S. and {Cassan}, A. and {Fouqu{\'e}}, P. and {Kervella}, P. and {Kubas}, D. and {Marquette}, J.~B.},
        title = "{A Low-Mass Planet with a Possible Sub-Stellar-Mass Host in Microlensing Event MOA-2007-BLG-192}",
      journal = {\apj},
         year = 2008,
        month = sep,
       volume = {684},
       number = {1},
        pages = {663-683},
          doi = {10.1086/589940},
archivePrefix = {arXiv},
       eprint = {0806.0025},
 primaryClass = {astro-ph},
       adsurl = {https://ui.adsabs.harvard.edu/abs/2008ApJ...684..663B}
}

@ARTICLE{nun25gal,
       author = {{Nunota}, Kansuke and {Sumi}, Takahiro and {Koshimoto}, Naoki and {Rattenbury}, Nicholas J. and {Abe}, Fumio and {Barry}, Richard and {Bennett}, David P. and {Bhattacharya}, Aparna and {Fukui}, Akihiko and {Hamada}, Ryusei and {Hamada}, Shunya and {Hamasaki}, Naoto and {Hirao}, Yuki and {Ishitani Silva}, Stela and {Itow}, Yoshitaka and {Matsubara}, Yutaka and {Miyazaki}, Shota and {Muraki}, Yasushi and {Nagai}, Tsutsumi and {Olmschenk}, Greg and {Ranc}, Clement and {Satoh}, Yuki K. and {Suzuki}, Daisuke and {Tristram}, Paul. J. and {Vandorou}, Aikaterini and {Yama}, Hibiki and {MOA Collaboration}},
        title = "{The Microlensing Event Rate and Optical Depth from MOA-II 9 Yr Survey Toward the Galactic Bulge}",
      journal = {\apj},
         year = 2025,
        month = feb,
       volume = {979},
       number = {2},
          eid = {123},
        pages = {123},
          doi = {10.3847/1538-4357/ada352},
archivePrefix = {arXiv},
       eprint = {2410.23553},
 primaryClass = {astro-ph.GA},
       adsurl = {https://ui.adsabs.harvard.edu/abs/2025ApJ...979..123N}
}

@ARTICLE{bes98,
       author = {{Bessell}, M.~S. and {Brett}, J.~M.},
        title = "{JHKLM Photometry: Standard Systems, Passbands, and Intrinsic Colors}",
      journal = {\pasp},
         year = 1988,
        month = sep,
       volume = {100},
        pages = {1134},
          doi = {10.1086/132281},
       adsurl = {https://ui.adsabs.harvard.edu/abs/1988PASP..100.1134B}
}

@ARTICLE{smi25,
       author = {{Smith}, Leigh C. and {Lucas}, Philip W. and {Koposov}, Sergey E. and {Gonzalez-Fernandez}, Carlos and {Alonso-Garc{\'\i}a}, Javier and {Minniti}, Dante and {Sanders}, Jason L. and {Bedin}, Luigi R. and {Belokurov}, Vasily and {Evans}, N. Wyn and {Hempel}, Maren and {Ivanov}, Valentin D. and {Kurtev}, Radostin G. and {Saito}, Roberto K.},
        title = "{VIRAC2: NIR astrometry and time series photometry for 500M+ stars from the VVV and VVVX surveys}",
      journal = {\mnras},
         year = 2025,
        month = feb,
       volume = {536},
       number = {4},
        pages = {3707-3738},
          doi = {10.1093/mnras/stae2797},
archivePrefix = {arXiv},
       eprint = {2501.06295},
 primaryClass = {astro-ph.GA},
       adsurl = {https://ui.adsabs.harvard.edu/abs/2025MNRAS.536.3707S}
}

@ARTICLE{gou00,
       author = {{Gould}, Andrew},
        title = "{A Natural Formalism for Microlensing}",
      journal = {\apj},
         year = 2000,
        month = oct,
       volume = {542},
       number = {2},
        pages = {785-788},
          doi = {10.1086/317037},
archivePrefix = {arXiv},
       eprint = {astro-ph/0001421},
 primaryClass = {astro-ph},
       adsurl = {https://ui.adsabs.harvard.edu/abs/2000ApJ...542..785G}
}

@ARTICLE{gou04,
       author = {{Gould}, Andrew},
        title = "{Resolution of the MACHO-LMC-5 Puzzle: The Jerk-Parallax Microlens Degeneracy}",
      journal = {\apj},
         year = 2004,
        month = may,
       volume = {606},
       number = {1},
        pages = {319-325},
          doi = {10.1086/382782},
archivePrefix = {arXiv},
       eprint = {astro-ph/0311548},
 primaryClass = {astro-ph},
       adsurl = {https://ui.adsabs.harvard.edu/abs/2004ApJ...606..319G}
}

@ARTICLE{cho16,
       author = {{Choi}, Jieun and {Dotter}, Aaron and {Conroy}, Charlie and {Cantiello}, Matteo and {Paxton}, Bill and {Johnson}, Benjamin D.},
        title = "{Mesa Isochrones and Stellar Tracks (MIST). I. Solar-scaled Models}",
      journal = {\apj},
         year = 2016,
        month = jun,
       volume = {823},
       number = {2},
          eid = {102},
        pages = {102},
          doi = {10.3847/0004-637X/823/2/102},
archivePrefix = {arXiv},
       eprint = {1604.08592},
 primaryClass = {astro-ph.SR},
       adsurl = {https://ui.adsabs.harvard.edu/abs/2016ApJ...823..102C}
}

@ARTICLE{dot16,
       author = {{Dotter}, Aaron},
        title = "{MESA Isochrones and Stellar Tracks (MIST) 0: Methods for the Construction of Stellar Isochrones}",
      journal = {\apjs},
         year = 2016,
        month = jan,
       volume = {222},
       number = {1},
          eid = {8},
        pages = {8},
          doi = {10.3847/0067-0049/222/1/8},
archivePrefix = {arXiv},
       eprint = {1601.05144},
 primaryClass = {astro-ph.SR},
       adsurl = {https://ui.adsabs.harvard.edu/abs/2016ApJS..222....8D}
}

@ARTICLE{yan22,
       author = {{Yang}, Hongjing and {Zang}, Weicheng and {Gould}, Andrew and {Yee}, Jennifer C. and {Hwang}, Kyu-Ha and {Christie}, Grant and {Sumi}, Takahiro and {Zhang}, Jiyuan and {Mao}, Shude and {Albrow}, Michael D. and {Chung}, Sun-Ju and {Han}, Cheongho and {Jung}, Youn Kil and {Ryu}, Yoon-Hyun and {Shin}, In-Gu and {Shvartzvald}, Yossi and {Cha}, Sang-Mok and {Kim}, Dong-Jin and {Kim}, Hyoun-Woo and {Kim}, Seung-Lee and {Lee}, Chung-Uk and {Lee}, Dong-Joo and {Lee}, Yongseok and {Park}, Byeong-Gon and {Pogge}, Richard W. and {Drummond}, John and {Maoz}, Dan and {McCormick}, Jennie and {Natusch}, Tim and {Penny}, Matthew T. and {Zhu}, Wei and {Bond}, Ian A. and {Abe}, Fumio and {Barry}, Richard and {Bennett}, David P. and {Bhattacharya}, Aparna and {Donachie}, Martin and {Fujii}, Hirosane and {Fukui}, Akihiko and {Hirao}, Yuki and {Itow}, Yoshitaka and {Kirikawa}, Rintaro and {Kondo}, Iona and {Koshimoto}, Naoki and {Li}, Man Cheung Alex and {Matsubara}, Yutaka and {Muraki}, Yasushi and {Miyazaki}, Shota and {Olmschenk}, Greg and {Ranc}, Cl{\'e}ment and {Rattenbury}, Nicholas J. and {Satoh}, Yuki and {Shoji}, Hikaru and {Silva}, Stela Ishitani and {Suzuki}, Daisuke and {Tanaka}, Yuzuru and {Tristram}, Paul J. and {Yamawaki}, Tsubasa and {Yonehara}, Atsunori and {MOA Collaboration}},
        title = "{KMT-2021-BLG-0171Lb and KMT-2021-BLG-1689Lb: two microlensing planets in the KMTNet high-cadence fields with followup observations}",
      journal = {\mnras},
         year = 2022,
        month = oct,
       volume = {516},
       number = {2},
        pages = {1894-1909},
          doi = {10.1093/mnras/stac2023},
archivePrefix = {arXiv},
       eprint = {2205.12584},
 primaryClass = {astro-ph.EP},
       adsurl = {https://ui.adsabs.harvard.edu/abs/2022MNRAS.516.1894Y}
}

@ARTICLE{miy25,
       author = {{Miyazaki}, Shota and {Kawahara}, Hajime},
        title = "{microJAX: A Differentiable Framework for Microlensing Modeling with GPU-Accelerated Image-Centered Ray Shooting}",
      journal = {\apj},
         year = 2025,
        month = dec,
       volume = {994},
       number = {2},
          eid = {144},
        pages = {144},
          doi = {10.3847/1538-4357/ae1005},
archivePrefix = {arXiv},
       eprint = {2510.02639},
 primaryClass = {astro-ph.EP},
       adsurl = {https://ui.adsabs.harvard.edu/abs/2025ApJ...994..144M}
}

\end{document}